\documentclass[11pt]{article}

\usepackage{jheppub}
\usepackage[usenames,dvipsnames,table]{xcolor}
\usepackage{graphicx,amsmath,amssymb,amsthm,multirow,array,bm,bbm,esint,slashed,hyperref}
\usepackage{xfrac,faktor}
\usepackage[mathscr]{eucal}
\usepackage[bbgreekl]{mathbbol}
\usepackage{subfig}
\usepackage{comment}
\usepackage{soul}

\newcommand\beq{\begin{equation}}
\newcommand\eeq{\end{equation}}
\newcommand{\nn}{\nonumber}
\newcommand{\sech}{\text{sech}}

\title{Low-dimensional de Sitter quantum gravity}

\author[a]{Jordan Cotler,}
\author[b]{Kristan Jensen,}
\author[c]{and Alexander Maloney}
\affiliation[a]{Stanford Institute for Theoretical Physics, Stanford University, Stanford, CA 94305, USA}
\affiliation[b]{Department of Physics and Astronomy, San Francisco State University, San Francisco, CA 94132, USA}
\affiliation[c]{Department of Physics, McGill University, Montr\'eal, Quebec H3A 2T8, Canada}

\emailAdd{jcotler@stanford.edu}
\emailAdd{kristanj@sfsu.edu}
\emailAdd{alex.maloney@mcgill.ca}

\preprint{\today}

\abstract{We study aspects of Jackiw-Teitelboim (JT) quantum gravity in two-dimensional nearly de Sitter (dS) spacetime, as well as pure de Sitter quantum gravity in three dimensions.
These are each theories of boundary modes, which include a reparameterization field on each connected component of the boundary as well as topological degrees of freedom. In two dimensions, the boundary theory is closely related to the Schwarzian path integral, and in three dimensions to the quantization of coadjoint orbits of the Virasoro group. Using these boundary theories we compute loop corrections to the wavefunction of the universe, and investigate gravitational contributions to scattering.

Along the way, we show that JT gravity in dS$_2$ is an analytic continuation of JT gravity in Euclidean AdS$_2$, and that pure gravity in dS$_3$ is a continuation of pure gravity in Euclidean AdS$_3$.  We define a genus expansion for de Sitter JT gravity by summing over higher genus generalizations of surfaces used in the Hartle-Hawking construction. Assuming a conjecture regarding the volumes of moduli spaces of such surfaces, we find that the de Sitter genus expansion is the continuation of the recently discovered AdS genus expansion. Then both may be understood as coming from the genus expansion of the same double-scaled matrix model, which would provide a non-perturbative completion of de Sitter JT gravity.}

\begin{document}
\maketitle
%***********************************************

%***********************************************

%***********************************************
\section{Introduction}
%***********************************************

Our goal is to study simple models of quantum gravity with a positive cosmological constant, where some of the deep questions about quantum cosmology can be addressed in a simplified -- and potentially solvable -- setting.  We will focus on two theories, in 1+1 and 2+1 dimensions, respectively, where it is possible to make concrete progress on problems which are intractable in higher dimensions. The first theory is Jackiw-Teitelboim gravity with a positive cosmological constant, which describes a metric coupled to a scalar dilaton~\cite{jackiw1985lower, teitelboim1983gravitation, jackiw1992gauge}. The second is pure Einstein gravity in three dimensions with a positive cosmological constant.   

These two theories share several features in common which make them ripe targets for an attack on the difficult problems of quantum cosmology.  The first is that they possess no local degrees of freedom: in both cases the dynamics can be completely reduced to a theory of constant positive curvature metrics.  
But nevertheless interesting non-local degrees of freedom remain. In particular, we will study these theories in de Sitter space, which possess asymptotic boundaries in the far future and the far past. There are boundary modes living on each asymptotic boundary, and the resulting boundary dynamics can be regarded as the gravitational analogue of the edge dynamics which appear in Chern-Simons description of quantum Hall systems.  In addition, these theories possess global degrees of freedom associated with the moduli space of constant curvature metrics on spacetimes of non-trivial topology.  We will discuss these dynamics both in Lorentzian signature as well as in Euclidean signature, where (with a certain analytic continuation prescription) they can be related to dynamics on the moduli space of metrics with constant negative curvature.

The topological nature of low dimensional gravity leads to a second important feature -- they can both be formulated in terms of topological gauge theories.  Roughly speaking, the gauge theory formulation arises when one thinks of a theory of gravity in a ``first order" formulation, where the metric and connection coefficients are taken to be independent degrees of freedom, as opposed to the more traditional ``second order" formulation where the connection is defined in terms of the metric.  In a first order formulation, JT gravity can be formulated as a $BF$ theory with $PSL(2,\mathbb{R})$ gauge group, and three dimensional gravity can be formulated as a Chern-Simons theory with $PSL(2;\mathbb{C})$ gauge group.  In both cases the equations of motion imply that the gauge connection is flat, and the resulting topological gauge theories therefore have no local dynamics.  The boundary degrees of freedom can then be formulated in terms of the dynamics of large gauge transformations.  The degrees of freedom associated with spacetime topology parameterize the moduli space of flat connections.   This provides a useful route into the quantization of these theories, and one which we will exploit.

It is important, however, to understand that the gauge theory formulation of a theory of gravity can only be taken so far.  In particular, while the equations of motion of the gravitational theory match those of a gauge theory, this does not mean that the two theories are the same at the quantum mechanical level.  This is particularly clear in the path integral formulation, where the quantum mechanical definition of a theory requires a choice of integration domain in the space of field configurations.  In a gauge theory, the path integral is typically formulated as an integral over gauge connections on a manifold of fixed topology.  This is not, however, what is typically meant by a gravitational path integral, where the topology of spacetime is allowed to fluctuate.  The result is that, while the gauge theory and metric formulations agree classically -- and indeed it is expected that they agree at all orders in perturbation theory -- they should not be taken to be the same thing at the non-perturbative level.  Our point of view, therefore, is that a gauge theory formulation is a useful tool in attempting to construct a full quantum mechanical theory, but does not provide a definition of one.  Rather, we will use gauge theory as a tool in our study of perturbation theory, but include a sum over topologies as instructed by the metric formulation.

One of our goals is to compute the wavefunction of the universe in these simple theories.  Our approach will be motivated by Hartle and Hawking \cite{hartle1983wave}, who argued that the wavefunction of the universe can be computed using a path integral. Lorentzian signature path integrals naturally compute time evolution operators which evolve between states, but do not in general specify a specific state without some additional prescription, such as the imposition of boundary conditions at some point in the past. A central observation of Hartle and Hawking is that there is a natural state which is prepared by a path integral in Euclidean, rather than Lorentzian signature:
\begin{equation}\label{hh}
\Psi[h] \propto \int_{g_{\partial M}=h} [Dg]  \,e^{-S_E}\,.
\end{equation}
In this expression the wavefunction, regarded as function of some data $h$ defined on a particular spacelike slice, is computed by performing an integral over Euclidean metrics $g$ which match with $h$ on this slice.  The essential point is that, whereas in Lorentzian signature there is no natural boundary condition to impose in the far past which singles out a preferred state, in Euclidean signature there is a clear natural prescription: one integrates over metrics and field configurations which smoothly ``cap off" the geometry.  This naturally generalizes to matter configurations as well; one simply integrates over smooth field configurations on the Euclidean geometry. In this picture one specifies a state via a choice of contour of integration through the space of (appropriately complexified) metrics.

We then need to understand which metrics contribute to the Euclidean path integral \cite{hartle1983wave}, and how this contour of integration can be constructed.  In the original approach of Hartle and Hawking, one starts with the de Sitter metric
$$
	{ds^2\over L_{\text{dS}}^2} = 
-dt^2 +\cosh^2 t \,d\Omega_d^2\,,
$$
and makes the complex change of coordinates $\tau=it$ to obtain the sphere metric
$$
	{ds^2 \over {L_{\text{dS}}^2}} = d\tau^2 + \cos^2 \tau \,d\Omega_d^2\,.
$$
This Euclidean geometry naturally matches on to the Lorentzian de Sitter geometry at the moment of time symmetry: $t=0$, so this Euclidean continuation can be used to compute the wavefunction on this slice.

This is not the only possibility, however.  We can also consider the approach inspired by Maldacena \cite{Maldacena:2010un}, where one analytically continues $L_{\text{dS}}$ as well.  Taking $L_{\text{dS}} =- i L_{\text{AdS}}$ and $r=t+i\pi/2$, the de Sitter metric becomes the metric in Euclidean Anti-de Sitter (AdS) space (i.e. the hyperboloid):
$$
{ds^2 \over {L_{\text{AdS}}}} = dr^2 + \sinh^2 r \, d\Omega_d^2
$$
This metric naturally matches on to the Lorentzian de Sitter geometry at $t\to \infty$, rather than at $t=0$.  Thus this naturally computes the wavefunction at future infinity ${\cal I}^+$, which coincides with the boundary $r\to \infty$ of Euclidean AdS space.  In this sense the wavefunction of the universe should coincide with the partition function of a CFT dual to Euclidean AdS, provided one can sensibly analytically continue $L_{\text{AdS}} \to i L_{\text{dS}}$.  In this formulation, one should interpret the dS/CFT correspondence \cite{Strominger:2001pn} as the conjecture that the wavefunction of an asymptotically future de Sitter space universe is the partition function of a suitably defined Euclidean CFT.\footnote{See also \cite{Witten:2001kn, Alishahiha:2004md, mcfadden2010holography, harlow2011operator, anninos2016cosmic, Castro:2011xb, Castro:2012gc,  hertog2012holographic, banerjee2013topology, anninos2013wave, anninos2014higher, anninos2018infrared, anninos2018sitter} for further comments related to this interpretation of the dS/CFT correspondence, as well as related work.}
The CFT under consideration, however, might be quite strange: for example, in a two-dimensional CFT the continuation $L_{\text{AdS}} \to i L_{\text{dS}}$ amounts to taking the central charge $c = {3L_{\text{AdS}} \over 2G} $ to be pure imaginary.  Similar oddities appear in other dimensions \cite{Anninos:2011ui}.

At first sight the two analytic continuation prescriptions described above appear to be quite different.  Our emphasis in this paper, however, is that in certain contexts they are identical: Maldacena's contour is {\it exactly} the Hartle-Hawking prescription.  The essential distinction is whether one considers a theory of gravity in first or second order formalism.  The original Hartle-Hawking prescription describes a theory of gravity in the second order formalism, where the metric is consider as the only dynamical variable.  In a first order formulation, however, the analytic continuation $\tau=it$ automatically takes us to Euclidean AdS, rather than dS.  This can be seen very easily in the gauge theory formulation described above.
We just need to observe that the gauge group of the gauge theory is the isometry group of Lorentzian dS$_d$ (i.e. $SO(d,1)$), which is the same as the isometry group of Euclidean AdS$_d$.  Thus, for example, in the formulation of three dimensional de Sitter gravity as a Chern-Simons theory the Lorentzian path integral is
$$
Z = \int {[DA]\over \text{vol}(\text{gauge})} \exp\left\{ ik_3 \, S_{CS}[A] \right\}
$$
with gauge group $PSL(2,\mathbb{C})$ and coupling constant $k_3={L_{\text{dS}} \over 16\pi G_3}$.  The usual Hartle-Hawking prescription then instructs us to evaluate the $PSL(2,\mathbb{C})$ Chern-Simons theory on a Euclidean manifold.  This is precisely the Chern-Simons formulation of three dimensional gravity in Euclidean AdS, evaluated with an imaginary Chern-Simons level.   Similarly, if we interpret JT gravity in de Sitter space as a $BF$ theory,
$$
Z = \int {[DA]\over \text{vol}(\text{gauge})} \exp\left\{i k_2 \int {\rm tr}( B F) \right\}
$$
with $k_2 = \frac{1}{4\pi G_2}$ and gauge group $SO(2,1)=PSL(2,\mathbb{R})$, then the analytic continuation again takes us to Euclidean AdS. 

The result is that, in our theories of gravity in two and three dimensions, the computation of the wavefunction of the universe is much clearer than in more complicated theories of gravity.  We will begin with a study of this wavefunction in JT gravity, where many of the techniques that have proven useful in AdS gravity can be naturally generalized to de Sitter.  In \textbf{Section~\ref{S:dS2}} we will discuss the formulation of JT gravity in a de Sitter background, and study the Schwarzian theory which governs the boundary dynamics at the conformal boundary.  We will study the gauge theory formulation, which will allow us to understand the relationship between the traditional Hartle-Hawking construction and the analytic continuation to Euclidean AdS (i.e. the hyperbolic disk).  We will discover that they are the same to all orders in perturbation theory.  Moreover we will compute the Hartle-Hawking wavefunction of the no-boundary state to all loop-order in the gravitational interaction $G_2$, and to leading order in a genus expansion. We will also study global dS$_2$ where the gravitational path integral computes a transition amplitude rather than a wavefunction. This integral is more intricate, and is related to the double hyperbolic cone, and the final result is in a sense the propagator for a closed universe to leading order in a genus expansion. We also study the gravitational contribution to scattering in both the Hartle-Hawking geometry and in global dS$_2$.

We then turn to non-perturbative considerations in \textbf{Section~\ref{S:genus}}, and define a genus expansion for nearly dS$_2$ gravity. This genus expansion is rather subtle in either the Lorentzian dS metric formulation or in Euclidean AdS. There are no smooth, Lorentzian higher genus spacetimes of constant positive curvature; meanwhile, in the Euclidean AdS continuation, the boundary conditions inherited from dS guarantee that the metric will in general develop conical singularities in the interior. We proceed by exploiting the gauge theory formulation, where, once we specify the topology of the spacetime, we simply integrate over flat connections on that surface subject to boundary conditions. Translated into the metric language, we integrate over those singular metrics that have a smooth gauge theory description. 

From here our discussion closely mirrors the analogous discussion in~\cite{Saad:2019lba}, which obtained the genus expansion of JT gravity in Euclidean AdS and showed that it arises from a double-scaled matrix integral. That expansion depends on the volumes of moduli spaces of bordered Riemann surfaces. We find that the de Sitter coefficients are similarly related to the volumes of moduli spaces of Riemann surfaces with cone points. For cone angles less than $\pi$, these volumes are known to be the continuation of the volumes of moduli spaces of bordered surfaces. We conjecture that this remains true past cone angles of $\pi$. If so, then we find that the de Sitter genus expansion is, on the nose, the continuation of the AdS expansion. Furthermore, our expansion comes from a double-scaled matrix integral, the same double-scaled model discovered for AdS, probed with different operators. Interpreting the matrix model as a non-unique, non-perturbative completion of the genus expansion, this gives a fully non-perturbative theory of de Sitter quantum gravity. It allows us in principle to compute de Sitter transition amplitudes between universes with an arbitrary number of boundary components.

In \textbf{Section~\ref{S:dS3}} we will discuss the analogous considerations in dS$_3$.  While we do not have a full genus expansion in this case, the discussion of the wavefunction in the Hartle-Hawking state and the transition amplitude in global de Sitter space is similar to that in dS$_2$.  The resulting boundary theory is richer and can be regarded as an analytic continuation of the theory of boundary gravitons relevant for AdS$_3$.  We will discuss the Chern-Simons formulation and its uses in computing loop corrections. Given the length of the manuscript, we conclude with a detailed summary of our results and a discussion in \textbf{Section~\ref{S:discuss}}.

{\it Note}: While this work was in progress, the paper \cite{Maldacena:2019cbz} appeared, which has some overlap with our analysis in Sections~\ref{S:dS2} and~\ref{S:genus} -- they study the Schwarzian dynamics on the boundary of two-dimensional nearly de Sitter spacetime in the metric formulation, as well as the prospect of a matrix model completion.

%***********************************************
\section{Nearly dS$_2$}
\label{S:dS2}
%***********************************************
   
Many near-extremal black holes in string theory have a near-horizon geometry of the form AdS$_2\times X$. Unlike near-horizons of the form AdS$_{d>2}\times X$, AdS$_2$ throats never decouple in the infrared. There are several ways to understand this. One is that, if AdS$_2$ throats did decouple, then string theory in the near-horizon geometry would be dual to a CFT$_1$. However, a CFT$_1$ is almost by definition topological, since the Hamiltonian vanishes by the trace Ward identity. See e.g.~\cite{Jensen:2011su} for a related argument using the density of states. This fact is mirrored in AdS$_2$ gravity, where backreaction of matter fields destroys the AdS$_2$ asymptotics~\cite{Maldacena:1998uz}. In the general relativity literature this fact is well-known and is called the Aretakis instability (see e.g.~\cite{Aretakis:2012ei}).

The resolution to all of these difficulties is quite natural~\cite{Almheiri:2014cka}. One simply includes the leading irrelevant deviation from criticality in the infrared. As far as $s$-wave physics on the near-horizon geometry is concerned, the two-derivative approximation to the gravitational sector, the analogue of the Einstein-Hilbert action, is dilaton gravity which can be parameterized in terms of the action
\beq
	S = \frac{1}{16\pi G_2} \int d^2x \sqrt{-g}\left( \varphi R + U(\varphi) \right) \,, 
\eeq
up to a boundary term. Here $\varphi$ is the dilaton, morally the size of the transverse space $X$, and $U$ is its potential. Near the AdS$_2$ throat the gravitational sector universally asymptotes~\cite{Jensen:2016pah,Nayak:2018qej} to an even simpler theory,
\beq
	S_{\rm JT} = \frac{1}{16\pi G_2} \int d^2x \sqrt{-g} \,\varphi_0  R + \frac{1}{16\pi G_2}\int d^2x \sqrt{-g}\,\bar{\varphi} \left( R + \frac{2}{L^2}\right) + (\text{bdy term})\,.
\eeq
Here $\varphi_0$ (which is a root of $U(\varphi)$) is the value of $\varphi$ on the extremal horizon, $\bar{\varphi}= \varphi - \varphi_0$, and $L$ is the radius of the AdS$_2$. The first term in the action is topological, just giving the Euler characteristic of the spacetime, and the second term describes Jackiw-Teitelboim gravity~\cite{jackiw1985lower, teitelboim1983gravitation, jackiw1992gauge}. In a slight abuse of terminology, we refer to the action functional above, including the first term, as Jackiw-Teitelboim theory. It has so-called nearly AdS$_2$ solutions, where the spacetime is asymptotically AdS$_2$ and the dilaton grows near the boundary, including
\beq
	ds^2 = -\frac{r^2}{L^2}dt^2 + \frac{L^2dr^2}{r^2}\,, \qquad \varphi = \frac{r}{\ell}\,.
\eeq
Backreaction may be consistently studied in this context, and in fact the gravitational path integral for this simple theory reduces to a boundary path integral, often called the Schwarzian theory\footnote{In fact the endpoint of the Aretakis instability may be found and studied using this Schwarzian theory~\cite{Hadar:2018izi}.}~\cite{Jensen:2016pah,Maldacena:2016upp,Engelsoy:2016xyb}.

In this Section we consider the de Sitter version of Jackiw-Teitelboim theory, which we expect to universally govern the low-energy physics of a near-extremal solution with dS$_2\times X$ near-horizon geometry. Its action is given by
\beq
	\label{E:dSJT}
	S_{\rm JT} = \frac{\varphi_0}{16\pi G_2} \int d^2x \sqrt{-g} \,R + \frac{1}{16\pi G_2} \int d^2x \sqrt{-g}\,\bar{\varphi}\left( R - \frac{2}{L^2}\right) + (\text{bdy term})\,.
\eeq
This theory has ``nearly dS$_2$'' solutions, much like the nearly AdS$_2$ solutions previous studied in the literature. We will discuss the boundary term shortly.

In this Section we have two main goals. The first is to find the analogue of the Schwarzian path integral, and to use it to compute loop-level contributions to the gravitational path integral and to gravitational scattering. This may be done straightforwardly when integrating over simple geometries with the topology of the disc or the annulus, like the capped-off geometry discussed in the Introduction or global dS$_2$.

The second main goal is to deduce the analytic continuation between nearly dS$_2$ gravity and Euclidean nearly AdS$_2$ gravity, using the simple Hartle-Hawking and global nearly dS$_2$ geometries as prototypes. With this analytic continuation in hand, we can relate the higher genus contributions to the nearly dS$_2$ path integral to the recent Euclidean results of~\cite{Saad:2019lba}, which we do in the next Section.

%***********************************************
\subsection{Basic solutions}
\label{S:basicdS2}
%***********************************************

To begin, let us write down the basic solutions of Jackiw-Teitelboim gravity that will be of interest to us throughout this Section. These are the global dS$_2$ spacetime, and the geometry used in the Hawking-Hartle construction.

Setting the de Sitter radius to unity, the field equations are
\beq
	R = 2\,, \qquad (D_{\mu}D_{\nu}+g_{\mu\nu})\bar{\varphi} = 0\,.
\eeq
The first equation implies that the spacetime has constant positive curvature, and the second that $\partial^{\mu}\bar{\varphi}$ is a conformal Killing vector. This in turn implies that $\epsilon^{\mu\nu}\partial_{\nu}\bar{\varphi}$ is a Killing vector. 

\begin{figure}[t]%
	\centering
	\includegraphics[width=5.2in]{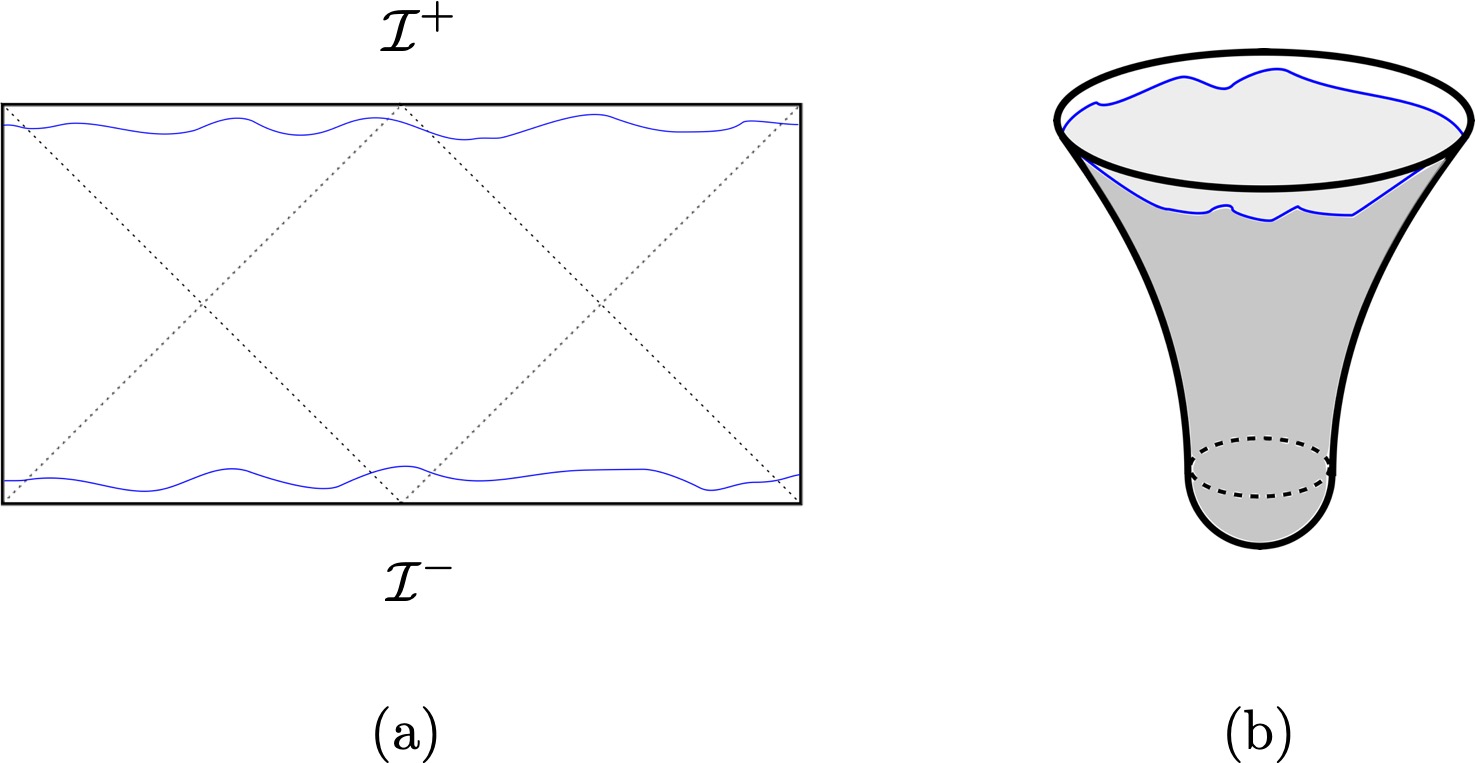}
	\caption{
			\label{F:dS2Penrose}
			(a) The Penrose diagram for global dS$_2$ with $\alpha = 1$.  In our analysis we consider fluctuating Cauchy surfaces (depicted as blue curves) near past and future infinity. The left and right boundaries are identified. The geometries with $\alpha <1$ are ``squashed'' horizontally, while those with $\alpha>1$ are ``stretched.'' 
			\label{F:HH} 
			(b) A cartoon of the geometry used in the Hartle-Hawking construction. The Lorentzian and Euclidean segments are glued together across the circle. The fluctuating boundary is indicated in blue.
		}%
\end{figure}

The first solution of interest is the global version of nearly dS$_2$ spacetime, which has two circle boundaries $\mathcal{I}^{\pm}$. See Fig.~\ref{F:dS2Penrose}(a) for its Penrose diagram. The background is specified by
\beq
\label{E:globaldS}
	ds^2 = -dt^2 + \cosh^2t \, d\theta^2\,, \qquad \bar{\varphi}= \frac{\sinh t}{\ell}\,,
\eeq
where $\ell$ is a free parameter. Note that the dilaton is necessarily large and negative near one boundary, and large and positive near the other. The global geometry may be understood in embedding coordinates. Let $X^{\mu}$ be coordinates on $\mathbb{R}^{1,2}$ with Minkowski metric $-(dX^0)^2 + (dX^1)^2 + (dX^2)^2$. The geometry above is the surface $X^2 = 1$, with
\beq
	X^0 = \sinh t\,, \qquad X^1 = \cosh t \cos\theta\,, \qquad X^2 = \sinh t\sin\theta\,.
\eeq
As for the dilaton, we see that
\beq
	\bar{\varphi} = \frac{X^0}{\ell}\,.
\eeq
So, in the embedding space, $\partial_{\mu} \bar{\varphi}$ is a constant timelike vector.

Following the standard practice in AdS/CFT, we introduce cutoff slices near conformal infinity. The relevant boundary conditions here are that on the cutoff slice the induced metric and dilaton are
\beq
\label{E:dS2BC}
	dS^2 = \left(\frac{\beta_{\pm}}{2\pi}\right)^2 \frac{d\theta^2}{\varepsilon_{\pm}^2}\,, \qquad \bar{\varphi} = \pm \frac{1}{\varepsilon_{\pm}J}\,,
\eeq
where the limit $\varepsilon_{\pm}\to 0$ corresponds to taking the cutoff to conformal infinity. (This process can be equivalently stated in the language of defining functions and conventional holographic renormalization.) Here the future and past circles have length $\beta_+$ and $\beta_-$ respectively.

In fact, there is a two-parameter family of solutions consistent with these boundary conditions, all of which have $\beta_+ = \beta_-$. These backgrounds are parameterized by
\beq
	ds^2 = -dt^2 + \alpha^2 \cosh^2 t \left( d\theta + \gamma \delta(t) dt\right)^2 \,, \qquad \bar{\varphi} = \frac{\alpha \sinh t}{\ell}\,,
\eeq
where $(\alpha,\gamma)$ label the phase space of solutions.   What is the interpretation of these solutions? The circle at $t=0$ is a geodesic, and its length is $2\pi |\alpha|$. As for $\gamma$, it is a ``twist,'' and the geometry with nonzero $\gamma$ can be arrived at by acting with an axial rotation $\gamma$ on the geometry with $\gamma = 0$.

It is important to note that, although these metrics appear non-smooth, this is an artifact of the way that we have chosen to write them. Indeed, we can write the metric as
\beq
	ds^2 = -dt^2 + \alpha^2 \cosh^2t \,d\psi^2 \,, \qquad \psi = \theta + \gamma \,\Theta(t)\,,
\eeq
with $\Theta$ the Heaviside theta function.  
This coordinate system also has the advantage of allowing us to deduce the periodicity condition on $\gamma$: because $\psi$ is periodic with period $2\pi$, so is  $\gamma$. Further, it is clear that we may restrict ourselves to positive $\alpha$. Indeed, the geometries we have written here are the natural Lorentzian versions of the metric on the hyperbolic cylinder, with $\alpha$ and $\gamma$ playing the role of the length and twist parameters which describe the moduli of constant curvature metrics on the cylinder.

These geometries are all hyperboloids, and can be thought of as the result of gluing a future half-hyperboloid with $t>0$ to a past hyperboloid across a geodesic of length $2\pi \alpha$ with a twist $\gamma$. From this point of view, we expect $\alpha$ and $\gamma$ to be conjugate to each other, and that the measure over $\alpha$ and $\gamma$ is simply the Weil-Petersson measure. This is indeed the case, as we demonstrate in Subsection~\ref{S:dS2measure}.

The $\alpha = 1$ geometry is what is usually meant by global dS$_2$. It has an analytic continuation $t =- i \tau$ to the unit sphere, and so a stable Bunch-Davies vacuum for matter fields. The geometries with $\alpha \neq 1$ analytically continue to the space $d\tau^2 + \alpha^2 \cos^2(\tau) d\theta^2$, which has conical singularities at the poles $\tau = 0,\pi$. It is not clear if there is a stable vacuum for matter fields propagating in the Lorentzian continuation. However, in this Section we are largely interested in the gravitational path integral, which seems to be sensible for general $\alpha$. 

While the metric with $\alpha = 1$ has isometry group $PSL(2;\mathbb{R}$), this isometry group is broken at generic $\alpha$ to the $U(1)$ subgroup that generates rotations in $\theta$. At integer $\alpha = n$, the isometry group is enhanced again to what is sometimes called $PSL^{(n)}(2;\mathbb{R})$, the group generated by the vector fields $\partial_{\theta}$ and $e^{\pm i n \theta} \partial_{\theta}$. Note however that while these are symmetries of the metric, the physical group of symmetries is always broken by the dilaton profile to the $U(1)$ rotation subgroup.

In ordinary de Sitter spacetime there is a static patch, a region of the global geometry analogous to a Rindler wedge of Minkowski space, on which one of the boost isometries of the de Sitter isometry group acts as time translation. Given that nearly dS$_2$ spacetime is only invariant under rotations, it should not be a surprise that the analogue of the static patch is slightly different. The would-be static patch of the $\alpha = 1$ geometry is parameterized in the embedding coordinates by
\beq
	X^0 = \sin\tilde{\theta} \sinh\tilde{t}\,, \qquad X^1 = \sin\tilde{\theta}\cosh\tilde{t}\,, \qquad X^2 = \cos\tilde{\theta}\,,
\eeq
with $\tilde{\theta}\in (0,\pi)$ and $\tilde{t}\in \mathbb{R}$. The metric in this patch is time-independent, but the dilaton is not,
\beq
	ds^2 = d\tilde{\theta}^2 - \sin^2\tilde{\theta} \,d\tilde{t}^2 \,, \qquad \bar{\varphi} = \frac{\sinh\tilde{t}\sin\tilde{\theta}}{\ell}\,.
\eeq
This geometry has a cosmological horizon at the two points $\tilde{\theta} = (0,\pi)$. We would like to assign an entropy to this horizon. Ordinarily there is no well-defined notion of entropy in a time-dependent setting, however in this case we are helped by the fact that the fluctuating part of the dilaton $\bar{\varphi}$ vanishes smoothly on the horizon. As a result the area of the horizon is constant, and taking the standard Bekenstein-Hawking entropy at face value we find the cosmological entropy
\beq
\label{E:Scosmo}
	S_{\rm cosmo} = \frac{\varphi_0}{2G_2}\,,
\eeq
where we have used that the area of the horizon is the value of the dilaton on the horizon $\tilde{\theta}=0$ plus the value on the other horizon $\tilde{\theta}=\pi$. Note that this entropy is independent of the parameter $\ell$ controlling the asymptotic behavior of the dilaton. This is in contrast with the thermal entropy of nearly AdS$_2$ black holes, whose extremal value is controlled by $\varphi_0$ and whose leading low-temperature correction is of the form $\frac{1}{G_2 \beta \ell}$. 

There is also another geometry of interest, namely the one used in the Hartle-Hawking construction. See Fig.~\ref{F:HH}(b). It is the union of a Lorentzian segment, the future half of global dS$_2$ with $\alpha=1$,
\begin{subequations}
\label{E:HHmetric}
\beq
	ds^2 = -dt^2 + \cosh^2t \,d\theta^2\,, \qquad \bar{\varphi} = \frac{\sinh t}{\ell}\,,\qquad t\geq 0\,,
\eeq
glued at $t=0$ to a Euclidean hemisphere
\beq
	ds^2 = d\tau^2 + \cos^2\tau \,d\theta^2 \,,  \qquad \bar{\varphi} = -i \,\frac{\sin\tau}{\ell}\,, \qquad \tau \in [0,\pi/2]\,,
\eeq
\end{subequations}
along the circle at $\tau = 0$. Note that the dilaton goes from being real in the Lorentzian section to being imaginary in the Euclidean section, crossing zero at the interface.

%***********************************************
\subsection{Schwarzian boundary action}
%***********************************************

There are various ways to derive the Schwarzian boundary action for nearly AdS$_2$ gravity~\cite{Jensen:2016pah,Maldacena:2016upp,Engelsoy:2016xyb,Cvetic:2016eiv}. All require the complete Jackiw-Teitelboim action, including the boundary term. Here we consider several complementary approaches to derive the Schwarzian boundary action for nearly dS$_2$ gravity.

In Eq.~\eqref{E:dS2BC} we described the boundary conditions on the metric and dilaton near conformal infinity. We must supplement the Jackiw-Teitelboim action~\eqref{E:dSJT} with a boundary term in order to ensure a variational principle consistent with the boundary condition, as well as to have a finite on-shell action. Including this boundary term, the complete action reads
\beq
\label{E:totalJT}
	S_{\rm grav} = \frac{\varphi_0}{4G_2}\chi +\frac{1}{16\pi G_2}\int_{\mathcal{M}} d^2x \sqrt{-g} \bar{\varphi}(R-2)- \frac{1}{8\pi G_2}\int_{\partial\mathcal{M}} dx \sqrt{h} \,\bar{\varphi} \left( K -1\right)\,.
\eeq
Here $\chi$ is
\beq
	\chi = \frac{1}{4\pi} \int d^2x \sqrt{-g} \,R - \frac{1}{2\pi} \int dx\sqrt{h}\,K\,,
\eeq
 $h$ is the induced metric on the boundary of the cutoff slice, and $K$ is its extrinsic curvature. On a Lorentzian manifold with real metric, $\chi$ is the Euler characteristic. The Hartle-Hawking geometry has a complex metric and so we have to be more careful. In any case, we follow the standard procedure in holographic renormalization where one introduces the cutoff slices near conformal infinity, evaluates the action including the boundary terms, and then takes the cutoff to infinity.

Using this action we compute the on-shell action of global dS$_2$ and the mixed Lorentzian/ Euclidean solution we discussed above. First consider global dS$_2$. The backgrounds obeying the boundary condition~\eqref{E:dS2BC} are
\beq
	ds^2 = - dt^2 + \alpha^2 \cosh^2t (d\theta+\gamma\delta(t)dt)^2 \,, \qquad \bar{\varphi} = \frac{2\pi \alpha}{\beta J}\sinh t\,.
\eeq
The Euler characteristic term vanishes, as does the bulk part of the action. We readily find for all $\alpha$ and twist $\gamma$
\beq
\label{E:onshellGlobal}
	S_{\rm global} =0\,.
\eeq

For the mixed Lorentzian/Euclidean geometry employed in the Hartle-Hawking construction, the total geometry has the topology of a disk, and so one might think that we have $\chi =1$. This is not quite true. The point is that the topological and metric definitions of the Euler characteristic differ by a phase on spaces with complex metric. For the Hartle-Hawking geometry~\eqref{E:HHmetric}, we have\footnote{We thank D.~Stanford for pointing this out to us.}
\begin{align}
\nonumber
	\chi &= \left( \frac{1}{4\pi}\int_0^{\Lambda} dt \int d\theta \,2\cosh t - \frac{1}{2\pi}\int d\theta\, \cosh\Lambda\tanh\Lambda\right) + \left( \frac{1}{4\pi} \int_0^{\pi/2} (-id\tau) \int d\theta \,2\cos\tau  \right)
	\\
	& = (0) + (-i) = -i\,,
\end{align}
where the first term comes from the Lorentzian segment and the second from the Euclidean. The Lorentzian segment is an annulus and so gives zero contribution. We then see that
\beq
	\chi = -i \chi_T\,,
\eeq
where $\chi_T$ is the topological Euler characteristic. As for the bulk part of the action, it vanishes, and after computing the boundary term we then find
\beq
\label{E:onshellHH}
	S_{\rm HH} = -i\frac{\varphi_0}{4G_2} + \frac{\pi }{4G_2 \beta J}\,.
\eeq

Let us now consider the quantum theory. Integrating out the dilaton, we have a residual integral over the moduli space of constant curvature metrics on either the annulus or the disk, depending on whether we are integrating over fluctuations around global dS$_2$, or the Hartle-Hawking geometry. This moduli space of metrics may be parameterized near conformal infinity in terms of asymptotic diffeomorphisms acting on the solutions presented in the last Subsection, as in~\cite{Jensen:2016pah}. It may also be parameterized in terms of a fixed $R=2$ solution with a fluctuating boundary, as in~\cite{Maldacena:2016upp}. We take the latter approach.

%***********************************************
\subsubsection{The Hartle-Hawking geometry}
%***********************************************

Consider the Hartle-Hawking geometry. Let us work in conformal coordinates for the Lorentzian patch, with $\sec(T) = \cosh(t)$, and $T\in [0,\pi/2)$. The future boundary is reached as $T\to\pi/2$. The metric now reads
\beq
	ds^2 =\sec^2(T) \left( -dT^2 + d\theta^2\right)\,.
\eeq
Let $T= T(u)$, $\theta = f(u)$ parameterize the future boundary, with $u$ a periodic variable of periodicity $2\pi$. Clearly $f(u+2\pi) = f(u)+2\pi$. Imposing the boundary condition~\eqref{E:dS2BC},
\beq
	g_{uu} = \left(\frac{\beta}{2\pi}\right)^2\frac{1}{\varepsilon^2} = \frac{-T'(u)^2 + f'(u)^2}{\cos^2(T(u))}\,,
\eeq
we find
\beq
	T(u) = \frac{\pi}{2} - \epsilon \,\frac{2\pi f'(u)}{\beta} + O(\epsilon^2)\,.
\eeq
The extrinsic curvature of the boundary is
\beq
	K = 1 - \epsilon^2\left(\frac{2\pi}{\beta}\right)^2\left( \{f(u),u\} + \frac{f'(u)^2}{2}\right) + O(\epsilon^3)\,,
\eeq
where
\beq
	\{f(u),u\} = \frac{f'''(u)}{f'(u)}-\frac{3}{2}\frac{f''(u)^2}{f'(u)^2}
\eeq
is the Schwarzian derivative of $f(u)$ with respect to $u$. Using the expression for the Schwarzian and plugging in $h = \frac{2\pi}{\beta\epsilon}$ and $\bar{\varphi} = \frac{1}{J\epsilon}$, the action~\eqref{E:totalJT} evaluates to
\beq
\label{E:HHschwarzian}
	S =-i \frac{\varphi_0}{4G} + \frac{1 }{4G \beta J} \int_0^{2\pi} du\left( \{f(u),u\} + \frac{f'(u)^2}{2}\right)\,.
\eeq
This is the Schwarzian action for the single boundary graviton degree of freedom of the chosen background. The integrand of the action may be alternatively expressed as
\beq
	\{f(u),u\} + \frac{f'(u)^2}{2}=\left\{\tan\left(\frac{f(u)}{2}\right),u\right\}\,.
\eeq
Notice that the Euclidean cap of the geometry completely drops out of the computation.

The $PSL(2;\mathbb{R})$ isometries of dS$_2$ leave the metric invariant but act non-trivially on the boundary, and thus on the Diff$(\mathbb{S}^1)$ field $f(u)$, by
\beq
	\tan\left(\frac{f(u)}{2}\right) \to \frac{a \tan\left(\frac{f(u)}{2}\right)+b}{c\tan\left(\frac{f(u)}{2}\right)+d}\,.
\eeq
The above field transformations preserve the geometry including the boundary, and so we identify these field configurations in the remaining path integral over $f(u)$. That is, $f(u) \in \text{Diff}(\mathbb{S}^1)/PSL(2;\mathbb{R})$. The Schwarzian derivative is projective-invariant, and so the action~\eqref{E:HHschwarzian} is manifestly invariant under these transformations, which may be viewed as a gauge symmetry of the model.

The classical trajectory of the model (modulo the $PSL(2;\mathbb{R})$ quotient) is simply 
\beq
	f(u) = u\,,
\eeq
which has on-shell action
\begin{equation*}
	S_{\rm HH} = -i\frac{\varphi_0}{4G} + \frac{\pi}{4G\beta J}\,,
\end{equation*}
reproducing our earlier result~\eqref{E:onshellHH}.

%***********************************************
\subsubsection{Global dS$_2$}
\label{SS:globalSchwarzian}
%***********************************************

The computation for global dS$_2$ spacetime proceeds similarly. The Euler characteristic term now vanishes, as does the bulk part of the action after integrating out the dilaton. It only remains to find the perturbation in the extrinsic curvature near conformal infinity. The analogue of conformal coordinates is simply
\beq
	ds^2 = \sec^2(T)\left( -dT^2 + \alpha^2 d\theta^2\right)\,.
\eeq
Near future infinity, we introduce a boundary $t=T_+(u), \theta = f_+(u)$, along with a boundary near past infinity $t = T_-(u), \theta = f_-(u)$. Here $f_{\pm}(u)$ are two diffeomorphisms of the circle, with $f_{\pm}(u+2\pi) =f_{\pm}(u)+2\pi$. The only effect of general $\alpha$ on the computation of the future extrinsic curvature is to replace $f_+\to \alpha f_+$, and similarly in the past.

Because the dilaton goes from $1/(\epsilon J)$ on the future boundary to $-1/(\epsilon J)$ on the past boundary, we then arrive at the boundary effective action
\begin{align}
\begin{split}
\label{E:globalschwarzian}
	S = &\frac{1}{4G_2J}\left[ \frac{1}{\beta_+} \int_0^{2\pi} du \left( \{ f_+(u),u\} +\frac{\alpha^2}{2}f'_+(u)^2 \right) \right.
	\\
	& \qquad \qquad\qquad \qquad  \left. - \frac{1}{\beta_-} \int_0^{2\pi} du \left( \{f_-(u),u\}+ \frac{\alpha^2}{2}f_-'(u)^2\right)\right]\,.
\end{split}
\end{align}
Note that the fields $f_{\pm}$ do not directly couple to each other, consistent with the fact that there are independently conserved future and past stress tensors. However, past/future couplings do arise on account of $\alpha$, as follows.  For general $\alpha$, the above action is no longer invariant under fractional linear transformations of $f_{\pm}$.  The action is only invariant under $U(1)\times U(1)$ transformations, $f_{\pm}(u)\to\ f_{\pm}(u) + \delta_{\pm}$. The diagonal subgroup corresponds to the $U(1)$ isometry of global dS$_2$, and so it is effectively gauged. Said another way, the reparameterization fields $f_{\pm}$ and $\alpha$ are together an element of the quotient space
\beq
	\faktor{\left( \text{Diff}(\mathbb{S}^1)\times \text{Diff}(\mathbb{S}^1)\times \mathbb{R}_+ \right)}{U(1)}\,.
\eeq
The axial $U(1)$ symmetry $f_+(u) \to f_+(u) + \delta, \,f_-(u) \to f_-(u) - \delta$ is physical and generates the parameter $\gamma$. Indeed, the solutions to the equations of motion of the model consistent with the boundary conditions and modulo the $U(1)$ quotient are simply
\beq
	f_+(u) = u  + \gamma\,, \qquad f_-(u) = u\,.
\eeq
The on-shell action is 
\beq
\label{E:globalSonshell}
	S_{\rm global} = \frac{\pi \alpha^2}{4GJ}\left( \frac{1}{\beta_+}-\frac{1}{\beta_-}\right)\,,
\eeq
which seemingly contradicts our earlier result that $S_{\rm global} = 0$. However, recall that these geometries solve all of the equations of motion of Jackiw-Teitelboim gravity only when $\beta_+ = \beta_- $, in which case $S_{\rm global}$ vanishes. (When $\beta_+ \neq \beta_-$ we still satisfy  $R=2$, but there is no solution for the dilaton.)

At integer $\alpha=n$ the isometry group is enhanced from $U(1)$ to $PSL^{(n)}(2;\mathbb{R})$, and correspondingly the symmetry group of the doubled model~\eqref{E:globalschwarzian} is enhanced to $PSL^{(n)}(2;\mathbb{R})\times PSL^{(n)}(2;\mathbb{R})$. The diagonal part is effectively gauged, in the sense that we identify $f_{\pm}$ modulo
\beq
	\tan\left( \frac{n f_+(u)}{2}\right)\sim \frac{a\tan\left(\frac{nf_+(u)}{2}\right)+b}{c\tan\left(\frac{nf_+(u)}{2}\right) + d}\,, 
	\quad 
	-\cot\left(\frac{n f_-(u)}{2}\right)\sim \frac{a \left( - \cot\left( \frac{n f_-(u)}{2}\right)\right)+b}{c\left( -\cot\left(\frac{n f_-(u)}{2}\right)\right) + d}\,. 
\eeq
Na\"{i}vely the axial part is physical, giving a non-compact moduli space of solutions. This is not quite the case, as we will see later.

Since $\alpha$ indexes a geometric quantity, the length of a minimal geodesic around the circle, one may work in the sector of geometries with fixed $\alpha$ and in particular $\alpha = 1$.  For this special case the boundary model is precisely
\beq
	S = \frac{1}{8G_2 J}\int_0^{2\pi} du \left( \frac{1}{\beta_+}\left\{\tan\left(\frac{f_+(u)}{2}\right),u\right\}  -\frac{1}{\beta_-} \left\{ \tan\left(\frac{f_-(u)}{2}\right),u\right\}\right)\,,
\eeq
with $f_{\pm}$ parameterizing
\beq
	\faktor{\left( \text{Diff}(\mathbb{S}^1)\times \text{Diff}(\mathbb{S}^1)\right)}{PSL(2;\mathbb{R})}\,.
\eeq

We would like to perform the path integral over these degrees of freedom, as well as the moduli $(\alpha, \gamma)$ described above. To do so we require the measure over the reparameterization degrees of freedom, and in particular the measure for $\alpha$. In this second-order formalism this is difficult to obtain, so we will instead use the description of JT theory as a topological gauge theory.  This will have the advantage of allowing us to efficiently study Jackiw-Teitelboim gravity on higher-genus surfaces.

%***********************************************
\subsection{Topological gauge theory for nearly dS$_2$ gravity}
%***********************************************

Jackiw-Teitelboim gravity with positive cosmological constant may be classically recast as a $PSL(2;\mathbb{R})$ topological gauge theory~\cite{Isler:1989hq,Chamseddine:1989yz}. The starting point is to pass to a first-order description. To do so in higher-dimensional Einstein gravity, one simply rewrites the Einstein-Hilbert term in terms of first-order variables. The spin connection appears quadratically and can be integrated out, enforcing the torsion-free constraint and leaving behind the ordinary Einstein gravity action.

In dilaton gravity this is no longer the case, and one must introduce two additional fields $t^a$ for $a=0,1$ to enforce the torsion-free constraint. These may be grouped into a single object in the following way. We introduce the generators
\beq
	J_A = (P_0,P_1, \Omega)\,,
\eeq
in the fundamental representation of the algebra $\mathfrak{sl}(2;\mathbb{R})$ with
\beq
	[P_a,P_b] = \epsilon_{ab} \Omega\,, \qquad [\Omega,P_a] = \epsilon_{ab} P^b\,, 
\eeq
and
\beq
	\text{tr}(J_AJ_B) = -\frac{1}{2}\eta_{AB} \,, \qquad \eta_{AB} = \begin{pmatrix} -1 \\ & 1 \\ & & -1\end{pmatrix}\,.
\eeq
Here $\epsilon^{01}=1$ and we raise and lower the indices $a=0,1$ with the Minkowski metric. The signature of the Killing-Cartan metric is $(+-+)$, i.e. we are dealing with $\mathfrak{so}(2,1)$. We then group the fluctuating part of the dilaton and the Lagrange multiplier fields $t^a$ as
\beq
	B = t^a P_a + \bar{\varphi} \,\Omega\,.
\eeq
We similarly group the zweibein $e^a$ and abelian spin connection $\omega = -\frac{1}{2}\epsilon^{ab} \omega_{ab}$ into an $\mathfrak{sl}(2;\mathbb{R})$-valued one-form,
\beq
\label{E:dSgauge}
	A = e^a P_a + \omega \,\Omega\,.
\eeq
In this convention the scalar curvature is given by
\beq
	d^2x \sqrt{-g} \,R = 2 d\omega\,.
\eeq
One of the insights of~\cite{Isler:1989hq,Chamseddine:1989yz} is that, on a solution of Jackiw-Teitelboim gravity, infinitesimal local Lorentz transformations and diffeomorphisms act on $A$ and $B$ in the same way as infinitesimal $\mathfrak{sl}(2;\mathbb{R})$ gauge transformations, and it is in this sense that $A$ is a connection. The field strength is
\beq
	F = \left( de^a + \omega \epsilon^a{}_b e^b\right)P_a +\left(d\omega +\frac{\epsilon_{ab}}{2} e^a \wedge e^b\right) \Omega = T^a P_a + \frac{1}{2}(R-2) \text{vol}\, \Omega\,,
\eeq
where in passing to the second equality we have introduced the torsion $T^a$ and rewritten the derivative of $\omega$ in terms of the scalar curvature $R$ and volume form $\text{vol}=d^2x \sqrt{-g}$. Observe that $F=0$ is equivalent to the torsion-free constraint and the dilaton equation of motion $R = 2$. 

As a result we may write the first-order version of~\eqref{E:dSJT} as
\begin{align}
\begin{split}
\label{E:bulkTopo}
	S_{\rm JT} &= -\frac{\varphi_0}{8\pi G_2}\int d\omega +\frac{1}{16\pi G_2}\int d^2x \sqrt{-g} \left( \bar{\varphi}(R-2) + 2t^a T_a\right)
	\\
	& = \frac{\varphi_0}{4G_2}\chi +\frac{1}{4\pi G_2}\int \text{tr}\left( B F\right) \,,
\end{split}
\end{align}
where $\chi$ is the Euler characteristic. We recognize the second term as a topological gauge theory with algebra $\mathfrak{sl}(2;\mathbb{R})$. The equation of motion for $B$ simply sets $F=0$, and the equation of motion for $A$ sets $D_{\mu}B = \partial_{\mu}B + [A_{\mu},B] = 0$.

One approach to quantizing Jackiw-Teitelboim gravity on nearly dS$_2$ spacetime is to start in this $BF$ formulation and integrate out $B$. There is then a residual integral over flat connections, subject to boundary conditions at conformal infinity. We will carefully investigate this residual integral shortly, but first, let us map the nearly dS$_2$ backgrounds discussed in the last Subsection to their gauge theory avatars.

%***********************************************
\subsubsection{The gauge theory description of nearly dS$_2$ spacetime}
%***********************************************

Let us consider global dS$_2$ with $\gamma=0$ for simplicity. Picking the zweibein to be
\beq
	e^0 = dt\,, \qquad e^1 = \cosh t\,d\theta\,,
\eeq
the corresponding gauge configuration is
\beq
	A = dt P_0 - \alpha \cosh t d\theta\, P_1 + \alpha \sinh t d\theta \,\Omega\,,
\eeq
or, in the explicit representation
\beq
	P_0 = \frac{1}{2}\begin{pmatrix} 1 & 0 \\0 & -1\end{pmatrix}\,, \qquad P_1 = \frac{1}{2}\begin{pmatrix} 0 & 1 \\ -1 & 0 \end{pmatrix}\,, \qquad \Omega = -\frac{1}{2}\begin{pmatrix} 0 & 1 \\ 1 & 0 \end{pmatrix}\,,
\eeq
we find
\beq
\label{E:dSA}
	A = \begin{pmatrix} \frac{dt}{2} & \frac{\alpha e^{-t}d\theta }{2} \\ -\frac{\alpha e^td\theta}{2} & -\frac{dt}{2}\end{pmatrix}\,.
\eeq
For the dilaton, one may find a complete solution for $B$ which includes a profile for $t^1$,
\beq
\label{E:dSB}
	B = \frac{\alpha}{\ell}\left( \cosh t P_1 + \sinh t \Lambda\right) = \begin{pmatrix} 0 & \frac{\alpha e^{-t}}{2\ell} \\ -\frac{\alpha e^t}{2\ell} & 0 \end{pmatrix}\,.
\eeq
Observe that $B =\frac{1}{\ell} A_{\theta}$.

The field configuration corresponding to the mixed Lorentzian/Euclidean geometry is the same for $t>0$. At $t=0$, it is glued to the continuation with $t =- i \tau$,
\beq
	A = \begin{pmatrix}- \frac{id\tau}{2} & \frac{e^{i\tau} d\theta}{2} \\ -\frac{e^{-i\tau}d\theta}{2} & \frac{id\tau}{2}\end{pmatrix}\,,
\eeq
for $\tau \in [0,\pi/2)$. 

%***********************************************
\subsubsection{Boundary conditions and on-shell action}
%***********************************************

We would also like to verify that the on-shell action of these solutions matches the on-shell action we computed in the second-order formalism in Eqs.~\eqref{E:onshellGlobal} and~\eqref{E:onshellHH}. To do so we must address the question of boundary terms and boundary conditions. The asymptotically dS$_2$ boundary conditions are, in the gauge theory variables, the statement that near future infinity $t\to \infty$ $A$ and $B$ behave as
\beq
\label{E:dSfutureBC}
	A = \begin{pmatrix} \frac{dt}{2} +O(e^{-t})& O(e^{-t}) \\ -\frac{\beta_+ e^t d\theta}{2\pi} + O(e^{-t}) & -\frac{dt}{2} + O(e^{-t})\end{pmatrix}\,, \qquad B = \frac{2\pi}{\beta_+ J} A_{\theta} + O(e^{-t})\,,
\eeq
and near past infinity $t\to-\infty$ as
\beq
	A = \begin{pmatrix} \frac{dt}{2} + O(e^{t}) & \frac{\beta_- e^{-t} d\theta}{2\pi} + O(e^{t}) \\ O(e^{t}) & -\frac{dt}{2} + O(e^t)\end{pmatrix}\,, \qquad B = \frac{2\pi}{\beta_- J}A_{\theta} + O(e^t)\,,
\eeq
and where the fields are allowed to fluctuate at the indicated orders in $e^{-t}$ or $e^t$. These boundary conditions are modeled upon global dS$_2$, which clearly respects them, upon performing the appropriate change of coordinate $t\to t \mp \ln \left(\frac{\pi}{\beta_{\pm}}\right)$ near $t\to \pm\infty$.

In order to ensure a variational principle consistent with these boundary conditions we supplement the bulk part of the action~\eqref{E:bulkTopo} with a boundary term, so that the total action reads
\beq
\label{E:dSBFS}
	S_{\rm grav} = \frac{\varphi_0}{4G}\chi + \frac{1}{4\pi G_2}\int_{\mathcal{M}} \text{tr}(BF) - \frac{1}{8\pi  G_2}\int_{\partial\mathcal{M}} d\theta \, n \text{tr}(B A_{\theta})\,,
\eeq
where $n = +1$ on the future boundary and $n=-1$ on the past boundary. We can check that this is the right boundary term by computing the on-shell variation, which is\footnote{Our orientation is such that $\epsilon^{t\theta} = \frac{1}{\sqrt{-g}}$.}
\beq
	\delta S_{\rm grav} = \frac{1}{4\pi G_2}\left[ \int_{\partial\mathcal{M}_+} d\theta\, \text{tr}\left( \left(\frac{2\pi}{\beta_+J}A_{\theta}-B)\right)\delta A_{\theta}\right) - \int_{\partial\mathcal{M}_-} d\theta \,\text{tr}\left(\left( \frac{2\pi}{\beta_-J}A_{\theta}-B\right)\delta A_{\theta}\right)\right]\,.
\eeq
The future term vanishes as $A_{\theta}$ fluctuates at $O(e^{-t})$ and $\frac{2\pi}{\beta_+J}A_{\theta}-B$ is fixed to be of the same order. The past term vanishes by an analogous argument.

Now we can reproduce the on-shell action of the Hartle-Hawking and global backgrounds. To find the on-shell action of the former, we employ a change of coordinates near future infinity $t\to t+\ln \left( \frac{\beta}{\pi}\right)$ so that $A$ respects the boundary condition at large $t$,
\beq
	A = \begin{pmatrix} \frac{dt}{2} & \frac{\pi}{\beta}\frac{e^{-t}d\theta}{2}\\ - \frac{\beta e^td\theta}{2\pi} & -\frac{dt}{2}\end{pmatrix}\,.
\eeq
The Euler characteristic term gives $\chi = -i$, the bulk term in the action vanishes since $A$ is flat, and using the boundary condition on $B$ the boundary term can be written as
\beq
	 - \frac{1}{4G \beta J} \int_0^{2\pi} d\theta \, \text{tr}(A_{\theta})^2\,.
\eeq
We then find
\begin{equation*}
	S_{\rm HH} =-i \frac{\varphi_0}{4G_2} + \frac{\pi}{4G_2\beta J}\,,
\end{equation*}
reproducing the earlier result~\eqref{E:onshellHH}.

The same sort of argument in global dS$_2$, now with $t\to t + \ln \left(\frac{\beta_+}{\pi \alpha}\right)$ near future infinity and $t\to t - \ln \left( \frac{\beta_-}{\pi \alpha}\right)$ near past infinity, gives
\begin{equation*}
	S_{\rm global} = \frac{\pi\alpha^2}{4G_2J}\left( \frac{1}{\beta_+} - \frac{1}{\beta_-}\right)\,,
\end{equation*}
which matches the result~\eqref{E:globalSonshell} we obtained from the Schwarzian boundary action.

%***********************************************
\subsubsection{Holonomy and minimal length geodesics}
%***********************************************

Let us reconsider the Hartle-Hawking geometry. The $\theta$-circle is contractible in the total space. Then the flatness condition implies that the holonomy of $A$ around that circle is trivial in all representations. In the fundamental representation, we find
\beq
	\mathcal{P} \exp\left( \int_0^{2\pi} d\theta \, A_{\theta}\right) = -I\,.
\eeq
This configuration is then singular when viewed as an $SL(2;\mathbb{R})$ connection, but non-singular as a $PSL(2;\mathbb{R}) = SL(2;\mathbb{R})/\mathbb{Z}_2$ connection, wherein we identify $U\in SL(2;\mathbb{R})$ with $-U$. We conclude that the global form of the gauge group of Jackiw-Teitelboim gravity with positive cosmological constant is $PSL(2;\mathbb{R})$, which is also the same form one finds with negative cosmological constant in Euclidean signature.

As for global dS$_2$, the holonomy in the fundamental representation is
\beq
\label{E:ellipticHolonomy}
	\text{tr}\left( \mathcal{P}\exp\left( \int_0^{2\pi}d\theta\,A_{\theta}\right) \right)= 2\cos(\pi\alpha)\,,
\eeq
which is in general non-trivial. The holonomy is in the elliptic conjugacy class of $PSL(2;\mathbb{R})$. Earlier we saw that the physical interpretation of $\alpha$ is that the minimal length geodesic around the circle has length $2\pi \alpha$. (This is also true in the Hartle-Hawking geometry even though in that case the circle is contractible; there, the circle shrinks to a minimal size $2\pi$ at the end of the Lorentzian segment, and only contracts further in the Euclidean section.) A similar computation shows that a Lorentzian geometry with hyperbolic holonomy is singular, with a Milne-like singularity. 

The lesson is that smooth Lorentzian geometries have minimal length spacelike geodesics around the circle. That length is encoded through an elliptic holonomy in the $PSL(2;\mathbb{R})$ gauge theory description. This is in contrast with smooth Eulicdean geometries of constant negative curvature. The gauge theory description of those geometries is characterized by hyperbolic holonomies around non-contractible cycles.

There is a minor puzzle at this stage, namely what is the domain of $\alpha$? From the spacetime analysis it is clear that $\alpha$ is valued on the real positive line, since it is the minimal length around the global dS$_2$ bottleneck. However from the holonomy~\eqref{E:ellipticHolonomy} (and recalling that we identify $U\sim - U$) it would seem that $\alpha$ is a periodic variable $\alpha \sim \alpha + 1$.

The resolution to this puzzle is that, as the geometry instructs us, $\alpha$ is valued on the positive real line. What happens in the gauge theory description is that, in addition to holonomies, flat connections are also characterized by a winding number. Field configurations with trivial holonomy and different winding number live in different topological sectors. Indeed, for the Hartle-Hawking geometry we integrate over the winding number 1 sector. In global dS$_2$ integrating over positive $\alpha$ means that we sum over all topological sectors, and within each sector integrate over elliptic holonomies.

In other words, for the gauge theory description of global dS$_2$ to match the second order description, we must integrate over the universal cover of $PSL(2;\mathbb{R})$, but only divide by gauge transformations valued in $PSL(2;\mathbb{R})$. 

%***********************************************
\subsection{The map to Euclidean nearly AdS$_2$ gravity}
\label{S:EAdS2}
%***********************************************

Jackiw-Teitelboim gravity with negative cosmological constant may also be recast as a topological gauge theory. In Lorentzian signature the algebra is $\mathfrak{so}(1,2)$, but in Euclidean signature it is $\mathfrak{so}(2,1)$, precisely the same as in our Lorentzian analysis with positive cosmological constant. The action for nearly AdS$_2$ gravity in Euclidean signature in the second-order formalism is
\beq
	S_E = -\frac{\varphi_0'}{4G_2} \chi- \frac{1}{16\pi G_2} \int d^2x\sqrt{g'} \,\bar{\varphi}'(R'+2) - \frac{1}{8\pi G_2}\int dx\sqrt{h'}\bar{\varphi}'(K'-1)\,,
\eeq
where we are distinguishing the dilaton, metric, and curvatures relative to our dS$_2$ analysis by priming them.

In this Subsection we derive a map from the gauge theory description of nearly dS$_2$ gravity to that of Euclidean nearly AdS$_2$ gravity. A part of this map is $L_{\rm AdS} \to i L_{\rm dS}$, but we must also sort out what happens to the dilaton and holonomies. Our strategy is to exploit the quantization around genus 0 and 1 surfaces in both Lorentzian dS$_2$ and Euclidean AdS$_2$ and learn how they map to each other. From this data we find a general map which defines higher genus contributions to the dS$_2$ path integral by continuation from Euclidean AdS$_2$, which we discuss later.

Let us pass to the first-order formalism of the Euclidean AdS$_2$ theory. From the spin connection we define $\omega' = \frac{1}{2}\epsilon^{ij} \omega'_{ij}$, which is related to the scalar curvature by 
\beq
	d^2x \sqrt{g'} \, R' = 2d\omega'\,.
\eeq
Flat indices are raised and lowered with the Euclidean metric, and we take $\epsilon^{12} = 1$. We also denote the Lagrange multiplier fields enforcing the torsion-free condition as $t'^i$. 

Using the same generators introduced earlier, the map between the first-order and the gauge theory variables may be expressed as
\beq
\label{E:AdS2gauge}
	A' = e'^1 \Omega + e'^2 P_0 + \omega' P_1\,, \qquad B' = t'^1 \Omega + t'^2 P_0 + \bar{\varphi}' P_1\,.
\eeq
The curvature of $A'$ is
\begin{align}
\begin{split}
	F' &= (de'^1 + \omega' \wedge e'^2)\Omega +(de'^2-\omega' \wedge e'^1) P_0 + (d\omega' + \epsilon_{ij} e'^i\wedge e'^j) P_1
	\\
	&= T'^1 \Omega + T'^2 P_0 + \frac{1}{2}(R'+2)\text{vol} \,P_1\,.
\end{split}
\end{align}
Up to the boundary term, we then arrive at the first-order action
\beq
	S_E = - \frac{\varphi_0'}{4G_2} \chi + \frac{1}{4\pi G_2}\int\text{tr}(B'F')\,.
\eeq

%***********************************************
\subsubsection{Poincar\'e disk and boundary conditions}
%***********************************************

Now consider the Poincar\'e disk, described by the metric and dilaton
\beq
	ds'^2 = d\rho^2 + \sinh^2\rho\,d\theta^2\,, \qquad \bar{\varphi}' = \frac{\cosh\rho}{\ell'}\,.
\eeq
The asymptotically Euclidean AdS$_2$ boundary conditions may be imposed by cutting off the bulk geometry on a boundary close to the conformal boundary, on which the induced metric and dilaton are
\beq
\label{E:EAdS2BC}
	dS'^2= \left(\frac{\beta'}{2\pi}\right)^2 \frac{d\theta^2 }{\epsilon^2} \,, \qquad \bar{\varphi}' = \frac{1}{J'\epsilon}\,,
\eeq
with the cutoff going to the conformal boundary as $\epsilon\to 0$. The Poincar\'e disk respects these boundary conditions, with the cutoff on a constant-$\rho$ slice $e^{\rho}= \frac{\beta'}{\pi \epsilon}$ and $\ell' = \frac{\beta' J'}{2\pi}$.

In the gauge theory variables, using the zweibein
\beq
	e'^1 = \sinh\rho\,d\theta\,,\qquad e'^2 = d\rho\,.
\eeq
we find 
\beq
\label{E:H2A}
	A' =  \begin{pmatrix} \frac{d\rho}{2} & \frac{e^{-\rho}d\theta}{2} \\ - \frac{e^{\rho}d\theta}{2} & - \frac{d\rho}{2}\end{pmatrix}\,, 
\eeq
and we can also solve the equation of motion for the $t'^i$ to give
\beq
\label{E:H2B}
	B' = \begin{pmatrix}  0 & \frac{e^{-\rho}d\theta}{2\ell'} \\ -\frac{e^{\rho}d\theta}{2\ell'} & 0 \end{pmatrix} = \frac{1}{\ell'}A'_{\theta}\,.
\eeq
In these gauge theory variables, the asymptotically Euclidean AdS$_2$ boundary conditions are, as $\rho\to\infty$,
\beq
\label{E:H2BC}
	A' = \begin{pmatrix} \frac{d\rho}{2} + O(e^{-\rho}) & O(e^{-\rho}) \\ -\frac{\beta' e^{\rho}d\theta}{4\pi} + O(e^{-\rho}) & -\frac{d\rho}{2} + O(e^{-\rho}) \end{pmatrix}\,, \qquad B' = \frac{2\pi}{\beta' J'} A'_{\theta} + O(e^{-\rho})\,.
\eeq
Adding a boundary term to ensure that there is a variational principle consistent with this boundary condition, we arrive at the complete form of the first-order action
\beq
\label{E:H2BFS}
	S_E = -\frac{\varphi_0'}{4G_2}\chi + \frac{1}{4\pi G_2}\int_{\mathcal{M}} \text{tr}(B'F') + \frac{1}{8\pi G_2} \int_{\partial\mathcal{M}} d\theta \,\text{tr}(B'A'_{\theta})\,.
\eeq

In the Poincar\'e disk the $\theta$-circle is contractible. Non-singularity of the gauge configuration then mandates that the holonomy around the circle is trivial in all representations. In the fundamental we find
\beq
	\mathcal{P}\exp\left( \int_0^{2\pi} d\theta\,A'_{\theta}\right) = -I\,,
\eeq
and so we find as before that the global topology of the gauge group is fixed to be $PSL(2;\mathbb{R})$. 

Now for the punchline. Comparing the expressions~\eqref{E:H2A} and~\eqref{E:H2B} for the gauge field and scalar parameterizing the Poincar\'e disk with~\eqref{E:dSA} and~\eqref{E:dSB} for the Lorentzian segment of the Hartle-Hawking geometry (with $\alpha = 1$), we see that they match exactly upon the replacement $\rho  \leftrightarrow   t$. The same correspondence holds for the boundary conditions~\eqref{E:H2BC} near the boundary of the Poincar\'e disk and those~\eqref{E:dSfutureBC} near future infinity in dS$_2$. In fact, even the action~\eqref{E:H2BFS} may be mapped to that of the Lorentzian dS$_2$ model~\eqref{E:dSBFS}. After integrating out the scalar, the action is just the topological term plus the boundary term. Demanding 
\beq
	\exp(-S_E) = \exp(i S)\,,
\eeq
one finds that with a single future boundary, we must analytically continue 
\beq
	\varphi_0' \to  \varphi_0 \,, \qquad B' \to i B\,.
\eeq
The first follows from the fact that the Euler term for Euclidean AdS$_2$ gravity gives the topological characteristic $\chi_T$, while the Euler term for Lorentzian dS$_2$ gives $-i \chi_T$.

Comparing the boundary conditions Eq.~\eqref{E:H2BC} for $B'$ and Eq.~\eqref{E:dSfutureBC} for $B$ we see that we must continue
\beq
	J' \to -i J\,.
\eeq
This is expected. Restoring the AdS radius $L_{\rm AdS}$ and dS radius $L_{\rm dS}$, the natural boundary conditions on the AdS and dS dilatons are that, on the cutoff slice near conformal infinity, the dilatons go to
\beq
	\bar{\varphi}' = \frac{L_{\rm AdS}}{J'\epsilon}\,, \qquad \bar{\varphi} =\frac{L_{\rm dS}}{J \epsilon}\,,
\eeq
and indeed under $L_{\rm AdS} \to -i L_{\rm dS}$ and $J' \to -i J$ the AdS dilaton is mapped to the dS dilaton.

If we instead mapped the disk to a time-reversed Hartle-Hawking geometry with a single \textit{past} boundary, then the continuation acts as $J'  \to iJ$. 

%***********************************************
\subsubsection{Double trumpet and global dS$_2$}
%***********************************************

Having understood the mapping from Euclidean AdS$_2$ to the Lorentzian dS$_2$ with the Hartle-Hawking geometry, we now turn to the analogous story for global dS$_2$ which has both a future and past boundary.

In fact we may reinterpret global dS$_2$ as the ``double trumpet'' geometry of~\cite{Saad:2019lba}, whose analytic continuation computes the semiclassical ramp of nearly AdS$_2$ gravity in~\cite{Saad:2018bqo}. This double trumpet is a topological annulus described by the metric
\beq
\label{E:doubletrumpet}
	ds'^2 = d\rho^2 + b^2\cosh^2\rho\, d\theta^2\,, 
\eeq
where $\theta$ has periodicity $2\pi$. While this is a constant curvature metric, there is no dilaton profile on it that solves the dilaton equation of motion subject to the dilaton boundary condition, and so this background is not a classical trajectory for any size of the boundary circles. Recall that the field equation for the dilaton states that its derivative is a conformal Killing vector. The problem is that there is no conformal Killing vector on the annulus which is outward pointing on both boundaries.

Comparing the metric~\eqref{E:doubletrumpet} on the double trumpet with the Lorentzian metric on global dS$_2$, 
\begin{equation*}
	ds^2 = -dt^2 + \alpha^2 \cosh^2t\, d\theta^2\,,
\end{equation*}
it is reasonable to expect that they can be mapped to each other. Doing so requires some delicate analytic continuation.

The double trumpet has a minimal length geodesic around the $\theta$-circle at $\rho=0$, and the length is $2\pi b$. (This convention differs slightly from that of~\cite{Saad:2019lba}, with $2\pi b_{\rm us} = b_{\rm them}$.) This fact is encoded in the gauge theory description in terms of a hyperbolic holonomy around the circle, and in the fundamental representation one finds 
\beq
\label{E:hyperbolicHolonomy}
	\text{tr}\left( \mathcal{P}\exp\left( \int_0^{2\pi} d\theta\,A'_{\theta}\right) \right)= 2\cosh(\pi b)\,.
\eeq

The geometry has two asymptotic boundaries, reached as $\rho\to \pm\infty$, and near each of them one imposes the boundary conditions~\eqref{E:EAdS2BC}, with boundary circles of sizes $\beta_{R,L}$. In the right region, near $\rho \to\infty$, we take the coordinate transformation
\beq
	\rho \to\rho - \ln \left( \frac{\pi}{\beta_R b}\right)
\eeq
so that
\beq
	ds'^2 = d\rho^2 + \left( \frac{\beta_R}{2\pi}\right)^2\left( e^{\rho} + b^2 e^{-\rho}\right)^2  d\theta^2\,,
\eeq
and we may put the cutoff slice at $e^{\rho} = \frac{1}{\epsilon_R}$. Taking the zweibein to be
\beq
	e'^1 = \frac{\beta_R}{2\pi}(e^{\rho} + b^2 e^{-\rho})d\theta\,, \qquad e'^2 = d\rho\,,
\eeq
the connection $A'$ is
\beq
A' = \begin{pmatrix} \frac{d\rho}{2} & - \frac{\pi}{\beta_R }\frac{b^2e^{-\rho}d\theta}{2} \\ - \frac{\beta_R e^{\rho} d\theta}{2\pi} & - \frac{d\rho}{2}\end{pmatrix}\,.
\eeq

Comparing this with the $A$~\eqref{E:dSA} describing the future half of global dS$_2$ (after changing $t\to t - \ln \left( \frac{\pi}{\beta_+ \alpha}\right)$ so as to be consistent with the future boundary condition), we see that $A$ and $A'$ match exactly upon making the analytic continuation
\beq
	b =i \alpha\,, \qquad \rho = t\,, \qquad \beta_R = \beta_+\,.
\eeq
Similarly, in the left region of the Euclidean geometry, near $\rho \to - \infty$, we take the coordinate transformation
\beq
	\rho \to- \rho + \ln \left( \frac{\pi}{\beta_L b}\right)
\eeq
so that the cutoff slice is at $e^{-\rho} = \frac{1}{\epsilon_L}$, and take the zweibein to be
\beq
	e'^1 = \frac{\beta_L}{2\pi} (e^{-\rho} + b^2e^{\rho})d\theta\,, \qquad e'^2 = d\rho\,.
\eeq
Analytically continuing as above along with
\beq
	\beta_L = \beta_-\,,
\eeq
the connection $A'$ takes the same form~\eqref{E:dSA} appearing in the past half of global dS$_2$,
\beq
	A= \begin{pmatrix} \frac{dt}{2} & \frac{\beta_-e^{-t}d\theta}{2\pi} \\ - \frac{\pi}{\beta_-}\frac{\alpha^2 e^{t} d\theta}{2} & -\frac{dt}{2} \end{pmatrix}\,.
\eeq

\begin{figure}[t]
	\centering
	\includegraphics[width=5in]{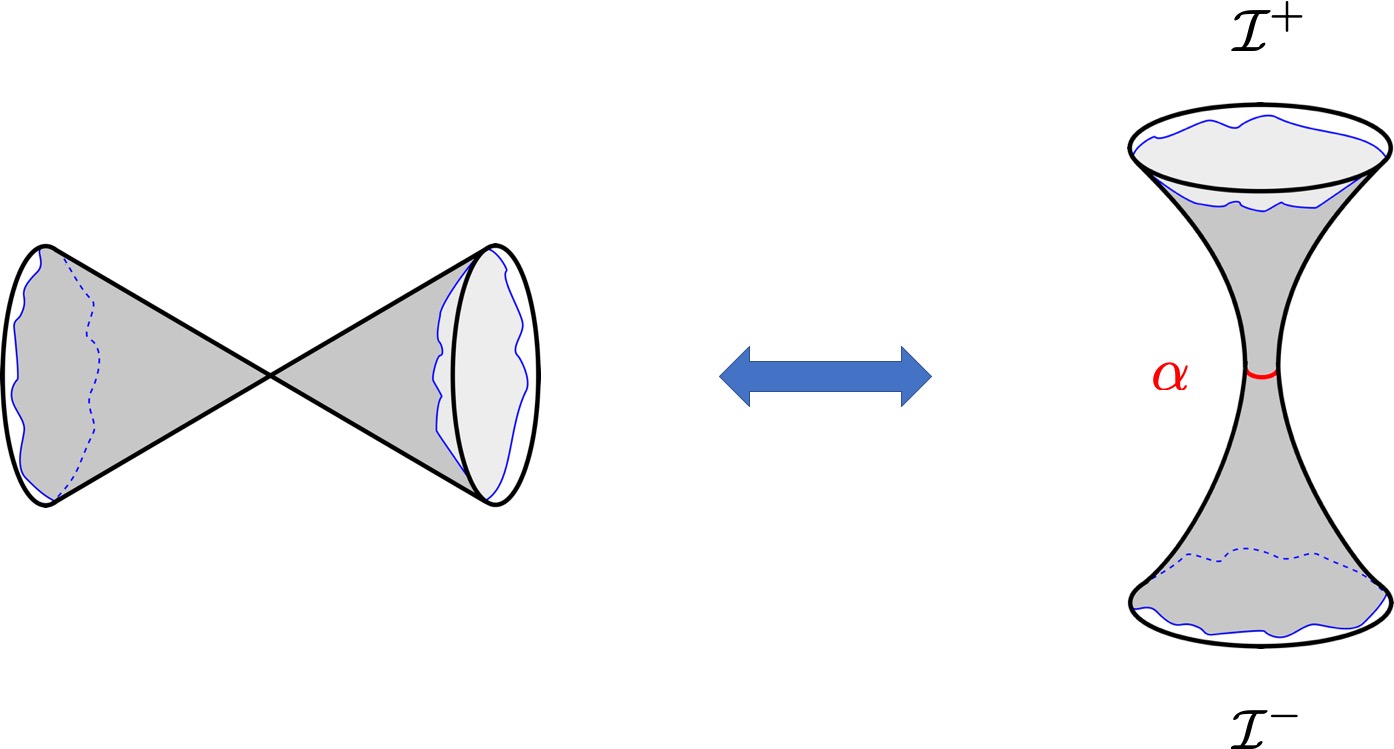}
	\caption{
		\label{F:trumpetTodS} Under the analytic continuation $b\to i \alpha$, $J'_R \to- i J$, $J'_L \to iJ$, the double trumpet of~\cite{Saad:2019lba} is mapped to nearly global dS$_2$.
		}
\end{figure}

Observe that the hyperbolic holonomy around the circle of the double trumpet,~\eqref{E:hyperbolicHolonomy}, is mapped under $b\to i \alpha$ to the elliptic holonomy~\eqref{E:ellipticHolonomy} characterizing global dS$_2$.  See Fig.~\ref{F:trumpetTodS}.

In nearly global dS$_2$ the dilaton goes from large and negative near past infinity, to large and positive near future infinity. In the ``double trumpet'' the dilaton is large and positive near both conformal boundaries. So, in mapping the double trumpet to nearly global dS$_2$, we must also analytically continue the dilaton boundary condition $J'$ as
\beq
	J'_R \to- i J\,, \qquad J'_L\to i J\,.
\eeq
Because we also continue $b\to i \alpha$, the Euclidean spacetime is actually a double hyperbolic cone rather than the hyperbolic cylinder. The cones are glued to each other at the tips, and the cone angles match and are given by $2\pi \alpha$.

%***********************************************
\subsubsection{Summary: the cosmological constant flipping map}
\label{S:ccflippingmap}
%***********************************************

For the disk and annulus we have found a continuation from Euclidean nearly AdS$_2$ geometries to Lorentzian nearly dS$_2$ geometries. The map has several parts. Firstly, the map takes the connection $A'$~\eqref{E:AdS2gauge} characterizing the Euclidean geometry to the connection $A$~\eqref{E:dSgauge} on the nose. It relates the first-order variables as
\beq
	(e'^1,e'^2,\omega') \quad \longleftrightarrow \quad  (\omega,e^0,e^1)\,, 
\eeq
with the convention that the ``radial'' component of the Euclidean zweibein is $e'^2$. Under this map the ``radial'' coordinate $\rho$ is mapped to time $t$. This is the ``trivial'' part of the map, and it boils down to $L_{\rm AdS} \to iL_{\rm dS}$. Secondly, the mapping specifies an analytic continuation of the dilaton boundary condition. In the Euclidean geometry the dilaton goes to $1/(J'\epsilon)$ on the boundary, and under the continuation we have
\beq
	J' \to \begin{cases} -i J\,, & \text{future} \\ i J \,, & \text{past} \end{cases}\,,
\eeq
depending on whether the asymptotic Euclidean AdS$_2$ region is mapped to a future or past asymptotically dS$_2$ region. Finally, the mapping specifies an analytic continuation of the holonomy around the boundary circles. The length $2\pi b$ around the ``bottleneck'' of the double trumpet was mapped to $2\pi i \alpha$, with $2\pi \alpha$ the length around the bottleneck of global dS$_2$.

What about higher genus spacetimes? In the next Section we discuss this question in detail, building upon our results for the disk and annulus above.

For now we note that, unsurprisingly, the boundary Schwarzian actions for the Poincar\'e disk and the double trumpet are directly mapped to those for the Hartle-Hawking geometry and global dS$_2$ respectively. In the conventions above, the Schwarzian boundary action for Jackiw-Teitelboim gravity on the disk reads
\beq
	S_E = -\frac{\varphi_0'}{4G_2} - \frac{1}{4G_2\beta' J'}\int_0^{2\pi} du \left( \{f(u),u\} + \frac{f'(u)^2}{2}\right)\,.
\eeq
Under $\varphi_0'\to \varphi_0$ and the map above to the future trumpet, $J' \to- i J$, we find
\begin{equation*}
	-S_E \to i S =  \frac{\varphi_0}{4G_2}+ \frac{i}{4G_2\beta J}\int_0^{2\pi} du\left(\{f(u),u\}+ \frac{f'(u)^2}{2}\right)\,,
\end{equation*}
which precisely matches the boundary action~\eqref{E:HHschwarzian} describing the Hartle-Hawking geometry. In fact, in both theories there is a $PSL(2;\mathbb{R})$ quotient under the action of fractional linear transformations on $\tan\left(\frac{f}{2}\right)$. Similarly, the Schwarzian boundary effective action for the double trumpet is~\cite{Saad:2019lba} in our conventions
\begin{align}
\begin{split}
	S_E = \frac{1}{4G_2}&\left[ \frac{1}{\beta_R J_R} \int_0^{2\pi} du\left( \{f_R(u),u\} - \frac{b^2}{2}f'_R(u)^2\right)\right.
	\\
	& \qquad \qquad \qquad \left.+ \frac{1}{\beta_L J_L}\int_0^{2\pi} du \left( \{f_L(u),u\} - \frac{b^2}{2}f'_L(u)^2\right)\right]\,,
\end{split}
\end{align}
which upon
\beq
	\beta_{R,L}\to \beta_{+,-}\,, \qquad f_{R,L} \to f_{+,-}\,, \qquad b \to i \alpha \,, \qquad J_R \to -i J\,, \qquad J_L\to i J\,,
\eeq
is mapped to the doubled action~\eqref{E:globalschwarzian} we derived for nearly global dS$_2$.

%***********************************************
\subsection{Boundary action from topological gauge theory}
%***********************************************

For completeness, let us briefly show how one can also get these Schwarzian boundary actions from the gauge theory formulation of Jackiw-Teitelboim gravity. The derivation in many respects resembles the one that two of us recently used~\cite{Cotler:2018zff} to rewrite the gravitational path integral on AdS$_3$ in terms of a boundary path integral. 

To start, let us parameterize global dS$_2$ in new coordinates,
\beq
	ds^2 = -\frac{dt^2}{t^2+1} + \alpha^2(t^2+1)d\theta^2\,,
\eeq
and pick a zweibein 
\beq
	e^0 = \frac{dt}{\sqrt{t^2+1}}\,, \qquad e^1 = \alpha\sqrt{t^2+1}d\theta\,.
\eeq
In these coordinates the connection $A$ describing global dS$_2$ reads
\beq
	A= \begin{pmatrix} \frac{dt}{\sqrt{t^2+1}} & \frac{\alpha}{2} (\sqrt{t^2 + 1} - t) \, d\theta \\
-\frac{\alpha}{2} (\sqrt{t^2 + 1} + t) \, d\theta & - \frac{dt}{2\sqrt{t^2 + 1}}\end{pmatrix}\,.
\eeq
The boundary conditions are now that at $t \to \infty$,
\beq
\label{E:newdSBC}
	A = \begin{pmatrix} \frac{dt}{2t} + O(t^{-2}) & O(t^{-1}) \\ -\frac{\beta_+ t}{2\pi} d\theta & -\frac{dt}{2t} + O(t^{-2}) \end{pmatrix}\,, \qquad B = \frac{2 \pi}{\beta J}t + O(t^0)\,,
\eeq
and as $t\to-\infty$,
\beq
	A = \begin{pmatrix} -\frac{dt}{2t} + O(t^{-2}) & -\frac{\beta_- t}{2\pi}d\theta \\ O(t^{-1}) & \frac{dt}{2t}+ O(t^{-2}) \end{pmatrix}\,, \qquad B = \frac{2\pi}{\beta J}t + O(t^0)\,.
\eeq
The boundary condition on future infinity of the Hartle-Hawking geometry is the same as that on future infinity of global dS$_2$.

Since $A$ is a flat connection it may be written
\beq
	A = \tilde{U}^{-1} d\tilde{U}\,,
\eeq
with $\tilde{U}\in PSL(2;\mathbb{R})$. One representative is
\beq
	\tilde{U} = \begin{pmatrix} \rho \cos\left( \frac{\alpha \theta}{2}\right) & \rho^{-1} \sin\left( \frac{\alpha \theta}{2}\right) \\ -\rho \sin\left( \frac{\alpha \theta}{2}\right) & \rho^{-1} \cos\left( \frac{\alpha \theta}{2}\right)\end{pmatrix}\,, \qquad \rho = \sqrt{\sqrt{t^2+1}+t}\,.
\eeq
This $\tilde{U}$ is in general multi-valued around the circle. We may write it in terms of a monodromy and a $U$ whose logarithm is single-valued
\beq
	\tilde{U} = \exp(\lambda \theta) U\,,
\eeq
with 
\beq
	\lambda = \begin{pmatrix} 0 & \alpha \\ -\alpha & 0 \end{pmatrix} \,, \qquad U = \begin{pmatrix} \rho \\ & \rho^{-1}\end{pmatrix}\,.
\eeq

For the Hartle-Hawking geometry, with $\alpha = 1$, $\tilde{U}$ exhibits a winding property. The gauge group is $PSL(2;\mathbb{R})$, which is contractible to a circle with $\pi_1(PSL(2;\mathbb{R})) = \mathbb{Z}$. The map $\tilde{U}$ takes the $\theta$-circle to this group circle. 

%***********************************************
\subsubsection{Hartle-Hawking}
%***********************************************

When quantizing around the Hartle-Hawking geometry, the strategy is to integrate out $B$ so that one is left with the residual action (after imposing the future boundary condition on $B$),
\beq
	S =-i \frac{\varphi_0}{4G_2} - \frac{1}{4G_2\beta J}\int_0^{2\pi} d\theta \,\text{tr}(A_{\theta}^2)\,.
\eeq
One then explicitly parameterizes flat connections as $A = U^{-1} dU$ with 
\beq
\label{E:HHU}
	U = \begin{pmatrix} \cos\left( \frac{f}{2}\right) & \sin\left(\frac{f}{2}\right) \\ -\sin\left(\frac{f}{2}\right) & \cos\left(\frac{f}{2}\right)\end{pmatrix} \begin{pmatrix} \Lambda & 0 \\ 0 & \Lambda^{-1} \end{pmatrix} \begin{pmatrix} 1 & \Psi \\ 0 & 1 \end{pmatrix}\,,
\eeq
where $\Lambda>0$, $\Psi \in \mathbb{R}$, and $f\sim f + 2\pi$. The winding property above implies that $f$ obeys the unconventional boundary condition
\beq
	f(\theta+2\pi) = f(\theta) + 2\pi\,.
\eeq
In terms of the component functions $(\Lambda,\Psi,f)$, the action reads
\beq
	S = -i\frac{\varphi_0}{4G_2} -\frac{1}{4G_2\beta J} \int_0^{2\pi} d\theta\left( \frac{2\Lambda'^2}{\Lambda^2} - \frac{f'^2}{2}  - \Lambda^2 f' \Psi'\right)\,,
\eeq
where $' = \partial_{\theta}$ and all fields are evaluated as $t\to\infty$. 

Imposing the future boundary conditions~\eqref{E:newdSBC} we find that $\Lambda$ and $\Psi$ are constrained as $t\to\infty$ in terms of $f$ via
\beq
\label{E:futureConstraint}
	\Lambda = \sqrt{\frac{\beta t}{\pi f'}} \,, \qquad \Psi = \frac{\pi}{\beta t}\frac{f''}{f'}\,,
\eeq
and $f$ is finite. Plugging these constrained values into the action above, we find
\beq
	S= -i\frac{\varphi_0}{4G_2} - \frac{1}{8G_2\beta J}\int_0^{2\pi} d\theta\left( \frac{f''^(\theta)2}{f'(\theta)^2}  - f'(\theta)^2 \right) = -i\frac{\varphi_0}{4G_2} + \frac{1}{4G_2\beta J} \int_0^{2\pi} d\theta\left( \{f(\theta),\theta\} + \frac{f'(\theta)^2}{2}\right)\,,
\eeq
which matches the Schwarzian action obtained in~\eqref{E:HHschwarzian}.

In the gauge theory formulation the $PSL(2;\mathbb{R})$ quotient arises rather naturally. In decomposing $A = U^{-1} dU$ we have introduced a $PSL(2;\mathbb{R})$ redundancy. Both $U(\theta,t)$ and $hU(\theta,t)$ give the same connection $A$ for any $h\in PSL(2;\mathbb{R})$. So in the residual integral over $U$ we identify these configurations. This left-action may be absorbed into a redefinition of the component fields $(\Lambda,\Psi,f)$. Letting $h = \begin{pmatrix} d & -c \\-b & a\end{pmatrix}$, we find that it acts on $f$ as
\beq
	\tan\left( \frac{f}{2}\right) \to \frac{a \tan\left( \frac{f}{2}\right) + b}{c\tan\left(\frac{f}{2}\right) + d}\,,
\eeq
with some more complicated action on $(\Lambda,\Psi)$. So we find the same Schwarzian boundary action along with the same $PSL(2;\mathbb{R})$ quotient.

%***********************************************
\subsubsection{Global dS$_2$}
\label{S:globalSchwarzianBF}
%***********************************************
\begin{figure}[t]%
	\centering
    \includegraphics[width=2.5in]{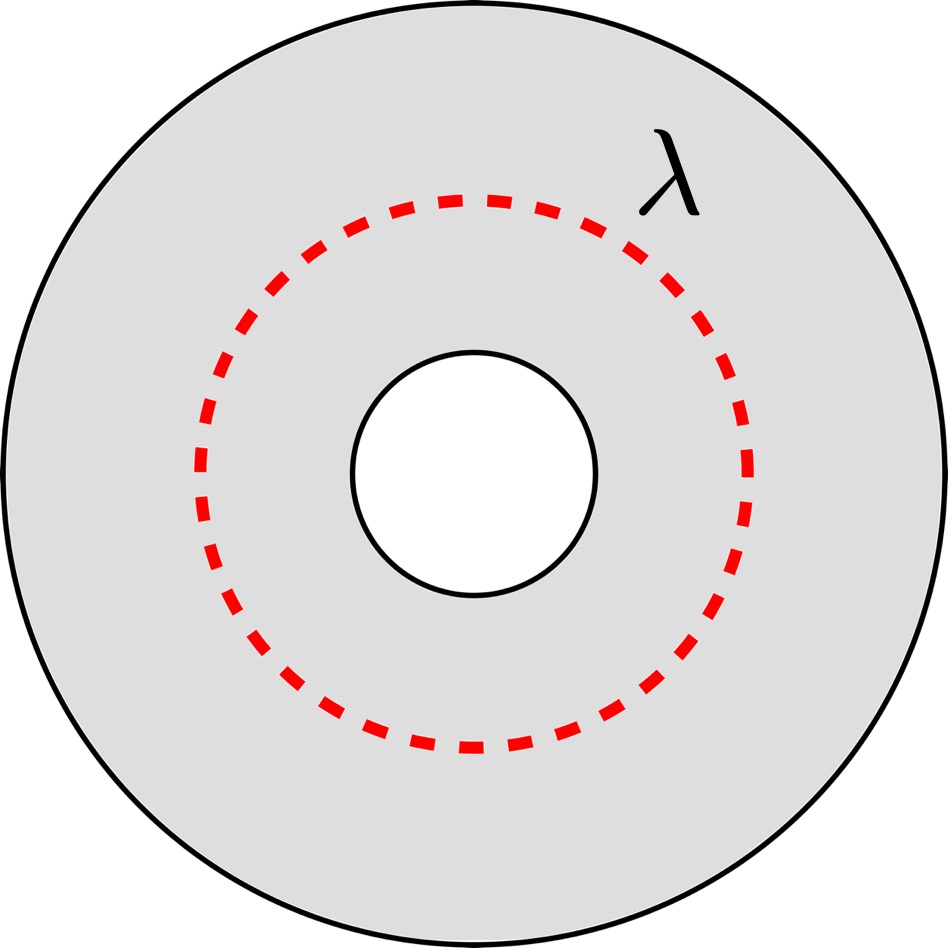}
	\caption{
		\label{F:annulus} A topological annulus, with a Wilson loop depicted by a dotted curve.  The monodromy is parameterized by $\lambda$.
	}
\end{figure}

The quantization on global dS$_2$ proceeds similarly, except now the space is a topological annulus and thus the moduli space of flat connections is parameterized by a monodromy. See Fig.~\ref{F:annulus}.  For dS$_2$ spacetimes with future and past trumpets, this monodromy is elliptic. We write
\beq
	\tilde{U}_+ = \exp\left( \lambda \theta\right) U_+\,,
\eeq
with $U_+$ and its logarithm single-valued, and the residual integral over flat $A$ becomes an integral over $\lambda$ and $U$. In this parameterization there is a $PSL(2;\mathbb{R})$ redundancy under
\beq
	\lambda \to h\lambda h^{-1} \,, \qquad U \to hU\,,
\eeq
which we may partially fix by taking
\beq
\label{E:fixedMonodromy}
	\lambda= \begin{pmatrix} 0 & \alpha \\ -\alpha & 0 \end{pmatrix} \,.
\eeq 
Now there is only a residual $U(1)$ redundancy under transformations of the form $h = \begin{pmatrix} \cos\left(\frac{\gamma}{2}\right) & \sin\left(\frac{\gamma}{2}\right) \\ -\sin\left(\frac{\gamma}{2}\right) & \cos\left(\frac{\gamma}{2}\right)\end{pmatrix}$. Parameterizing $U_+$ in the same way as $U$ above, we see that
\beq
	\tilde{U}_+ = \begin{pmatrix} \cos\left( \frac{\alpha \theta + f_+}{2}  \right) & \sin\left( \frac{\alpha \theta + f_+}{2}\right) \\ -\sin\left(\frac{\alpha \theta + f_+}{2}\right) & \cos\left(\frac{\alpha\theta+f_+}{2}\right)\end{pmatrix} \begin{pmatrix} \Lambda_+ & 0 \\ 0 & \Lambda_+^{-1}\end{pmatrix} \begin{pmatrix} 1 & \Psi_+ \\ 0 & 1 \end{pmatrix}\,.
\eeq
Here the component fields $(\Lambda_+,\Psi_+,f_+)$ are all periodic around the circle. Clearly we can absorb the monodromy into a redefinition of $f_+$ with
\beq
	\alpha \theta + f_+ \to \alpha f_+\,,
\eeq
where the redefined field obeys the Diff$(\mathbb{S}^1)$ boundary condition
\beq
	f_+(\theta+2\pi) = f_+(\theta)+2\pi\,.
\eeq
The future boundary conditions then fix $\Lambda_+$ and $\Psi_+$ in the same way as in~\eqref{E:futureConstraint}.

Near past infinity it is convenient to pass over to a different parameterization of $PSL(2;\mathbb{C})$ elements,
\beq
\label{E:globalPastU}
	U_- = \begin{pmatrix} \cos\left( \frac{f_-}{2}\right) & \sin\left(\frac{f_-}{2}\right) \\ -\sin\left(\frac{f_-}{2}\right) & \cos\left(\frac{f_-}{2}\right) \end{pmatrix} \begin{pmatrix} \Lambda_-^{-1} & 0 \\0 & \Lambda_-\end{pmatrix} \begin{pmatrix} 1 & 0 \\ -\Psi_- & 1 \end{pmatrix}\,.
\eeq
As above we then absorb the monodromy into a redefinition of $f_-$ via
\beq
\label{E:globalRedef}
	\alpha \theta + f_- \to \alpha f_-\,,
\eeq
which obeys
\beq
	f_-(\theta+2\pi) = f_-(\theta)+2\pi\,.
\eeq
The past boundary conditions fix $\Lambda_-$ and $\Psi_-$ near $t\to-\infty$ as
\beq
	\Lambda_- = \sqrt{-\frac{\beta t}{\pi f_-'} } \,, \qquad \Psi_- = -\frac{\pi}{\beta t}\frac{f_-''}{f_-'}\,,
\eeq
with $f_-$ finite. Because the past boundary term is flipped relative to the future one (this is accomplished by the ``$n$'' in the boundary term~\eqref{E:dSBFS}), we then find the action
\begin{equation*}
	S = \frac{1}{4G_2J}\left[ \frac{1}{\beta_+} \int_0^{2\pi} d\theta \left( \{f_+(\theta),\theta\} + \frac{\alpha^2}{2}f_+'(\theta)^2\right)- \frac{1}{\beta_-} \int_0^{2\pi}d\theta \left( \{f_-(\theta),\theta\} + \frac{\alpha^2}{2}f_-'(\theta)^2\right)\right]\,,
\end{equation*}
which matches what we obtained previously~\eqref{E:globalschwarzian} in the metric formulation. By fixing the monodromy to take the form in~\eqref{E:fixedMonodromy}, we only have a $U(1)$ gauge symmetry of our reparameterization, which simultaneously acts on the $f_{\pm}$ as
\beq
	f_{\pm}(\theta) \to f_{\pm}(\theta) + \delta\,.
\eeq

At integer $\alpha$ the derivation above breaks down, because the $PSL(2;\mathbb{R})$ redundancy is no longer partially fixed by~\eqref{E:fixedMonodromy}. As such, the case of integer $\alpha$ must be treated separately, and we do so in Subsection~\ref{S:integern}. One way to see that integer $\alpha$ is special is that the action above obtains an emergent axial $PSL^{(n)}(2;\mathbb{R})$ global symmetry for $\alpha = n$, and so na\"{i}vely the path integral is infinite. Another observation is that, at $\alpha = n$, $\tilde{U}$ is single-valued in $PSL(2;\mathbb{R})$. 

%***********************************************
\subsubsection{Aside: fixing $\alpha = 1$}
\label{S:fixingAlpha}
%***********************************************

In the derivation above it is clear that we can fix $\alpha$ to unity in the following way. In the original path integral over the monodromy $\lambda$ and group-valued field $U$, we can introduce a parameter $\mu$ and add the term to the action
\beq
	S \to S +\mu\left( \text{tr} (\lambda^2) -\mathfrak{L}\right)\,,
\eeq
which fixes the monodromy $\lambda$ up to similarity transform. If $\mathfrak{L} = -1/4$ then we arrive precisely at the $\alpha = 1$ doubled Schwarzian theory discussed at the end of Subsection~\ref{SS:globalSchwarzian}. From the gauge theory formulation we see that the only degrees of freedom of that model, after integrating out the Lagrange multiplier $\mu$, are the reparameterization fields $f_{\pm}$. The gauge symmetry of the model is the simultaneous $PSL(2;\mathbb{R})$ quotient discussed at the end of~\ref{SS:globalSchwarzian}.

%***********************************************
\subsection{Symplectic measure}
\label{S:dS2measure}
%***********************************************

One advantage of the topological gauge theory description is that we can use it to efficiently compute the measure for the degrees of freedom living on the boundary of nearly dS$_2$ spacetimes. We require this measure to compute the gravitational path integral over fluctuations around the Hartle-Hawking and global dS$_2$ geometries, and with it in hand we will be able to use existing results to compute the path integral to all orders in perturbation theory.

For a recent discussion of the derivation of the measure for boundary degrees of freedom in the context of AdS$_3$ gravity see~\cite{Cotler:2018zff}, and for Jackiw-Teitelboim theory in nearly AdS$_2$ spacetime see~\cite{Saad:2019lba}. Our discussion here is closely related to that of~\cite{Saad:2019lba}, and we arrive at the same results. Indeed the starting point of our derivation of the boundary measure, Eq.~\eqref{E:BFmeasure}, is the same as that of~\cite{Saad:2019lba}. We also refer the reader to the very nice discussion of~\cite{Kim:2015qoa} in the context of pure AdS$_3$ gravity.

To warm up, consider $BF$ theory with compact, connected gauge group $G$ and level $k$ on the disk,
\beq
	S_E = \frac{ik}{2\pi}\int \text{tr}(BF)\,.
\eeq
This theory reduces to the boundary model of a particle propagating in Euclidean time on $G$. The integration space of the model is the trivial coadjoint orbit of a Kac-Moody group at level $k$~\cite{Alekseev:1988ce}. (For recent introductions to coadjoint orbits see e.g.~\cite{Oblak:2016eij,Cotler:2018zff}.) That coadjoint orbit is a symplectic space by the Kirillov-Kostant theorem (for instance, see \cite{woodhouse1997geometric, khesin2008geometry}), and its symplectic form is inherited from the gauge field measure as follows.

After integrating out the adjoint scalar $B$ one is left with a residual integral over the moduli space of flat spatial connections $A= U^{-1} dU$. This space is symplectic, with symplectic form~\cite{Witten:1991we}
\beq
\label{E:BFmeasure}
	\omega= \frac{k}{2\pi} \int d^2x \, \epsilon^{ij} \text{tr}\left( dA_i\wedge dA_j\right)\,.
\eeq
The $d$ appearing in $dA_i$ is not the exterior derivative. It refers to a formal one-form in the space of flat connections, satisfying $d\partial_i A_j = \partial_i dA_j$. One may readily verify that this $\omega$ is gauge-invariant. Parameterizing a general variation of $A$ through a variation of $U$, $dU = (dX) U$ with $X\in\mathfrak{g}$ a vector, we have
\beq
\label{E:BFmeasure2}
	\omega = \frac{k}{2\pi} \int d^2x \,\epsilon^{ij} \text{tr}\left( d\partial_i X \wedge d \partial_j X\right) = \frac{k}{2\pi} \int_0^{2\pi} d\theta \,\text{tr}(dX \wedge dX')\,,
\eeq
where $' = \partial_{\theta}$ is the angular derivative on the disc. Note that the symplectic form only depends on the boundary value of the field $U$, and that morally $X$ is canonically conjugate to $X'$. This is precisely the Kirillov-Kostant symplectic form on the basic coadjoint orbit of Kac-Moody~\cite{Alekseev:1988ce} at level $k$ whose quantization leads to the vacuum representation.

We have not yet accounted for a gauge redundancy. By parameterizing the connection as $U^{-1}dU$ we have introduced a redundancy under $U(\theta) \sim hU(\theta)$, or equivalently under $X(\theta) \sim h X(\theta)h^{-1}$ and $U(\theta)\sim h U(\theta)$ for $h\in G$. Vectors on the orbit may be understood as vector fields $X(\theta)$ modulo this identification. It may be fixed in a local way, e.g. $U(0) = 1$, or equivalently $X(0)=0$. 

The corresponding gauge-fixed measure may be denoted as $[dX]\text{Pf}(\omega)$ or more precisely $\left( \prod_{\theta>0} dX(\theta)\right) \text{Pf}(\omega)$. In fact it is equivalent to $\prod_{\theta>0} d\mu(U(\theta))$ with $d\mu(U)$ the Haar measure on $G$. To motivate this equivalence, note that the symplectic form above is invariant both under the left action (here $X\to X, U \to Uh^{-1}$) and the right action ($X \to h X h^{-1}, U \to hU$). The quotient breaks the right-invariance, but since we have fixed the quotient in a local way, as long as we are away from $\theta=0$ the measure is local and both left- and right-invariant. Parameterizing the integral over $U$ as one over $X$, the Pfaffian is proportional to $\prod_{\theta>0} \sqrt{g}$ with $g$ the Killing-Cartan metric on $\mathfrak{g}$, and so we find the Haar measure.

There is another point to note before going on. A priori the normalization of the symplectic form is a choice of convention, but in $BF$ theory there is a very natural choice. The integration space is a coadjoint orbit of a Kac-Moody group, and we simply fix our normalization to coincide with that of the Kirillov-Kostant symplectic form. It has the property that the symplectic volume is very large in the weak coupling limit $k \gg 1$. Further, the symplectic flux is quantized in this convention with $\oint \omega \in 2\pi k \mathbb{Z}$.

Now let us treat the $PSL(2;\mathbb{R})$ topological gauge theory corresponding to nearly dS$_2$ gravity. On the Hartle-Hawking background we have the same expression as above for the symplectic form on the moduli space of flat connections
\beq
\label{E:PSLomega}
	\omega = \frac{1}{4\pi G_2} \int_0^{2\pi} d\theta \,\text{tr}(dX \wedge dX')\,.
\eeq
(The normalization here is fixed by the value of $k$ we find for the $BF$ description of nearly dS$_2$ gravity, $k = \frac{1}{2G_2}$.) Taking $U$ in~\eqref{E:HHU}, plugging in the constrained values for $\Lambda$ and $\Psi$~\eqref{E:futureConstraint} in terms of $f$, and computing the variation $dX$ induced by a variation $df$, we find a rather nontransparent result. However, plugging it into the expression above and after integrating by parts, we arrive at the familiar symplectic form
\beq
	\omega = \frac{1}{8\pi G_2}\int_0^{2\pi} d\theta\left( \frac{df'\wedge df''}{f'^2} - df\wedge df'\right)\,.
\eeq
This is nothing more than the Kirillov-Kostant symplectic form on the integration space for $f$, $\text{Diff}(\mathbb{S}^1)/PSL(2;\mathbb{R})$, which may be understood as a coadjoint orbit of the Virasoro group~\cite{Witten:1987ty,Alekseev:1988ce} at central charge $C = 6/G_2$. This measure was conjectured to be the correct one for the Schwarzian model as obtained from Jackiw-Teitelboim gravity and from SYK~\cite{Stanford:2017thb}, and shown to be the right measure for nearly AdS$_2$ gravity in~\cite{Saad:2019lba}.

The $PSL(2;\mathbb{R})$ quotient may be locally fixed by e.g. $f(0) = 0, f'(0) = 1, f''(0) = 0$, in which case the measure above, $\prod_{\theta} [df] \delta(f(0))\delta(f'(0)-1)\delta(f''(0))\text{Pf}(\omega)$, may be expressed as $\prod_{\theta>0} \frac{df}{f'}$~\cite{Stanford:2017thb}, and this form of the measure has been put to great use~\cite{bagrets2016sachdev} in the study of the Schwarzian path integral. This is also what one gets from the Haar measure for $PSL(2;\mathbb{R})$ at each $\theta>0$, $\prod_{\theta>0} d\mu(U(\theta))$, plugging in the constrained values for $\Lambda$ and $\Psi$, and then reducing the Haar measure to an integral over the component $f$.

It is a useful fact that Diff$(\mathbb{S}^1)/PSL(2;\mathbb{R})$ is not only symplectic, but K\"ahler (see e.g.~\cite{Witten:1987ty}). This has the practical consequence here that the measure $\frac{ [df]\text{Pf}(\omega)}{PSL(2;\mathbb{R})}$ is positive-definite. 

The analysis for global dS$_2$ proceeds in basically the same way. The symplectic form now reads
\beq
\label{E:globalomega}
	\omega = \frac{1}{4\pi G_2} \int_0^{2\pi} d\theta\,\text{tr}\left( dX_+ \wedge dX_+' - dX_- \wedge dX_-'\right)\,,
\eeq
where $d\tilde{U}_{\pm} = (dX_{\pm})\tilde{U}_{\pm}$. Absorbing the monodromy into a redefinition of $f_{\pm}$ via~\eqref{E:globalRedef} we arrive at the measure
\beq
\label{E:alphaomega}
	\omega = \frac{1}{8\pi G_2} \int_0^{2\pi} d\theta \left( \frac{df_+'\wedge df_+''}{f_+'^2} - \alpha^2 df_+\wedge df_+'-\alpha (f'_+df_+-f_+df_+') \wedge d\alpha  - (+\to -)\right) \,.
\eeq

For non-integer $\alpha$ the first two terms, and their $+ \to -$ partners, each comprise the Kirillov-Kostant symplectic form on the orbit Diff$(\mathbb{S}^1)/U(1)$ of the Virasoro group (although the total integration space $(f_+(\theta),f_-(\theta),\alpha)$ is not two copies of this orbit, but instead the space $(\text{Diff}(\mathbb{S}^1)\times \text{Diff}(\mathbb{S}^1)\times \mathbb{R}_+)/U(1)$). For integer $\alpha$ our analysis above is corrected. The space Diff($\mathbb{S}^1)/U(1)$ is K\"ahler for $\alpha < 1$, while for $\alpha>1$ fluctuations of $f$ have negative directions. In what follows we address this by rotating the contour of field integration for the negative modes, effectively flipping the negative directions into positive ones.

What does the third term in~\eqref{E:alphaomega}, and its past partner, correspond to? Let us consider the fixed variation $df_+ = d\gamma$, $df_- = 0$, along with $d\alpha$. The fluctuations of the $f_{\pm}$ corresponds to a twist, and since the $f_{\pm}$ are Diff$(\mathbb{S}^1)$ fields we have $\gamma\sim \gamma+2\pi$. The part of the symplectic form sensitive to these fluctuations is the third term, and it gives
\beq
	\omega = \frac{1}{2 G_2} \,\alpha d\alpha \wedge d\gamma\,,
\eeq
and so we see that $\alpha^2$ and $\gamma$ are canonically conjugate. A similar result for the measure on the moduli space of the double trumpet was derived in~\cite{Saad:2019lba}. In fact our symplectic form is precisely theirs upon the analytic continuation $b_{\rm them}=2\pi b_{\rm us} \to 2\pi i \alpha$.

As in our discussion of $BF$ theory with compact gauge group, we fix our normalization of the symplectic form to be the same as that of the corresponding coadjoint orbit. As before, the symplectic volume is large when the gauge theory is weakly coupled, equivalently that gravitational interactions are weak, $G_2\to 0$. There is one important distinction, however, which arises with non-compact gauge group.  When the gauge group is compact there is a natural quantization condition on the coupling constant, which is just the usual statement that the symplectic structure is an element of integer cohomology.  In the present case, however, the phase space $\text{Diff} (\mathbb{S}^1)/ U(1)$ has no closed 2-cycles, so there is no quantization condition to impose.  We will therefore simply choose $\omega$ to be normalized with the factor of $\frac{1}{G_2}$ written above, although there does not appear to be a unique choice for this normalization.

%***********************************************
\subsection{The special case of integer $\alpha $}
\label{S:integern}
%***********************************************

In this Subsection we carefully treat the case where $\alpha = m$ integer and $\beta_+ = \beta_-=\beta$. We follow the same procedure as in Subsection~\ref{S:globalSchwarzianBF}, except that now we cannot partially fix the $PSL(2;\mathbb{R})$ symmetry. For a general monodromy
\beq
	\lambda = \lambda^a P_a + \lambda_2 \Omega\,,
\eeq
we parameterize $U$ near future and past infinity as in~\eqref{E:HHU} and~\eqref{E:globalPastU} respectively, and define the quantities
\beq
	\gamma_{\pm} = \lambda_1\pm  \lambda_0 \sin(f_{\pm})  \pm \lambda_2 \cos(f_{\pm} )\,.
\eeq
After integrating out $B$, from~\eqref{E:dSBFS} and the boundary condition on $B$ we find the boundary action
\beq
	S= -\frac{1}{4G_2 \beta J}\int_0^{2\pi} d\theta\left( \text{tr}\left( U_+^{-1} U_+' U_+^{-1} U_+' + 2 \lambda U_+'U_+^{-1} +\lambda^2\right) - (+\to -)\right)\,,
\eeq
which may be written in terms of the component fields as
\beq
	S = -\frac{1}{4G_2 \beta J}\int_0^{2\pi} d\theta\left( \frac{2\Lambda_+'^2}{\Lambda^2} - \frac{f_+'^2}{2} - \Lambda_+^2\Phi_+ \Psi_+' - \lambda_1 f_+' - \frac{2\gamma_+'}{f_+'} \frac{\Lambda_+'}{\Lambda_+} - (+\to-)\right)\,,
\eeq
where we have defined
\beq
	\Phi_{\pm} = f_{\pm}' + \gamma_{\pm}\,.
\eeq
The future boundary conditions fix $\Lambda_+$ and $\Psi_+$ as $t\to\infty$ to be
\beq
	\Lambda_{\pm} = \sqrt{\pm\frac{\beta t}{\pi \Phi_{\pm}}}\,, \qquad \Psi_{\pm} = \pm\frac{\pi}{\beta t}\left( \frac{\gamma_{\pm}'}{f_{\pm}'} + \frac{\Phi_{\pm}'}{\Phi_{\pm}}\right)\,,
\eeq
with $f_{\pm}$ finite. 

Plugging these values back into the action and after some integrations by parts we arrive at the boundary action
\begin{align}
\begin{split}
	S = -\frac{1}{8G_2\beta J}\int_0^{2\pi} d\theta \left( \frac{\Phi_+'^2}{\Phi_+^2} - f_+'^2 -2 \lambda_1 f_+' + \frac{2\gamma_+'}{f_+'} \frac{\Phi_+'}{\Phi_+} - (+\to -)\right)\,.
\end{split}
\end{align}
The $PSL(2;\mathbb{R})$ redundancy of our description can be shown to act on the various fields as
\beq
	\lambda \to h \lambda h^{-1}\,, \quad  h = \begin{pmatrix} d & -c\\ -b & a\end{pmatrix}\,,
\eeq
along with
\beq
\tan\left( \frac{f_+}{2}\right) \to \frac{a\tan\left(\frac{f_+}{2}\right)+b}{c\tan\left(\frac{f_+}{2}\right)+d}\,, \quad -\cot\left(\frac{f_-}{2}\right) \to \frac{a \left(-\cot\left(\frac{f_-}{2}\right)\right) + b}{c\left(-\cot\left(\frac{f_-}{2}\right)\right)+d}\,.
\eeq

The crucial feature of this action is that, whereas our boundary action for general $\alpha$ suggests the appearance of zero modes at integer $\alpha$, the would-be zero modes are lifted by the monodromy. To see this let us compute the quadratic action around the critical point corresponding to dS$_2$. A convenient presentation of that critical point is
\beq
	f_{\pm}(\theta) = m\theta\,, \qquad \lambda = 0\,.
\eeq
Allowing for fluctuations of the reparameterization fields,
\beq
	f_{\pm}(\theta) =m \theta + \sum_{n} \epsilon^{\pm}_n e^{in\theta}\,,
\eeq
and taking $\lambda$ of the same order as the $\epsilon^{\pm}$, we find a quadratic effective action
\begin{align}
\begin{split}
	S =  -\frac{\pi}{4G_2\beta J}\left( \sum_n \frac{n^2}{m^2}(n^2-m^2)(|\epsilon^+_n|^2 -|\epsilon_n^-|^2)+ \frac{4}{m}\left(\lambda^{(+)} \epsilon^{(-)} + \lambda^{(-)} \epsilon^{(+)}\right)\right)\,,
\end{split}
\end{align}
where we have defined the combinations
\beq
	\lambda^{(\pm)} = \lambda^0 \pm i \lambda^2\,, \qquad \epsilon^{(\pm)} = \frac{\epsilon^+_{\pm m} + \epsilon^-_{\pm m}}{2}\,, \qquad \epsilon^{(0)} = \epsilon^+_0 - \epsilon^-_0\,,
\eeq
and used that $\epsilon_{-n} = \epsilon_n^*$.

For $\alpha = m$, infinitesimal $PSL(2;\mathbb{R})$ transformations leave the $\lambda$'s invariant and act as
\beq
	\delta\epsilon^+_m = - \delta\epsilon^-_m \,, \qquad \delta\epsilon^+_{-m} = -\delta\epsilon^-_{-m}\,, \qquad \delta\epsilon^+_0 = \delta\epsilon^-_0\,.
\eeq
The quadratic action above is then manifestly invariant. The combinations $\epsilon_{\pm}$ are the would-be zero modes; they are lifted through their coupling to the monodromy. The only exact zero modes are $\epsilon^{(0)}$ and $\lambda_1$. Both are expected. The first is the Goldstone boson corresponding to spontaneously broken axial rotations. The $\lambda_1$ zero mode can be thought of as a fluctuation of $\alpha$. When  $\beta_+ = \beta_-$ the on-shell action vanishes, and so fluctuations of $\alpha$ are also a zero direction.

These facts have their counterparts in the symplectic measure, now including the complete monodromy. Evaluating the symplectic form~\eqref{E:globalomega} to quadratic order in fluctuations around the global dS$_2$ critical point described above, we find
\begin{align}
	\omega = \frac{1}{4 G_2}& \left( i \sum_n\ \frac{n}{m^2}(n^2-m^2)(d\epsilon^+_{-n} \wedge d\epsilon_n^+ - d\epsilon^-_{-n} \wedge d\epsilon_n^-)\right.
	\\
	\nonumber
	& \quad  \left. \phantom{\sum_n}+4 id\lambda^{(+)} \wedge d\epsilon^{(-)} - 4 id\lambda^{(-)} \wedge d\epsilon^{(+)} +4  d\epsilon^{(0)} \wedge d\lambda^1+\frac{3i}{2m} d\lambda^{(-)}\wedge d\lambda^{(+)} \right)\,.
\end{align}
The second line tells us that the gauge-invariant fluctuations $\epsilon^{(0),(+),(-)}$ of the past and future reparameterizations are conjugate to the monodromy. 

%***********************************************
\subsection{Gravitational scattering}
\label{S:scattering}
%***********************************************

In this manuscript we dedicate most of our attention to the study of the gravitational path integral in nearly dS$_2$ spacetime and in dS$_3$. However it is worth noting that the boundary path integrals obtained in this work can be used to compute gravitational scattering. This is well-known in nearly AdS$_2$ gravity~\cite{Jensen:2016pah,Maldacena:2016upp,Engelsoy:2016xyb}, and more recently in AdS$_3$~\cite{Cotler:2018zff}. Using a relation between the Schwarzian theory and two-dimensional Liouville theory the authors of~\cite{Mertens:2017mtv} obtained integral expressions for gravitational scattering in nearly AdS$_2$ which hold to all orders in perturbation theory in $G_2$. Here we sketch how gravitational scattering works in global dS$_2$ with $\alpha = 1$.

The basic idea is the following. Suppose we have a minimally coupled scalar field $\chi$ propagating in the global $\alpha = 1$ dS$_2$ spacetime with metric
\beq
	ds^2 = - dt^2 + \cosh^2t \, d\theta^2\,.
\eeq
Using the natural holographic dictionary appropriate for de Sitter spacetime, there is a dual operator $\mathcal{O}_{\pm}$ of dimension $\Delta$ on the future and past boundaries satisfying
\beq
	\Delta(1 -\Delta) = m^2\,,
\eeq
and let us restrict our attention to $m^2 \in [0,1/2]$ so that $\Delta \in [0,1]$.  For simplicity we will also take $\beta_{\pm} = 2\pi$. The isometries of dS$_2$ act as conformal transformations on the boundary, which constrain the boundary-to-boundary propagator for $\chi$. From that propagator one reads off the two-point function of boundary operators~\cite{Strominger:2001pn}
\begin{align}
\begin{split}
	\langle \mathcal{O}_+(\theta_1) \mathcal{O}_+(\theta_2) \rangle& =\langle \mathcal{O}_-(\theta_1)\mathcal{O}_-(\theta_2)\rangle^* =   \frac{1}{\left(2\sin\left( \frac{\theta_{12}}{2}\right)\right)^{2\Delta}}\,, 
	\\  \\
	\langle \mathcal{O}_+(\theta_1)\mathcal{O}_-(\theta_2) \rangle& = \langle \mathcal{O}_-(\theta_1)\mathcal{O}_+(\theta_2)\rangle^* = \frac{\cos(\pi \Delta)}{\left( 2 \cos\left( \frac{\theta_{12}}{2}\right)\right)^{2\Delta}}\,.
\end{split}
\end{align}
The latter is, up to the $\cos(\pi \Delta)$ proportionality factor, the former with $\theta_2\to \theta_2+\pi$. This property is sometimes called the ``KMS'' symmetry of de Sitter.

The two-point functions $\langle \mathcal{O}_+(\theta_1)\mathcal{O}_-(\theta_2) \rangle$ are comprised of an operator of dimension $\Delta$ in the future, and another operator of dimension $\Delta$ in the past. Denoting $\lambda_{\pm}(\theta)$ to be the sources conjugate to $\mathcal{O}_{\pm}$ via the bulk-to-boundary dictionary, the effect of integrating out $\chi$ at the classical level is to rewrite its action as a pure boundary term,
\beq
	iS_{\chi} = \frac{1}{2}\int_0^{2\pi} d\theta_1 d\theta_2 \sum_{i,j=+,-}\lambda_i(\theta_1)\lambda_j(\theta_2) \langle \mathcal{O}_i(\theta_1)\mathcal{O}_j(\theta_2)\rangle\,.
\eeq

At fixed $\alpha$, the moduli space of $R=2$ metrics on global dS$_2$ may be generated by simply acting on the basic geometry $-dt^2 + \alpha^2 \cosh^2t d\theta^2$ with asymptotic symmetries which act as independent local conformal transformations $f_{\pm}(\theta)$ on the future and past boundaries. The classical field theory of $\chi$ on any one of these metrics may be similarly rewritten as a boundary term, which is just the reparameterization of the action above,
\beq
	iS_{\chi} = \frac{1}{2}\int_0^{2\pi} d\theta_1 d\theta_2\sum_{i,j=+,-} \lambda_i(\theta_1)\lambda_j(\theta_2)\Big(( f_i'(\theta_1)f_j'(\theta_2))^{\Delta} \langle \mathcal{O}_i(f_i(\theta_1)) \mathcal{O}_j(f_j(\theta_2))\rangle\Big)\,.
\eeq
This gives us an effective coupling between sources for the boundary operators $\mathcal{O}_{\pm}$ and the reparameterization fields. The operator in question is a bilocal $\mathcal{B}_{ij}(\theta_1,\theta_2;\Delta)$ with
\begin{align}
\begin{split}
	\mathcal{B}_{++}(\theta_1,\theta_2;\Delta) &= \left(\frac{f_+'(\theta_1)f_+'(\theta_2)}{4\sin^2\left(\frac{f_+(\theta_1)-f_+(\theta_2)}{2}\right)}\right)^{\Delta}\,, 
	\\
	\mathcal{B}_{--}(\theta_1,\theta_2;\Delta) &=  \left(\frac{f_-'(\theta_1)f_-'(\theta_2)}{4\sin^2\left(\frac{f_-(\theta_1)-f_-(\theta_2)}{2}\right)}\right)^{\Delta}\,,
	\\
	\mathcal{B}_{+-}(\theta_1,\theta_2;\Delta) & = \cos(\pi \Delta) \left( \frac{f_+'(\theta_1)f_-'(\theta_2)}{4\cos^2\left( \frac{f_+(\theta_1)-f_-(\theta_2)}{2}\right)}\right)^{\Delta}\,.
\end{split}
\end{align}
One may readily verify that these bilocal operators are invariant under the simultaneous $PSL(2;\mathbb{R})$ transformation
\beq
	\tan\left(\frac{f_+}{2}\right) \to \frac{a \tan\left(\frac{f_+}{2}\right) + b}{c\tan\left(\frac{f_+}{2}\right)+d}\,, \qquad -\cot\left(\frac{f_-}{2}\right) \to \frac{a\left(-\cot\left(\frac{f_-}{2}\right)\right) + b}{c\left(-\cot\left(\frac{f_-}{2}\right)\right) + d}\,,
\eeq
which is a gauge symmetry of the $\alpha = 1$ model.

How do we interpret these bilocal operators? Suppose that we turn on a source $\lambda_{\pm}(\theta)$ on the boundary, say delta localized at $2n$ points. Within the free scalar theory of $\chi$ we would obtain a boundary $2n$-point function of $\mathcal{O}$'s which is just the generalized free-field result, computed from Witten diagrams with $n$ boundary-to-boundary propagators. Consider a single one of these diagrams, in which a boundary-to-boundary propagator attaches $\theta_1$ to $\theta_2$, $\theta_3$ to $\theta_4$ and so on. Now consider the correlation function of $n$ bilocals within the Schwarzian theory, stitched in the same order. 

\begin{figure}[t]%
	\centering
	\includegraphics[width=4.5in]{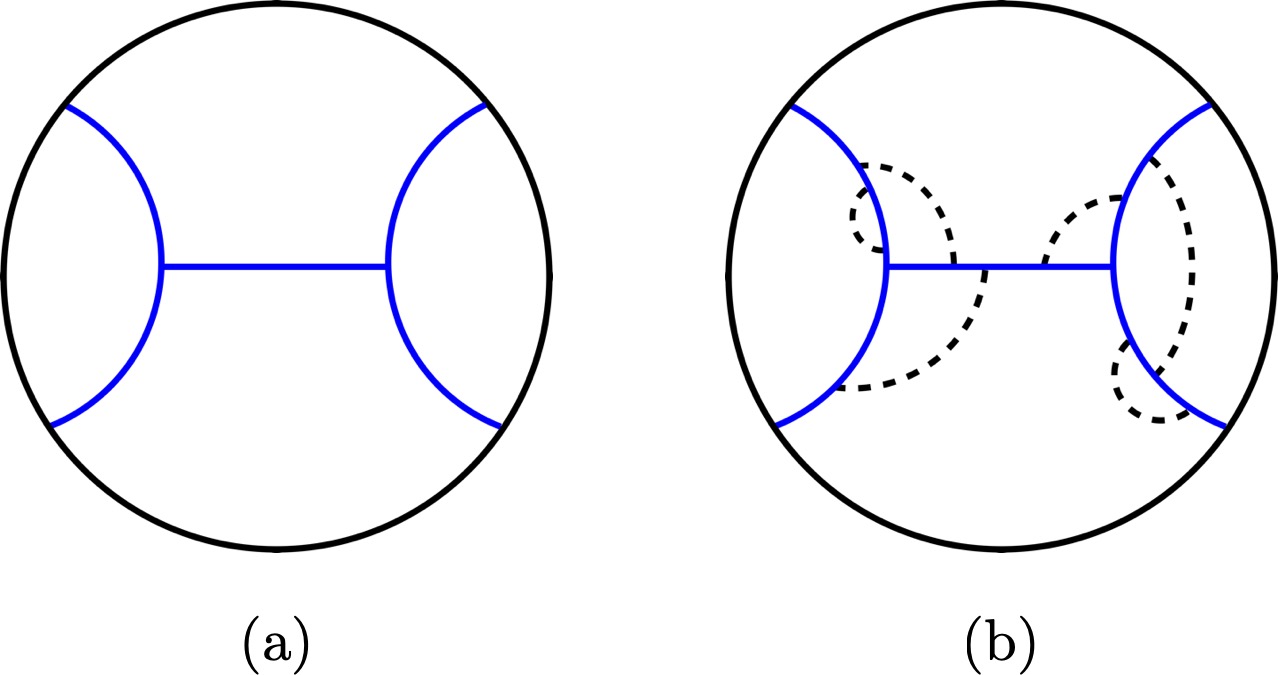}
	\caption{
		\label{F:dressing} (a) A Witten diagram, with a spatial slice of dS$_2$ drawn as a disc. The solid lines are scalars, and there is a single scalar exchange. This process is encoded in the boundary Schwarzian description as an operator, the reparameterized four-point function, endowing it with a coupling to the reparameterization field.  (b) The expectation value of the matter four-point function in the Schwarzian theory dresses the process with an infinite number of corrections involving exchanges of the reparameterization modes, indicated with dotted lines.
	}
\end{figure}

Since $G_2\to 0$ is the weak coupling limit of the Schwarzian theory, as $G\to 0$ this correlation function simply reduces to the product of $n$ bilocal operators evaluated on the classical trajectory, which is $f_{\pm} = \theta$. In this limit we then recover free field theory in dS. However, in perturbation theory in $G_2$ there are new diagrams in which the reparameterization field is exchanged between the various bilocals. These diagrams exactly correspond to the gravitational Witten diagrams in Jackiw-Teitelboim theory. One way to think about the matter is that the non-gravitational Witten diagram is ``dressed'' by an infinite sum of gravitational corrections which are all computed by the Schwarzian model. See Fig.~\ref{F:dressing}. 

There is one complication in global dS$_2$ which is not present in the usual nearly AdS$_2$ story. Namely, there is a $U(1)$ manifold of $\alpha=1$ de Sitter solutions with a relative twist between past and future. This twist is invisible in correlation functions of all past or all future operators, but appears in mixed future/past correlators. 

Given a diagram $\mathcal{D}$ which depends on $n_+$ future angles $\theta_i$ and $n_-$ past angles $\psi_m$, $\mathcal{D}(\theta_i,\psi_m)$, computed from global dS$_2$ with zero twist, the diagram evaluated in the background with twist $\gamma$ is $\mathcal{D}(\theta_i+\gamma,\psi_m)$. Integrating over the twist zero mode $\mathcal{D}$ contributes to an observable as
\beq
	\frac{1}{2\pi}\int_0^{2\pi}d\gamma\, \mathcal{D}(\theta_i+\gamma,\psi_m)\,.
\eeq

We expect that there is a similar story for $\alpha \neq 1$. We leave it for future study.

%***********************************************
\subsubsection{One-loop correction to two-point function}
%***********************************************

\begin{figure}[t]%
	\centering
	\includegraphics[width=4.5in]{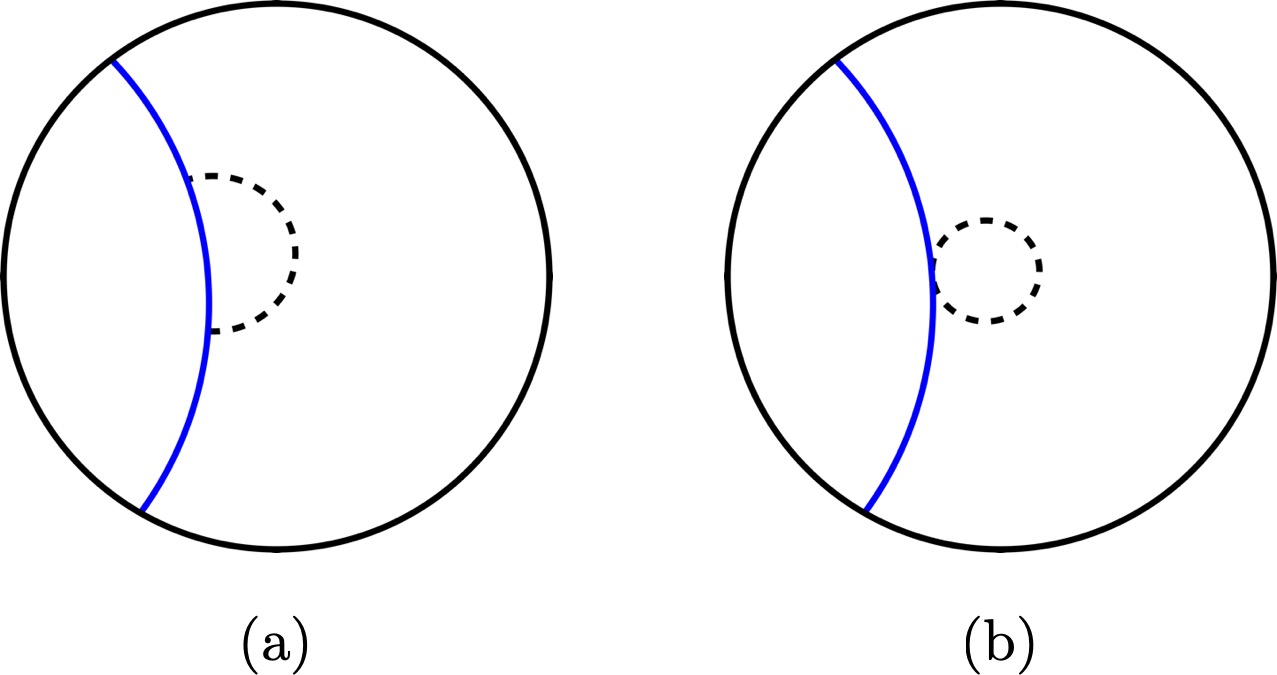}
	\caption{
			\label{F:oneloopB}
			The two one-loop diagrams that correct the two-point function at $O(1/C) = O(G_2)$.}%
\end{figure}

To illustrate the machinery let us compute the one-loop contribution to the two-point function of boundary operators. That is, we want $\langle \mathcal{B}_{ij}(\theta_1,\theta_2;\Delta)\rangle$ to $O(G_2)$. See the two diagrams in Fig.~\ref{F:oneloopB}. This correction to the past-past and future-future two-point function can be obtained from analytic continuation of existing results in the ordinary Schwarzian model, while the past-future correction cannot. 

To compute them we require the propagator of the reparameterization field as well as the linear and quadratic couplings of the bilocal to the reparameterization field. Expanding around the critical point 
\beq
	f_{\pm}(\theta) = \theta + \epsilon_{\pm}(\theta)\,,
\eeq
the quadratic effective action is when $\beta_+ = \beta_- = \beta$
\beq
	S = -\frac{1}{8G_2 \beta J}  \int_0^{2\pi} d\theta\, \Big(( \epsilon_+''^2 - \epsilon_+'^2) -(\epsilon_-''^2-\epsilon_-'^2)\Big) \,.
\eeq
Letting $C = \frac{\pi}{4  G_2 \beta J}$, the propagators are
\begin{align}
\nonumber
	\langle \epsilon_+(\theta) \, \epsilon_+(0) \rangle &=\langle \epsilon_-(\theta)\epsilon_-(0)\rangle^* = \frac{i}{2\pi C} \left(- \frac{(|\theta|-\pi)^2}{2} + (|\theta| - \pi)\sin|\theta| + 1 + \frac{\pi^2}{6} + \frac{5}{2} \, \cos(\theta) \right) 
	\\
	\langle \varepsilon_+(\theta) \, \varepsilon_-(0) \rangle &= \langle \epsilon_-(\theta) \, \epsilon_+(0) \rangle^*  = 0\,.
\end{align}
Note that $\epsilon_+$ and $\epsilon_-$ do not directly couple to one another.

The bilocal operators have linearized and quadratic couplings to the $\epsilon$ fields, e.g.
\beq
	\mathcal{B}_{++}(\theta_1,\theta_2;\Delta) =\left( \frac{1}{2\sin\left(\frac{\theta_{12}}{2}\right)}\right)^{\Delta}\exp\left(  B_1\cdot \epsilon_+ + B_2\cdot \epsilon_++O(\epsilon_+^3)\right)\,,
\eeq
with
\begin{align}
\begin{split}
	B_1\cdot\epsilon_+ &= \Delta\left( \epsilon_+'(\theta_1) + \epsilon_+'(\theta_2) - \cot\left( \frac{\theta_{12}}{2}\right)(\epsilon_+(\theta_1)-\epsilon_+(\theta_2))\right)\,,
	\\
	B_2 \cdot \epsilon_+ & = -\frac{\Delta}{2} \left( \epsilon_+'(\theta_1)^2 + \epsilon_+'(\theta_2)^2 - \frac{1}{2}\csc^2\left(\frac{\theta_{12}}{2}\right)(\epsilon_+(\theta_1)-\epsilon_+(\theta_2))^2\right)\,,
\end{split}
\end{align}
and similar couplings to the other components of the bilocal.

After a straightforward computation we obtain
\begin{align}
\begin{split}
\left\langle \left( \frac{f_+'(\theta_1)f_+'(\theta_2)}{4\sin^2\left( \frac{f_+(\theta_1)-f_+(\theta_2)}{2}\right)} \right)^\Delta \right\rangle &=\left(\frac{1}{ 2 \sin\left(\frac{\theta_{12}}{2}\right)}\right)^{2\Delta}\left( 1 + \frac{\Delta}{C} \, \mathcal{C}_1(\theta_1, \theta_2) + O(C^{-2}) \right)\,,
\\
\left\langle  \left(\frac{f_+'(\theta_1)f_-'(\theta_2)}{4\cos^2\left(\frac{f_+(\theta_1)-f_-(\theta_2)}{2}\right)}\right)^{\Delta}\right\rangle & = \left( \frac{1}{2\cos\left(\frac{\theta_{12}}{2}\right)} \right)^{\Delta}\left( 1 + \frac{\Delta}{C} \,\mathcal{C}_2(\theta_1,\theta_2) +O(C^{-2})\right)\,,
\end{split}
\end{align}
and in the second line we have not yet integrated over the twist. The corresponding formulae for the past-past and past-future two-point functions are just the complex conjugates of those above.
The corrections $\mathcal{C}_1$ and $\mathcal{C}_2$ are for $\theta_1>\theta_2$
\begin{align}
\nonumber
	\mathcal{C}_1(\theta_1, \theta_2) &= \frac{i}{2\pi} \frac{1}{4\sin^2\left(\frac{\theta_{12}}{2}\right)^2} \bigg( \theta_{12}(\theta_{12} - 2\pi)(\Delta + 1) + (\Delta \theta_{12} (\theta_{12} - 2\pi) - 4 \Delta - 2) \cos(\theta_{12})
	\\
& \qquad \qquad \qquad \qquad \qquad+2 + 4 \Delta + 2 (\pi - \theta_{12})(2 \Delta + 1) \sin(\theta_{12}) \bigg)\,,
	\\
	\nonumber
	\mathcal{C}_2(\theta_1,\theta_2)& = 0\,.
\end{align}
The correction to the future-past two-point function vanishes in the special case where $\beta_+ = \beta_-$ essentially because in this case the propagators for $\epsilon_+$ and $\epsilon_-$ are equal and opposite. However the correction is more generally nonzero.

Note that the leading quantum corrections break the ``KMS'' symmetry of de Sitter propagators, and further that the correction to the future-future two-point function breaks conformal invariance. Relatedly the corrections have a factor of $\beta J$ relative to the tree-level result. In the context of black holes with nearly dS$_2$ near-horizon, one has $\beta J \gg 1$ and so the quantum correction is enhanced.

%***********************************************
\subsubsection{Tree-level exchange}
%***********************************************

\begin{figure}
\centering
	\includegraphics[width=2.5in]{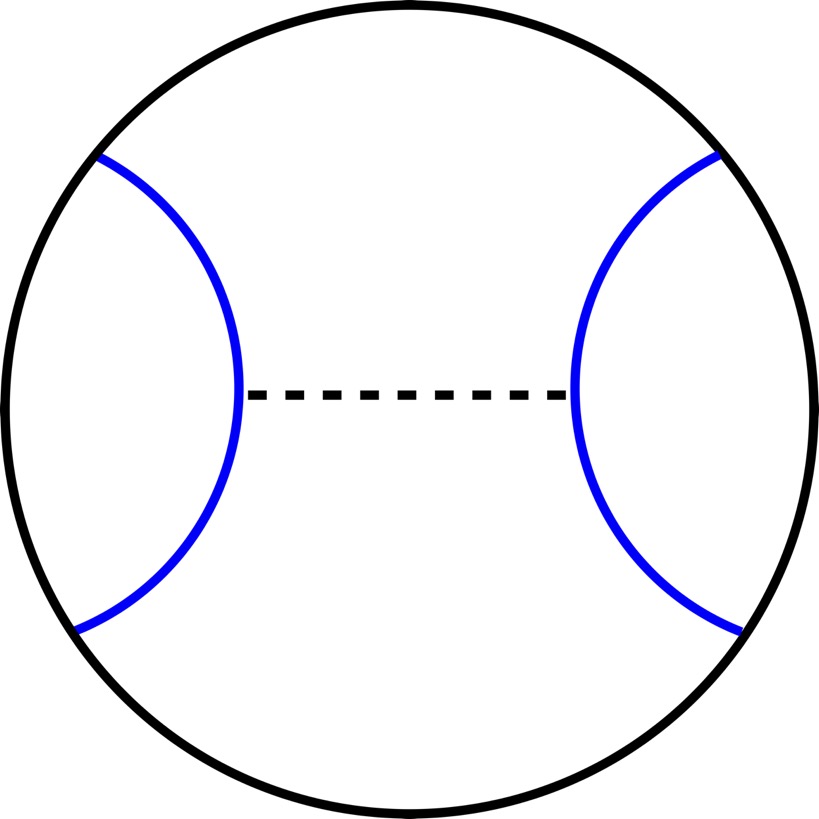}
	\caption{
			\label{F:tree4pt} The tree-level exchange diagram which contributes the $O(1/C)$ correction to the connected four-point function of two different scalar operators $V$ and $W$.
		}
\end{figure}

We can also compute the tree-level exchange diagram that appears in the gravitational scattering of two different scalars dual to operators we denote as $V$ and $W$. See Fig.~\ref{F:tree4pt}. That is, we compute to $O(1/C)$ the correction to the two-point function of bilocals,
\beq
	\langle V_+(\theta_1) V_-(\theta_2) W_+(\theta_3) W_-(\theta_4)\rangle = \langle \mathcal{B}_{+-}(\theta_1,\theta_2;\Delta_V)\mathcal{B}_{+-}(\theta_3,\theta_4;\Delta_W)\rangle\,,
\eeq
where $\Delta_V$ is the dimension of $V$ and $\Delta_W$ the dimension of $W$. Ignoring the integral over twist in both the numerator and denominator, we represent the $1/C$ correction to this four-point function as
\begin{align}
\frac{\langle \mathcal{B}_{+-}(\theta_1,\theta_2;\Delta_V)\mathcal{B}_{+-}(\theta_3,\theta_4;\Delta_W)\rangle}{\langle \mathcal{B}_{+-}(\theta_1,\theta_2;\Delta_V)\rangle \langle \mathcal{B}_{+-}(\theta_3,\theta_4;\Delta_W)\rangle} &= 1 + \frac{\Delta_V \Delta_W}{C} \,\mathcal{F}(u_1, u_2, u_3, u_4) + O(C^{-2})\,.
\end{align}
For $\theta_1 > \theta_3$ and $\theta_2 > \theta_4$,
\begin{align}
	\nonumber
	\mathcal{F}(\theta_1, \theta_2, \theta_3, \theta_4) &= \frac{i}{8 \pi} \frac{1}{\sin(\frac{\theta_{12}}{2}) \sin(\frac{\theta_{34}}{2})} \bigg( 4 (\cos(\theta_{13}) - \cos(\theta_{24})) \cos\left(\frac{\theta_{13} + \theta_{24}}{2}\right) 
\\
	\nonumber
	& + 2(2\pi - \theta_{13} - \theta_{24})\bigg[(\theta_{13} - \theta_{24}) \, \cos\left(\frac{\theta_{12}}{2}\right) \cos\left(\frac{\theta_{34}}{2}\right) - 2 \sin\left(\frac{\theta_{12} - \theta_{34}}{2}\right)\bigg]
	\\
	& + \cos\left(\frac{\theta_{12} - \theta_{34}}{2}\right) \big(6\cos(\theta_{13}) - 4 (\pi - \theta_{13}) \sin(u_{13}) - (13\to 24) \big)  
	\\
	\nonumber
	&+ \sin\left(\frac{\theta_{12} - \theta_{34}}{2}\right) \big(6\sin(\theta_{13})  + 4 (\pi - \theta_{13}) \cos(\theta_{13}) +(13\to 24) \big) \bigg)\,.
\end{align}

%***********************************************
\subsection{Partition functions}
\label{S:partitions}
%***********************************************

To complete this Section we compute the path integral for the Schwarzian theory on the boundary of the Hartle-Hawking geometry, as well as on global dS$_2$.  The former computes the Hartle-Hawking wavefunction of the universe in the no-boundary state, to leading order in the genus expansion, and the latter corresponds to a transition amplitude between one-universe states.

To begin let us recall what one finds for Jackiw-Teitelboim gravity on the hyperbolic disc, i.e. on Euclidean nearly AdS$_2$. There one finds an integral over a Diff$(\mathbb{S}^1)/PSL(2;\mathbb{R})$ field $f(\tau)$, with
\begin{align}
\begin{split}
\label{E:Zdisc}
	Z_{\rm disc}(\beta J') &= e^{S_0'} \int \frac{[df(\tau)]\text{Pf}(\omega)}{PSL(2;\mathbb{R})} \exp\left(- \int_0^{\beta} d\tau H\right)\,, 
	\\
	H& = -\frac{1}{8\pi G_2\beta J'} \left( \{f(\tau),\tau\} + \frac{2\pi^2}{\beta^2}f'(\tau)^2\right)\,, \quad S_0' = \frac{\varphi_0'}{4G_2}\,.
\end{split}
\end{align}
Here $H$ is the energy of the Schwarzian mode. This object is the thermal partition function of the Schwarzian model, which we would naively interpret as
\beq
\label{E:trschw1}
	Z_{\rm disc}(\beta J') = \text{tr}_{\rm Sch}(e^{-\beta H})\,.
\eeq

This path integral is one-loop exact~\cite{Stanford:2017thb}, with the result
\beq
\label{E:Zdisc2}
	Z_{\rm disc}(\beta J') = \frac{1}{\sqrt{2\pi}} \frac{1}{(2\beta J')^{3/2}} e^{S_0'+\frac{\pi}{4G_2 \beta J'}}\,,
\eeq
in our normalization of the symplectic form. The essential point is that the integration space Diff$(\mathbb{S}^1)/PSL(2;\mathbb{R})$ of the Schwarzian theory is symplectic, and so the integral is invariant under a BRST-like Grassmann-odd symmetry $Q$. The space is not only symplectic, but K\"ahler, with a K\"ahler metric that is invariant under the flow generated by the Hamiltonian of the model. There is a natural $Q$-exact positive-definite term (the term is $g_{ij} V^i V^j + \hdots$ for $g_{ij}$ the metric and $V^i = \omega^{ij}\partial_j H$ which generates Hamiltonian flow) which may be added to the action with large coefficient, and used to localize the model to its critical points.

Interpreting the disc partition function as a sum over states and performing the inverse Laplace transform, one finds a continuous density of states
\beq
\label{E:Zdisc3}
	Z_{\rm disc}(\beta J') = \int_0^{\infty} dE\, \rho(E) e^{-\beta E}\,, \qquad \rho(E) = \frac{e^{S_0'}\sqrt{G_2}}{2\pi^{3/2}J'}\sinh\left(\sqrt{\frac{\pi E}{G_2J'}}\right)\,.
\eeq
However as pointed out in~\cite{Stanford:2017thb} this result is not consistent with a Hilbert space interpretation of the trace. In quantum mechanics one can have a continuous density of states, as long as there is an additional parameter $X$ so that the density describes a number of states per unit $X$. There is no such parameter here.

For this and other reasons the authors of~\cite{Saad:2019lba} interpret the disc partition function differently. Instead they regard it as the leading approximation to the average thermal partition function within an ensemble of Hamiltonians,
\beq
	 Z_{\rm disc}  = \left\langle \text{tr}\left(e^{-\beta H}\right)\right\rangle_{\overline{\text{MM}}',0}  \,,
\eeq
where the average on the right-hand-side is taken within a matrix model of $L\times L$ Hermitian Hamiltonians $H$ as $L\to \infty$. This is notated by the subscript $\overline{\text{MM}}'$, where the MM stands for matrix model, and the bar denotes that we are taking a scaling limit. The prime is in keeping with our notation for Euclidean AdS quantities, in this case a matrix model.  The $0$ subscript is present since in matrix integrals there is a genus expansion in powers of $e^{-S_0'}$, and we are considering the genus $0$ term.  In this interpretation the density of states in~\eqref{E:Zdisc3} is the leading approximation in powers of $e^{-S_0'}$ to the density of states of the matrix model.

%***********************************************
\subsubsection{Hartle-Hawking}
%***********************************************

The path integral of JT gravity on the Hartle-Hawking geometry is rather similar. It is
\beq
	Z_{\rm HH} =e^{S_0} \int\frac{[df(\tau)]\text{Pf}(\omega)}{PSL(2;\mathbb{R})} \exp\left( \frac{i}{8\pi G_2 J} \int_0^{\beta} d\tau \left( \{ f(\tau),\tau\} + \frac{2\pi^2}{\beta^2}f'(\tau)^2\right)\right)\,, \quad S_0 = \frac{\varphi_0}{4G_2}\,, 
\eeq
where we have rescaled $\theta = \frac{2\pi\tau}{\beta}$ and $f(\theta) \to \frac{\beta}{2\pi} f(u)$ in order to make it clear that the model is on a circle of physical size $\beta$. The model is invariant under rotations $\tau\to \tau + \delta$, generated by
\beq
	R=- \frac{1}{8\pi G_2 \beta J}\left( \{f(\tau),\tau\} + \frac{2\pi^2}{\beta^2}f'(\tau)^2\right)\,,
\eeq
Then 
\beq
\label{E:preZHH}
	Z_{\rm HH}(\beta J) = e^{S_0} \int\frac{ [df(\tau)]\text{Pf}(\omega)}{PSL(2;\mathbb{R})} \exp\left(- i\int_0^{\beta}d\tau\,  R\right)\,.
\eeq

By comparing these expressions for those with the hyperbolic disc partition function~\eqref{E:Zdisc}, we see that the path integrals above is simply given by $Z_{\rm disc}$ under
\beq
	S_0' \to S_0\,, \quad J' \to -i J\,,
\eeq
under which $H \to i R$. Then (under $S_0'\to S_0$)
\beq
\label{E:fromDiscToHH}
	Z_{\rm HH}(\beta J) = Z_{\rm disc}(-i \beta J)\,.
\eeq
Naively comparing Eqs.~\eqref{E:preZHH} and~\eqref{E:Zdisc} we would then interpret this Schwarzian path integral as computing the trace of $\exp(-i \beta H)$, however this is not quite right. At the level of the partition function the analytic continuation $J' \to -i J$ can be absorbed into a redefinition of $\beta\to -i \beta$, since the two only appear together through the combination $\beta J$. At the level of the boundary Schwarzian model, $Z_{\rm HH}$ then computes the trace of $e^{i \beta H}$. 

In the next Section we will arrive at a genus expansion for nearly dS$_2$ gravity, which comes from a matrix model. As with nearly AdS$_2$ we will interpret the path integral $Z_{\rm HH}$ as computing
\beq
	Z_{\rm HH}=\left\langle \text{tr}\left( e^{i\beta H}\right)\right\rangle_{\overline{\text{MM}},0} \,.
\eeq
The subscript is $\overline{\text{MM}}$, rather than $\overline{\text{MM}}'$, to denote that we are a priori dealing with different scaled matrix models. Implicitly $\beta$ has some small positive imaginary part so that $e^{i\beta H}$ is well-behaved at high energy, and so that the gravitational path integral is convergent.

Given the continuation~\eqref{E:fromDiscToHH} we can simply read off $Z_{\rm HH}$ from $Z_{\rm disc}$. However for completeness let us obtain it at one-loop order by direct computation. We do so in terms of the original field $f(\theta)$. The model has a boundary condition and redundancy, 
\beq
	f(\theta+2\pi) = f(\theta)+2\pi\,, \qquad \tan\left( \frac{f}{2}\right) \sim \frac{a\tan\left(\frac{f}{2}\right)+b}{c\tan\left(\frac{f}{2}\right)+d}\,,
\eeq
and its field equation is simply $\left\{ \tan\left(\frac{f}{2}\right),\theta\right\}' = 0$. Modulo the quotient, there is a unique critical point of the model obeying the boundary condition,
\beq
	f_0 = \theta\,.
\eeq
Expanding in fluctuations around it,
\beq
	f = \theta + \sum_n \epsilon_n e^{in \theta}\,, \qquad \epsilon_{-n}= \epsilon_n^*\,,
\eeq
the $PSL(2;\mathbb{R})$ quotient may be used to fix
\beq
	\epsilon_{-1,0,1} = 0\,.
\eeq
The quadratic effective action and symplectic form are
\begin{align}
\begin{split}
	S &=-iS_0 + \frac{\pi}{g}  -\frac{2\pi }{g} \sum_{n>1} n^2(n^2-1) |\epsilon_n|^2\,,
	\\
	\omega & = \frac{2i\beta J}{g} \sum_{n>1} n(n^2-1) d\epsilon_n^* \wedge d\epsilon_n\,.
\end{split}
\end{align}
We then have
\begin{align}
\label{E:Z010}
\begin{split}
	Z_{\rm HH}&= e^{ S_0+ \frac{\pi i}{g}} \prod_{n>1}\int  \left(  d^2\epsilon_n\frac{4\beta J}{g} n(n^2-1)\right) \exp\left(  -\frac{2\pi i}{g} n^2(n^2-1)|\epsilon_n|^2\right)
	\\
	& = e^{ S_0+ \frac{\pi i}{g}}\prod_{n>1} \left( \frac{- 2i \beta J}{n}\right) = \frac{1}{\sqrt{2\pi}(-2i\beta J)^{3/2}}e^{ S_0+\frac{\pi i}{g}} \,.
\end{split}
\end{align}
This precisely matches the disc result~\eqref{E:Zdisc2} under $S_0'\to S_0$ and $J'\to -i J$.

This should be understood as the wavefunction of the universe in the Hartle-Hawking state, $\langle\beta J|\text{HH}\rangle$, to leading order in the genus expansion.  Interpreting $Z_{\rm HH}$ as $\text{tr}_{\rm Sch}(e^{i\beta H})$ and inverting, we find the \emph{same} density of states as from the disc,
\beq
	Z_{\rm HH}(\beta J) = \int_0^{\infty} dE\,\rho(E)e^{i\beta E}\,, \qquad \rho(E) = \frac{e^{S_0}\sqrt{G_2}}{2\pi^{3/2}J}\sinh\left(\sqrt{\frac{\pi E}{G_2J}}\right)\,.
\eeq

Similarly, we may consider the path integral on the time-reversed version of the Hartle-Hawking geometry with a single past boundary of size $\beta$. Using similar methods as above one finds that it is given by the analytic continuation $Z_{\rm disc}(i \beta J)$, i.e. the complex conjugate of $Z_{\rm HH}$, and we assign it a matrix model interpretation via
\beq
	 Z_{\rm HH}^*(\beta J) = \left\langle \text{tr}\left( e^{-i\beta H}\right)\right\rangle_{\overline{\text{MM}},0}\, .
\eeq
Here $\beta$ has small negative imaginary part.

%***********************************************
\subsubsection{Global dS$_2$}
%***********************************************

Now for global dS$_2$. As above, interpreting the path integral $Z_{\rm global}$ in terms of an ensemble average, we have that $Z_{\rm global}$ is the leading approximation to the connected two-point function,
\beq
\label{E:ZglobalMM}
	Z_{\rm global}(\beta_+J,\beta_-J)  = \left\langle \text{tr}\left( e^{i\beta_+H}\right)\text{tr}\left(e^{-i\beta_- H}\right) \right\rangle_{\overline{\text{MM}}\text{, conn, }0} \,,
\eeq
where $\text{Im}(\beta_+)>0$ and  $\text{Im}(\beta_-)<0$.  By connected, we mean we are computing the second cumulant with respect to the scaled matrix ensemble (and in particular, extracting the genus zero contribution).

The path integral for $Z_{\rm global}$ is again one-loop exact. It is
\beq
	Z_{\rm global}=\int_0^{\infty} d\alpha\int \frac{[df_+df_-] \text{Pf}(\omega) }{U(1)}\exp\left( i \left( \frac{1}{g_+}\int_0^{2\pi}  d\theta \left(\{ f_+,\theta\} + \frac{\alpha^2}{2}f_+'^2\right) - (+\to -)\right)\right)\,,
\eeq
where $g_{\pm} = 4G_2 \beta_{\pm} J$. The boundary conditions and quotient read
\beq
	f_{\pm}(\theta+2\pi) = f_{\pm}(\theta)+2\pi\,, \qquad f_{\pm}\sim f_{\pm} + \delta\,.
\eeq
There is a family of critical points modulo the quotient,
\beq
	f_{+,0} = \theta + \gamma\,, \qquad f_{-,0} = \theta\,,
\eeq
where $\gamma\sim \gamma+2\pi$. Expanding in fluctuations we write
\beq
	f_+ = \theta + \gamma + \sum_{n\neq 0} \epsilon^+_n e^{in\theta}\,, \qquad f_- = \theta + \sum_{n\neq 0} \epsilon^-_n e^{i n\theta}\,, \qquad \epsilon_{-n}^{\pm}  = (\epsilon_n^{\pm})^*\,,
\eeq
where we have used that the diagonal combination of $n=0$ modes may be set to zero by the quotient and the axial combination is already accounted for with $\gamma$. The effective action and symplectic form to quadratic order in fluctuations is
\begin{align}
\begin{split}
\label{E:globaldS2quadratic}
	S&= \pi \alpha^2\left( \frac{1}{g_+} - \frac{1}{g_-}\right) - 2\pi \sum_{n >0}n^2(n^2-\alpha^2) \left( \frac{|\epsilon_n^+|^2}{g_+}   - \frac{ |\epsilon_n^-|^2}{g_-}\right)\,,
	\\
	\omega &= \frac{1}{2G_2}\left(  i\sum_{n>0}n(n^2-\alpha^2) \left( d(\epsilon_n^+)^* \wedge d\epsilon_n^+ -d(\epsilon_n^-)^*\wedge d\epsilon_n^-\right) +  \frac{1}{2}\alpha d\alpha \wedge d\gamma\right)\,.
\end{split}
\end{align}

\begin{figure}
\centering
	\includegraphics[width=5in]{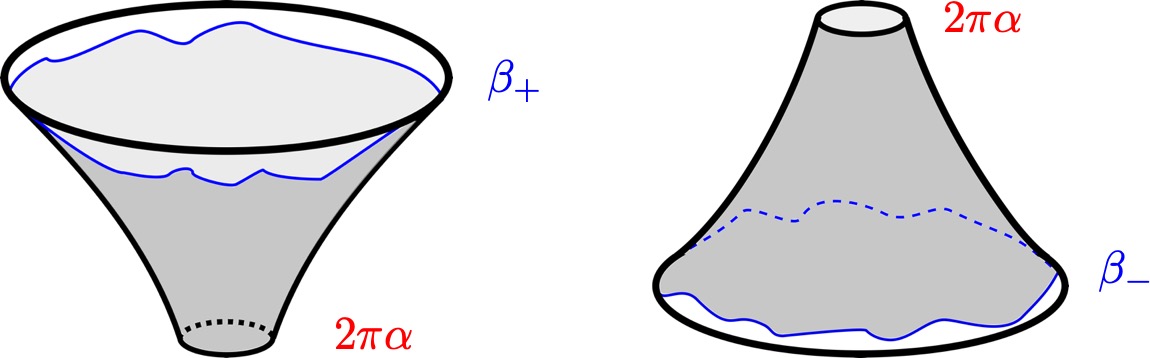}
	\vskip1cm
	\caption{
		\label{F:trumpets} Future and past dS$_2$ trumpets.
	}
\end{figure}

It is convenient to divide the integral into one over the reparameterization modes on the future circle, one over the modes on the past circle, and a residual integral over $(\alpha,\gamma)$. In this way one thinks of global dS$_2$ as the sewing of a future half of dS, a ``trumpet'' geometry in the language of~\cite{Saad:2019lba}, to a past ``trumpet.'' See Fig.~\ref{F:trumpets}. Each half or trumpet is characterized by the parameter $\alpha$, and the parameter $\gamma$ corresponds to the fact that this gluing may be performed with some twist. In the computation below we find the integrand at non-integer $\alpha$. This integrand is a smooth function of $\alpha$. This suggests that at the special points $\alpha = m$ we find the same result for the one-loop determinant as when taking $\alpha \to m$. Indeed this is the case as may be shown using the quadratic effective action and measure discussed in Subsection~\ref{S:integern}. As a result, the analysis below suffices for all $\alpha$.

The ``future trumpet'' path integral is the part of the integral for global dS$_2$ over the reparameterization modes on the future circle, and is given by
\beq
	Z_F(\beta J,\alpha) = \int \frac{[Df_+]}{U(1)} \exp\left( \frac{i}{g_+} \int_0^{2\pi} d\theta \left( \{f_+(\theta),\theta\} + \frac{\alpha^2}{2}f_+'(\theta)^2\right)\right)\,,
\eeq
where we identify $f_+(\theta) \sim f_+(\theta)+\delta$. Here $[Df_+]$ is the symplectic measure for $f_+$ we described above. This path integral $Z_F$ is also one-loop exact, and is given by
\begin{align}
\begin{split}
	Z_F &= e^{\frac{i \pi \alpha^2}{g_+}}\prod_{n>0} \int \left(d^2\epsilon^+_n  \frac{n(n^2-\alpha^2)}{G_+}\right)\exp\left( - \frac{2\pi i}{g_+}n^2(n^2-\alpha^2)|\epsilon_n^+|^2\right)
	\\
	& = e^{\frac{i\pi \alpha^2}{g_+}}\prod_{n>0} \left( \frac{-2i\beta J}{n}\right) = \frac{1}{\sqrt{2\pi}(-2i\beta J)^{1/2}}e^{\frac{i\pi \alpha^2}{g_+}}\,.
\end{split}
\end{align}
There is also a ``past trumpet'' factor given by its complex conjugate,
\beq
	Z_P(\beta J,\alpha) = Z_F^*(\beta J,\alpha) \,.
\eeq

The path integral for global dS$_2$ is a gluing together of these past and future trumpet factors. Now we come to an ambiguity. From the symplectic form~\eqref{E:globaldS2quadratic}, we see the measure for $\alpha$ and $\gamma$ has absolute value $\frac{\alpha \, d\alpha d\gamma}{2G_2}$. However it is not clear at this stage what the sign of the measure should be.

Allowing for both signs, we then have
\begin{align}
\begin{split}
\label{E:Zglobal}
	Z_{\rm global} &= \int_0^{\infty} \frac{\pm d\alpha \,\alpha}{2G_2}\int_0^{2\pi}d\gamma\, Z_F(\beta_+J,\alpha) Z_P(\beta_-J,\alpha) 
	\\
	& = \pm \int_0^{\infty} \frac{d\alpha\,\alpha}{\sqrt{g_+g_-}} \exp\left( \pi i \alpha^2\left( \frac{1}{g_+} - \frac{1}{g_-}\right)\right) = \mp \frac{i}{2\pi}\frac{\sqrt{\beta_+\beta_-}}{\beta_+-\beta_-}\,.
\end{split}
\end{align}
Interpreting the result as a transition amplitude between a past circle and a future circle, we see that the amplitude is enhanced when the two circles have similar size. 

A natural observable in a matrix model is the ``resolvent,''
\beq
	R(\lambda) = \text{tr}\left( \frac{1}{\lambda - H}\right)\,  = \int_0^{\infty} dE \frac{\rho(E)}{\lambda - E}\,.
\eeq
Using the matrix model interpretation of $Z_{\rm global}$ in~\eqref{E:ZglobalMM}, by an integral transform, we can extract the connected genus-0 two-point function of resolvents to be
\begin{align}
\begin{split}
\label{E:R02}
	R_{0,2}(\lambda_1,\lambda_2) &\equiv \int_0^{\infty} dE_1dE_2 \left\langle \frac{\rho(E_1)}{\lambda_1-E_1}\frac{\rho(E_2)}{\lambda_2-E_2}\right\rangle_{\overline{\text{MM}}\text{, conn, }0}
	\\
	&= \int_0^{\infty} d\beta_+d\beta_- \,e^{-i \beta_+\lambda_1+i \beta_-\lambda_2}Z_{\rm global}(\beta_+J,\beta_-J)
	\\
	& = \mp \frac{1}{4\sqrt{-\lambda_1}\sqrt{-\lambda_2}(\sqrt{-\lambda_1}+\sqrt{-\lambda_2})^2}\,.
\end{split}
\end{align}
Here we take $\lambda_i $ to be real and negative, and evaluate the contour integral by rotating $\beta_+$ and $\beta_-$ to the positive and negative imaginary axes respectively.

For a large $L$ matrix models with a single cut, the function $R_{0,2}$ only depends on the endpoints of the cut, and has a universal form. See e.g.~\cite{marino2005chern}. Further, $R_{0,2}$ simplifies in the \emph{double-scaled} limit, in which one end of the cut is at $E=0$ and the other is at positive infinity, in which case $R_{0,2}$ becomes precisely the expression above with the $+$ sign.

For this reason we fix our convention for the symplectic measure to give us this sign. That is, we take the integration measure over $\alpha$ and $\gamma$ to be $-\frac{\alpha \, d\alpha d\gamma}{2G_2}$. Then the path integral over global dS$_2$ takes exactly the form predicted by random matrix theory. Furthermore it is the analytic continuation of a computation in Euclidean AdS$_2$, the path integral over the double trumpet of~\cite{Saad:2019lba}, which they find to be
\beq
	Z_{0,2}(\beta_1J',\beta_2J') = \frac{1}{2\pi}\frac{\sqrt{\beta_1\beta_2}}{\beta_1+\beta_2}\,.
\eeq
Here $\beta_{1}$ and $\beta_2$ are the sizes of the asymptotic circles. Clearly with this choice of sign
\beq
\label{E:Zglobal1} 
	Z_{\rm global} = Z_{0,2}(\beta_1J' \to -i \beta_+J, \beta_2J' \to i \beta_-J) =  \frac{i}{2\pi}\frac{\sqrt{\beta_+\beta_-}}{\beta_+-\beta_-}\,,
\eeq
reproducing the result above. The two path integrals may even be related at intermediate steps. Matching their normalization of the symplectic measure to ours, the integral for $Z_{0,2}$ takes the form
\beq
	Z_{0,2} = \int_0^{\infty} \frac{db\, b}{2G_2 \times (2\pi)} Z_T(\beta_1J')Z_T(\beta_2J')\,, \qquad Z_T(\beta J') = \frac{1}{\sqrt{2\pi}(2\beta J')^{1/2}}e^{-\frac{b^2}{8\pi G_2\beta J'}}\,,
\eeq
where $Z_T$ is the path integral over the Schwarzian mode on the boundary of a Euclidean AdS$_2$ ``trumpet.'' Note that our trumpet is the analytic continuation of theirs,
\beq
	Z_F(\beta J) = Z_T(-i \beta J)\,, \qquad Z_P(\beta J) = Z_T(i \beta J)\,,
\eeq
and that under the continuation $b\to 2\pi i \alpha$ the measure is transformed into ours $-(2\pi)\frac{\alpha \, d\alpha}{2G_2}$ (after integrating over $\gamma$). So the integral for $Z_{0,2}$ maps exactly to ours Eq.~\eqref{E:Zglobal} for $Z_{\rm global}$.

The global de Sitter amplitude~\eqref{E:Zglobal1} diverges when the future and past boundaries have the same renormalized length $\beta_+=\beta_-$. But, recalling that global nearly de Sitter spacetimes always have $\beta_+ = \beta_-$, we see that this divergence is physical: it is exactly when we go on-shell. It is tempting to then regard the global amplitude as a propagator for a closed universe.

Finally, consider the $\alpha=1$ double Schwarzian theory discussed briefly in Subsection~\ref{S:fixingAlpha}. That case has a $PSL(2;\mathbb{R})\times PSL(2;\mathbb{R})$ symmetry, and the diagonal part is quotiented out. The axial $PSL(2;\mathbb{R})$ is a global symmetry of the model, and thus there is a $PSL(2;\mathbb{R})$ manifold of critical points. Up to the axial quotient they are characterized by
\beq
	\tan\left( \frac{f_+}{2}\right) = \frac{a \tan\left(\frac{\theta}{2}\right)+b}{c\tan\left(\frac{\theta}{2}\right)+d}\, , \qquad f_- = \theta\,.
\eeq
The path integral evaluates to
\beq
	Z_{\alpha = 1}= \frac{\text{vol}(PSL(2,\mathbb{R}))}{2\pi (2\beta_+J)^{3/2} (2\beta_-J)^{3/2} }\exp\left( \frac{i \pi}{4G_2 J}\left( \frac{1}{\beta_+} - \frac{1}{\beta_-}\right)\right)\,.
\eeq
Up to the infinite volume of the $PSL(2;\mathbb{R})$ moduli space, this is the result we found above for the future Hartle-Hawking geometry evaluated at $\beta_+$, times its complex conjugate evaluated at $\beta_-$ (corresponding to the past Hartle-Hawking geometry).

  %***********************************************
\section{A genus expansion}
\label{S:genus}
%***********************************************

%*********************************************** 
\subsection{Physical interpretation of higher genus nearly dS$_2$ spacetimes}
%***********************************************
 
Our goal is to understand whether there is de Sitter version of the topological expansion of \cite{Saad:2019lba}, where the path integral of JT gravity is organized as a genus expansion.  We will first recall a few features of the higher genus contributions to the Euclidean path integral of JT gravity, before discussing their Lorentzian analogues.  We will conclude that, with the proper analytic continuation, the Euclidean gravity path integral of \cite{Saad:2019lba} can be interpreted as the preparation of a Hartle-Hawking state for Lorentzian de Sitter universes with an arbitrary number of boundaries.

%***********************************************
\subsubsection{Quotients of $\mathbb{H}_2$ and dS$_2$}
%***********************************************

We begin by recalling the geometry of the Euclidean metrics with higher genus that contribute to the path integral of JT gravity with a negative cosmological constant.  By integrating out the dilaton, the integral over the space of metrics reduces to an integral over constant negative curvature Euclidean geometries with $R=-2$.  All such geometries are locally $\mathbb{H}_2$, and can be written as quotients of $\mathbb{H}_2$ by some subgroup $\Gamma$ of the isometry group $SO(2,1)$ of hyperbolic space.  This gives a constant negative curvature Euclidean surface $\Sigma = \mathbb{H}_2/\Gamma$.  As a group, $\Gamma$ is just the fundamental group $\pi_1(\Sigma)$ of our surface, and the choice of embedding of $\Gamma$ into $SO(2,1)$ parameterizes the moduli of the surface $\Sigma$.  In the gauge theory language, the embedding of $\pi_1(\Sigma)$ into $SO(2,1)$ is the holonomy map, which associates to each non-contractible cycle an element of the gauge group $SO(2,1)$.

In order to understand the generalization to de Sitter space, we note that $\mathbb{H}_2$ can be represented as a coset of $SO(2,1)$ group manifold, as
\beq
	\mathbb{H}_2 = SO(2,1) / SO(2)\,.
\eeq
In other words, we identify each point in $\mathbb{H}_2$ with an equivalence class of elements $g \in SO(2,1)$, with the identification
\beq
	g\sim g R
\eeq
where $R$ is an element of some fixed subgroup $SO(2) \subset SO(2,1)$.  One way of understanding this coset is to recall that $\mathbb{H}_2$ is a homogeneous symmetric space, which can be identified with its isometry group modulo the stabilizer of a point.  The isometry group of $\mathbb{H}_2$ is $SO(2,1)$, and the stabilizer of a point is $SO(2)$. To see this, we can just think of the hyperboloid $-T^2 + X^2 + Y^2 = -1$, and note that the point $(T,X,Y)=(1,0,0)$ is invariant under rotations in the $(X,Y)$ plane.  Our conclusion is that $\mathbb{H}_2$ is the quotient of $SO(2,1)$ by a one-dimensional subgroup of elliptic elements of $SO(2,1)$.

The advantage of this description is that it makes the $SO(2,1)$ isometries of $\mathbb{H}_2$ completely explicit -- they are given by left multiplication by an element $L\in SO(2,1)$, which takes
\beq
	g \longmapsto L g\,.
\eeq
The constant negative curvature metric on $\mathbb{H}_2$ is ${\rm tr} \left({\cal A} \otimes {\cal A}\right)$, where ${\cal A} = g^{-1} dg$ is the usual left-invariant one-form on the $SO(2,1)$ group manifold. We can now consider the quotient of $\mathbb{H}_2$ by a discrete subgroup $\Gamma$ of the isometry group.  This can be represented as a double quotient
\beq
	\Sigma = \Gamma \backslash SO(2,1) \slash SO(2)\,.
\eeq
This is a two-dimensional Euclidean geometry with constant negative curvature $R=-2$.
In order for $\Sigma$ to be smooth, we must require the action of every (non-identity) element $\gamma$ of $\Gamma$ to be fixed-point free.  This means we must have
\beq
	\gamma g \ne g R
\eeq
for any $g\in SO(2,1)$ and $R \in SO(2)$.  Since $R$ is an elliptic element, this means that $\gamma$ cannot be conjugate to an elliptic element of $SO(2,1)$: it must be either parabolic or hyperbolic.  Note that associated to each $\gamma \in SO(2,1)$ there is an element of the fundamental group $\pi_1(\Sigma)$.  If $\gamma$ is hyperbolic, then there is a minimum length geodesic homologous to this element, with length $b_\gamma = \cosh^{-1} {1\over 2}{\rm tr}(\gamma)$.\footnote{In this formula (and in later expressions below) we have used the fact that $SO(2,1)=PSL(2,\mathbb{R})$, so represent elements $\gamma\in SO(2,1)$ as a $2\times 2$ matrices in $PSL(2,\mathbb{R})$. }  If $\gamma$ is parabolic then our surface $\Sigma$ has a cusp singularity; we must remove this point from our surface, and $\gamma$ is the holonomy of the path around this point. In the double-coset description given here, the moduli space of surfaces is the moduli space of embeddings of $\Gamma$ into $SO(2,1)$.

In the topological gauge theory the geometry is characterized by a flat $PSL(2;\mathbb{R})$ bundle over $\Sigma$. The constant curvature is not manifest in this description. To recover it, one must use the map described in Subsection~\ref{S:EAdS2} to convert the $PSL(2;\mathbb{R})$ connection into the first order variables, and this in turn into metric data. What is manifest are the holonomies around the various cycles. The generators of $\Gamma$ are mapped to the holonomies around non-contractible cycles, and the underlying smoothness of $\Sigma$ to the statement that the holonomy around any trivial cycle is trivial.

In order to understand the analytic continuation of this construction to de Sitter space, we recall that dS$_2$ is characterized by a number $\alpha$. For $\alpha = 1$, dS$_2$ can be identified with the coset
\beq
	\text{dS}_2 = SO(2,1) / SO(1,1)\,.
\eeq
As before, we identify each point in dS$_2$ with an equivalence class of elements $g \in SO(2,1)$, with the identification
\begin{equation}
g\sim g R
\end{equation}
where $R$ now is an element of some fixed subgroup $SO(1,1) \subset SO(2,1)$.  This is because dS$_2$ is a homogeneous symmetry space with $SO(2,1)$ isometry.  The difference is that now the stabilizer of a point is $SO(1,1)$. This can be seen by representing de Sitter space as $-T^2 + X^2 + Y^2 = 1$, and noting that the point $(T,X,Y)=(0,1,0)$ is invariant under boosts in the $(T,Y)$ plane.  Our conclusion is that dS$_2$ is the quotient of $SO(2,1)$ by a one-dimensional subgroup of {\it hyperbolic} elements of $SO(2,1)$.

We can now construct quotients of dS$_2$ just as we did for hyperbolic space, by taking the left quotient by some subgroup of $SO(2,1)$.  In particular, for any subgroup $\Gamma$ of $SO(2,1)$ we may consider the space-time
\beq
	\Gamma \backslash SO(2,1) \slash SO(1,1)\,.
\eeq
This is a two-dimensional spacetime with Lorentzian signature and constant {\it positive} curvature $R=2$.  
There is one important difference with the hyperbolic construction, however.  
In order for the action of an element $\gamma\in SO(2,1)$ to be fixed-point free on dS$_2$, we must have
\beq
	\gamma g \ne g R
\eeq
where now $R \in SO(1,1)$ is hyperbolic.  This means that any hyperbolic element $\gamma$, when acting on dS$_2$, will have fixed points.   In order to obtain a smooth quotient without fixed points, each element of $\Gamma$ must be either a parabolic or elliptic element of $SO(2,1)$.

There is one important subtlety that arises which distinguishes the $\mathbb{H}_2=SO(2,1)/SO(2)$ case from that of dS$_2=SO(2,1)/SO(1,1)$.  The group $SO(2,1)$ has a non-contractible circle, and is homeomorphic to the solid torus.  $SO(2)$ similarly has a non-contractible circle, so when we perform the quotient to obtain $\mathbb{H}_2$ we obtain a smooth simply connected space.  $SO(1,1)$, on the other hand, is itself simply connected, which is why the resulting quotient  dS$_2=SO(2,1)/SO(1,1)$ is topologically non-trivial. However, we could instead consider the universal cover $\overline{SO(2,1)}$ of $SO(2,1)$.  This is not a matrix group, but it allows us to construct a more general family of solutions.  In this case we would interpret dS$_2$ itself as the double quotient
\beq
	\text{dS}_2 = \mathbb{Z} \backslash \overline{SO(2,1)} \slash SO(1,1)\,.
\eeq
The $\mathbb{Z}$ here is generated by an elliptic element $\gamma$ of $\overline{SO(2,1)}$ which introduces a non-contractible spatial cycle of length $a = \cos^{-1} {1\over 2} {\rm tr} (\gamma)$.
If we do not perform the left quotient by $\mathbb{Z}$, then we instead obtain the universal cover of dS$_2$, which has metric $-dt^2 + \cosh^2 t d\theta^2$ where the $\theta$ coordinate is uncompactified.  This is a perfectly reasonable Lorentzian geometry with $R=2$.  Indeed, from this point of view we can choose any subgroup ${\mathbb Z} \subset \overline{SO(2,1)}$ to obtain the family of locally de Sitter geometries with $\alpha \ne 1$ considered earlier, where $\theta \sim \theta + 2\pi \alpha$.  

In order to construct a quotient of de Sitter which is fixed point-free, one then needs to consider the double quotient 
\beq
	\Gamma\backslash \overline{SO(2,1)} \slash SO(1,1)
\eeq
where the subgroup $\Gamma$ of $\overline{SO(2,1)}$ contains only parabolic or elliptic elements. Because $\overline{SO(2,1)}$ is not a matrix group it is difficult to classify all such $\Gamma$. However, it is easy to show that if we replace $\overline{SO(2,1)}$ with $SO(2,1)$, i.e.~we consider quotients of the $\alpha = 1$ de Sitter space, then $\Gamma$ must be abelian.\footnote{This can be seen by noting that the commutator of an elliptic element $A={\left(\cos \theta~\sin\theta\atop -\sin \theta~\cos\theta\right)}$ with another element $B = {\left(a~b\atop c~d\right)}$ obeys ${\rm tr} \left(ABA^{-1}B^{-1}\right) = 2+ \sin^2\theta \left[(a-d)^2+(b+c)^2)\right]$.  This implies that the either $a=d$ and $b=-c$, in which case $B$ commutes with $A$, or else the commutator $ABA^{-1} B^{-1}$ is hyperbolic.} Thus the only smooth quotients of dS$_2$ are those by an abelian group: it is impossible to construct a smooth, Lorentzian $R=2$ metric on a quotient of dS$_2$ with non-abelian fundamental group. We expect that this is also true for the double quotients of $\overline{SO(2,1)}$. This is very much in contrast with the Euclidean case, where one can obtain a smooth $R=-2$ metric on an arbitrary higher genus Riemann surface $\Sigma$.  

The question, then, is can we still obtain a sensible Lorentzian interpretation of the higher genus contributions to the JT gravity path integral? We have ruled out the simplest possibility, that we sum over smooth Lorentzian $R=2$ surfaces of arbitrary genus. Those surfaces simply do not exist. We could imagine summing over Lorentzian surfaces with singularities, but it is not clear what kinds of singularities we should allow and which ones we should forbid.

One might wonder if we could instead work in Euclidean signature. We have seen that in the $BF$ formulation of JT gravity, the natural continuation of de Sitter JT gravity is not to a theory with positive cosmological constant, but rather one with negative cosmological constant, namely the Euclidean AdS$_2$ version of JT gravity. In that theory one has a genus expansion in which one sums over constant negative curvature surfaces with asymptotic regions. Perhaps we could \emph{define} a genus expansion by continuation from the AdS genus expansion, in which these hyperbolic surfaces play the role of higher genus versions of the Maldacena contour. This does not work either. The culprits are the elliptic elements $\gamma$ in the quotient construction above, which in the gauge theory language correspond to elliptic holonomies around the boundary circles. The would-be Euclidean AdS higher genus geometries are double quotients 
\begin{equation*}
	\Gamma \backslash SO(2,1) \slash SO(2)\,,
\end{equation*}
where $\Gamma$ contains precisely these elliptic elements. As we discussed above, this double quotient is fixed-point free only when $\Gamma$ contains parabolic or hyperbolic elements, and so these surfaces are necessarily singular.\footnote{The exception is when $\Gamma$ is trivial, which corresponds to the hyperbolic disk here or to global dS$_2$ with $\alpha = 1$.}

%***********************************************
\subsubsection{Gauge theory characterization}
%***********************************************

Since neither the Lorentzian dS nor the Euclidean AdS higher genus geometries are non-singular, we elect to define the de Sitter JT path integral on higher genus surfaces using the topological gauge theory description. The idea is to consider a surface of fixed topology, and to integrate over the moduli space of flat $PSL(2;\mathbb{R})$ connections on it subject to the asymptotically dS$_2$ boundary conditions. In general we may allow for genus $g$ surfaces with a number of asymptotically nearly dS$_2$ regions; some in the past, and some in the future. Non-singularity in this setting simply means that the holonomy around each contractible cycle is trivial. There is then no obstruction to performing the path integral at fixed genus.

One way to summarize the analytic continuation of Euclidean AdS$_2$ to dS$_2$ JT gravity is that, starting from the $PSL(2;\mathbb{R})$ topological theory, the same flat connection corresponds to two different metrics on the same spacetime. One is of Euclidean signature with $R = -2$, and the other Lorentzian with $R=2$. Non-singularity of the gauge configuration does not guarantee that the metric descriptions are everywhere smooth. Indeed at nonzero genus, the discussion in the last Subsection implies that both the Euclidean and Lorentzian metrics constructed from the same flat connection will develop singularities. These singularities are always conical in Euclidean signature, and the analogue of conical in Lorentzian.

Defining the path integral this way we are effectively allowing ourselves in the metric formulation to sum over some types of singularities but not others. To get some intuition for this recall that the continuation of global dS$_2$ was the hyperbolic cylinder under $b\to 2\pi i \alpha$, where $b$ is the length of the minimal geodesic around the cylinder. The ensuing space is a double hyperbolic cone, where two cones each with a cone point of angle $2\pi \alpha$ are joined at the tips. We allow this cone point, provided that the spacetime does not end at it but instead continues into another cone of the same opening angle. The crucial feature is that this background is completely non-singular in the gauge description. There, nothing is shrinking to zero size at the tips since there is no metric and so no way to measure size. There is simply a non-contractible cycle around which there is an elliptic holonomy. It is only when we extract a metric that one finds, from the point of view of said metric, that this cycle shrinks to zero size and then expands again. Of course in the Lorentzian description, one simply has global dS$_2$ which is smooth everywhere.

In the next Subsection we will evaluate the JT path integral over these spacetimes. Before doing so, let us further discuss some interpretation of these higher genus spacetimes.

Recall that we would like to interpret the Euclidean path integral a la Hartle-Hawking, as preparing a quantum state -- the wave function of the universe -- in the Hilbert space of the Lorentzian theory. In the metric formulation, we construct a state by gluing a Lorenztian geometry onto a Euclidean geometry.  The resulting geometry has mixed-signature, where Lorentzian and Euclidean geometries are glued together along a spacelike slice. In the original Hartle-Hawking procedure the future half of Lorentzian dS$_2$ is glued to a Euclidean hemisphere along the surface of vanishing extrinsic curvature, $t=0$. 

In the first-order formulation there are two seemingly different ways to glue in a Euclidean segment, both of which are different from the standard picture in the metric formulation. One can either construct the geometry by gluing a future region of Lorentzian dS onto a hyperbolic disk (with the asymptotic region removed), or one could simply glue ${\cal I}^+$ onto the Euclidean disk. Although apparently different, these constructions are completely equivalent. In Subsection~\ref{S:partitions} we saw that they lead to the same path integral. In the latter, the Euclidean gravity path integral on the disk computes a partition function $Z_{\rm disk}(\beta J)$, which is related to the de Sitter wave function via the analytic continuation $\Psi_{\rm HH}(\beta J) = Z_{\rm disk}(-i\beta J)$, which we computed directly from the Lorentzian description.

\begin{figure}[t!]
	\centering
	\includegraphics[width=4.5in]{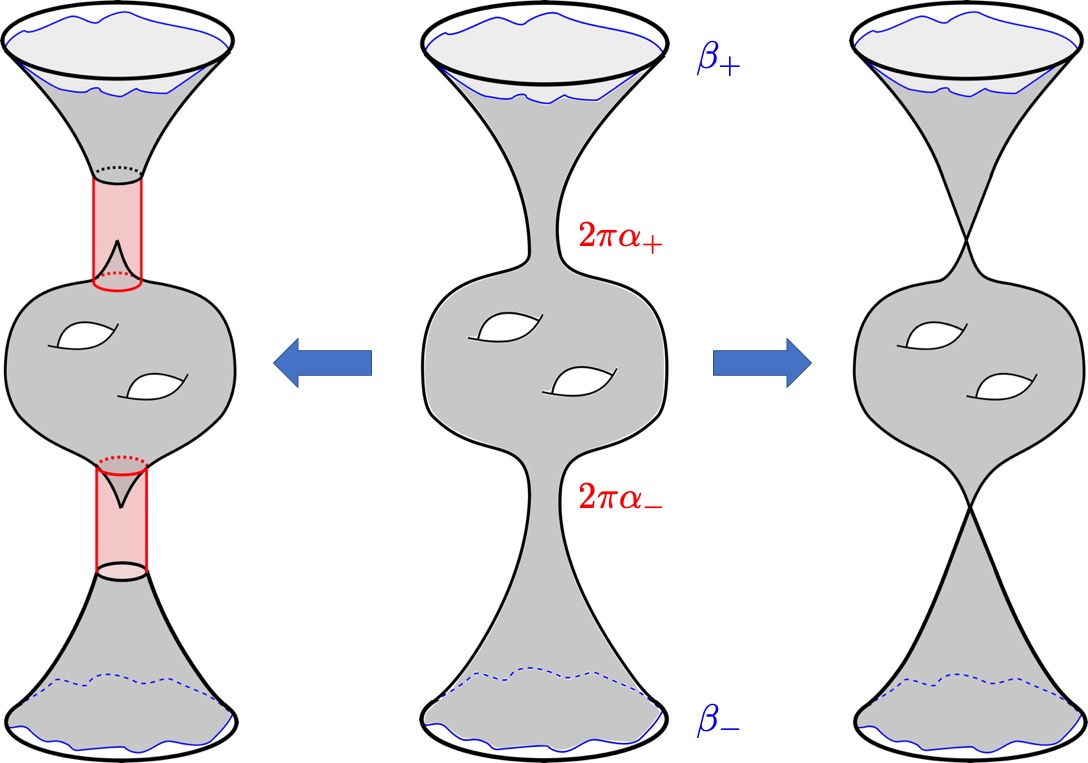}
	\vskip.5cm
	\caption{
		\label{F:singularGluing} Our higher genus geometries. We begin with some smooth surface equipped with a flat $PSL(2;\mathbb{R})$ connection, pictured in the middle. It has asymptotic boundaries, around which there are elliptic holonomies characterized by $\alpha_i$. From this connection we may obtain a mixed-signature metric, pictured on the left. It has Lorentzian dS$_2$ regions glued to an intermediate $R=-2$ surface $\Sigma$ with cone points. The Lorentzian regions have minimal geodesics of lengths $2\pi \alpha_i$, matching the cone angles of $\Sigma$. Equivalently, we can extract a Euclidean $R=-2$ geometry pictured on the right, in which the asymptotic regions are replaced with hyperbolic cones of opening angles $2\pi \alpha_i$ glued to the circles at $\mathcal{I}^{\pm}$.
		}
\end{figure}

The first of these constructions is the first-order analogue of the Hartle-Hawking contour, while the second is the analogue of the Maldacena contour. In the gauge theory these approaches are identical: both spacetimes are described by the same, real $PSL(2;\mathbb{R})$ connection, from which we can extract the two metrics. Note that, while in the metric formulation we had to perform the gluing across a surface of vanishing extrinsic curvature in order to ensure that the total geometry was smooth, in the gauge theory formulation we only require that the $PSL(2;\mathbb{R})$ connection is flat with trivial holonomy around trivial cycles.

From this point of view we may discuss metric interpretations for our higher genus surfaces as higher genus analogues of the Hartle-Hawking and Maldacena contours. In the former, we fill in $\mathcal{I}^+$ and $\mathcal{I}^-$ with $n$ asymptotically future and past Lorentzian dS$_2$ regions, where $n$ is the number of connected components of the boundary. They are characterized by $n$ elliptic elements of $\overline{SO(2,1)}$, which in the gauge theory description are exactly the holonomies around the asymptotic circles. We then glue all of these regions to an intermediate genus $g$ surface $\Sigma$. This surface is most naturally described in the Euclidean $R=-2$ continuation, i.e. as a quotient of $\mathbb{H}_2$, but from our point of view it is merely a surface equipped with a flat $PSL(2;\mathbb{R})$ bundle. In order for this surface to be glued to the asymptotic regions it must have the same elliptic holonomies around some cycles; that is, it must have $n$ cone points. In the total spacetime we glue the trumpets to $\Sigma$, and the whole construction is pictured on the left-most diagram of Fig.~\ref{F:singularGluing}. 

The other approach, the analogue of the Maldacena contour, is to fill in the $\mathcal{I}^+$ and $\mathcal{I}^-$ with $n$ asymptotically $\mathbb{H}_2$ disks. Due to the elliptic holonomies at infinity, these asymptotic regions will have conical singularities in their interior. We glue these to the same intermediate surface $\Sigma$ described above, as in the right-most diagram of Fig.~\ref{F:singularGluing}.

As we have emphasized these seemingly different approaches are really one and the same, since we are dealing with the same flat $PSL(2;\mathbb{R})$ bundle. As far as a metric interpretation goes, the ``Hartle-Hawking'' geometry glues Lorentzian $R=2$ regions to a Euclidean $R=-2$ surface, while in the ``Maldacena'' contour the geometry is Euclidean with negative curvature throughout.  We elaborate on this below.

%***********************************************
\subsubsection{Complex metrics}
%***********************************************

Here we elaborate on the Maldacena contour perspective on the metric formulation of nearly-dS$_2$ gravity. This discussion has significant overlap with that of~\cite{Maldacena:2019cbz}. That is, we can deform the time coordinate of the nearly-dS$_2$ metric configurations along a complex contour to arrive at Euclidean nearly-AdS$_2$ metrics in $(-,-)$ signature.  As such, we can view Lorentzian nearly-dS$_2$ gravity and Euclidean nearly-AdS$_2$ gravity as equivalent to one another under a deformation of the complex time contour.

We will carry out these contour deformations in the cases of the nearly-dS$_2$ Hartle-Hawking geometry which will become the Euclidean nearly-AdS$_2$ disk, and the global de Sitter geometry which will become the Euclidean nearly-AdS$_2$ double trumpet.

\begin{figure}[t]
	\centering
	\includegraphics[width=3.4in]{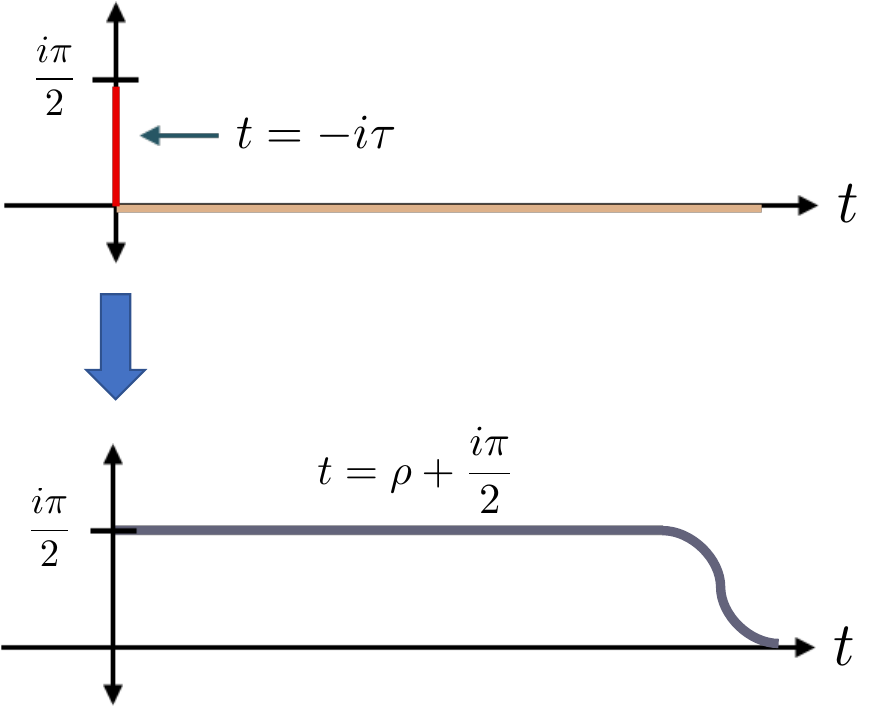}
	\vskip.5cm
	\caption{
		\label{F:complex1} 
Depiction of the change of complex time contour to go from the Hartle-Hawking geometry in Lorentzian nearly-dS$_2$ in $(-,+)$ signature to the disk in Euclidean nearly-AdS$_2$ in $(-,-)$ signature. For the $t = \rho + \frac{i \pi}{2}$ contour, we take it to dip down to the real axis at infinity, as sketched in the diagram. \\ }
\end{figure}

First, we consider the Hartle-Hawking geometry.  The complex metric is
\begin{align}
\tau \in [-\pi/2,0] : \,\, \begin{cases} ds^2 = d\tau^2 + \cosh^2(\tau) \, d\theta^2 \\
\overline{\varphi} = - \frac{2\pi i}{\beta J} \, \sinh(\tau)
\end{cases} ,\qquad  t \geq 0 : \,\, \begin{cases} ds^2 = - dt^2 + \cosh^2(t) \, d\theta^2 \\
\overline{\varphi} = \frac{2\pi}{\beta J} \, \sinh(t) \end{cases} \,.
\end{align}
The real-time segment is glued at $t=0$ to the imaginary time segment at $\tau=0$. In fact, the Euclidean part is just the Lorentzian part at imaginary time $t=-i\tau$, and so the total geometry may be understood as a complex time contour as in Fig.~\ref{F:complex1}. 
Deforming the time contour as $t$ as in that Figure, we obtain the disk geometry in Euclidean nearly-AdS$_2$ in $(-,-)$ signature:
\begin{equation}
ds^2 = - \left(d\rho^2 + \sinh^2(\rho) \,d\theta^2 \right)\,, \qquad \overline{\varphi} = - \frac{2\pi i}{\beta J} \, \cosh(\rho)\,.
\end{equation}
Note that no singularities obstruct this continuation. This mapping from the Hartle-Hawking geometry in dS to the Euclidean disk in AdS was in fact the original conception of the Maldacena contour \cite{Maldacena:2010un}. The only new ingredient in two dimensions is that we also have to keep track of the dilaton profile, which also continues smoothly, albeit with a factor of $i$. 

\begin{figure}[h]
	\centering
	\includegraphics[width=3.4in]{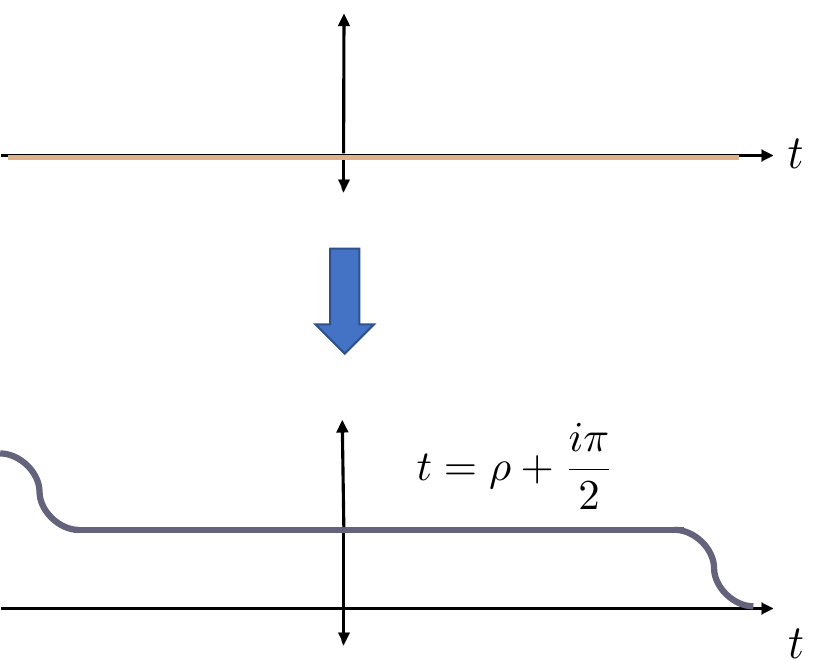}
	\vskip.5cm
	\caption{
		\label{F:complex2} 
The change of complex time contour to go from the global de Sitter geometry in Lorentzian nearly-dS$_2$ in $(-,+)$ signature to the double trumpet geometry in Euclidean nearly-AdS$_2$ in $(-,-)$ signature.  We take the complex contour to go to $i\pi$ as $t \to -\infty$ and $0$ as $t \to \infty$.  The global de Sitter geometry is the only classical solution to nearly-dS$_2$ JT gravity. \\}
\end{figure}

Next, let consider global de Sitter.  The metric and dilaton are
\begin{equation}
ds^2  = - dt^2 + \alpha^2 \cosh^2(t) \, d\theta^2\,, \qquad \overline{\varphi} = \frac{\alpha}{\beta J} \, \sinh(t) \,.
\end{equation}
This geometry is not complex, which is due to the fact that the global de Sitter geometry is the only solution to the equations of motion of nearly-dS$_2$ JT gravity.  Deforming the time contour as per Fig.~\ref{F:complex2}, we obtain 
\begin{equation}
ds^2 = - (d\rho^2 + \alpha^2 \sinh^2(\rho) \, d\theta^2)\,, \qquad \overline{\varphi} = - \frac{i \,\alpha}{\beta J} \, \cosh(\rho)\,.
\end{equation}
which is the union of two hyperbolic cones in $(-,-)$ signature.  Notice that the de Sitter geometry is non-singular, whereas the corresponding Euclidean nearly-AdS$_2$ geometry has a double cone point.  Thus the de Sitter continuation resolves the singularity of the double cone.

We expect that a more general version of the analysis above works for higher-genus spacetimes.

%[[ the original continuation of maldacena ]]

%[[ continuation for global dS: two hyperbolic discs ]]

%[[ continuation for general spacetimes ]]

%***********************************************
\subsection{Higher genus amplitudes}
\label{S:HigherGenusContinuation}
%***********************************************

In this Subsection we compute the path integral of JT gravity over the higher genus surfaces obtained above. First, let us develop some notation. Suppose we have a geometry where $\mathcal{I}^+$ is composed of $n_+$ circles, and $\mathcal{I}^-$ of $n_-$ circles. Let $n=n_++n_-$ count the connected components of the boundary. Let $\beta_a$, with $a=1,...,n_+$ denote the sizes of the future circles, and $\beta_m$ with $m=n_++1,...,n$ the sizes of the past circles. Each of these asymptotic circles is characterized by an elliptic holonomy. The bulk spacetime fills in these circles in the following way. We glue asymptotically dS$_2$ ``trumpets'' to each of these circles. If we let the trumpet go to a geodesic boundary, then the holonomy is encoded in the length $2\pi \alpha_i$ of this geodesic. We then glue in an intermediate genus $g$ surface $\Sigma$. Metrically, this surface is most naturally described as hyperbolic with conical singularities of angles $2\pi \alpha_i$. We then glue the trumpets to $\Sigma$. See Fig.~\ref{F:singularGluing}.

The other way of building the higher genus surface is the analogue of the Maldacena contour. We instead attach hyperbolic disks to $\mathcal{I}^+$ and $\mathcal{I}^-$. The elliptic holonomies around the asymptotic circles guarantees that these disks are hyperbolic cones with angles $2\pi \alpha_i$. We then glue to $\Sigma$. Importantly, both of these constructions correspond in the topological gauge theory description to the same $PSL(2;\mathbb{R})$ gauge field, and so the same path integral.

Since all of the boundaries are glued to the same intermediate surface, we are computing a \textit{connected} spacetime diagram. We parameterize this contribution to the transition amplitude as
\beq
	Z_{g,n_+,n_-}(\beta_aJ,\beta_mJ) / (e^{ S_0})^{2g + n - 2}\,.
\eeq
As defined, $Z_{g,n_+,n_-}$ does not depend on $S_0$. Now, if we want to consider the sum over all connected genus $g$ contributions to the $n$ boundary amplitude, we have
\beq
\label{E:Psiconn1}
	\Psi_{n_+, n_-}^{\text{conn}}(\beta_aJ,\beta_mJ) \simeq \sum_{g=0}^\infty \frac{Z_{g,n_+, n_-}(\beta_aJ,\beta_mJ)}{ (e^{S_0})^{2g + n - 2}}\,.
\eeq
The factor $(e^{S_0})^{-2g - n + 2}$ comes from the Euler characteristic term in the action, $S_0\chi$. Recall that for the Hartle-Hawking geometry, even though the geometry was a topological disk, we found $\chi = -i$. The Lorentzian part of the geometry, a topological annulus, gave zero contribution, while the Euclidean part, a topological disk, gave $-i$ times the topological characteristic of the disk. In these more general geometries the Lorentzian trumpets, each being a topological annulus, give zero contribution to the topological Euler characteristic. The only contribution comes from $\Sigma$, and so we have $\chi = - i \chi_T$, where $\chi_T$ is the topological Euler characteristic $\chi_T = 2 -2g -n$. In this way the contribution of genus $g$ surfaces with $n$ boundaries is weighted as above. This is the genus expansion of nearly dS$_2$ gravity.

The full $n$ boundary amplitude $\Psi_{n_+,n_-}(\beta_aJ,\beta_mJ)$  contains both connected and disconnected contributions.  Thus, we can express $\Psi_{n_+, n_-}(\beta_aJ,\beta_mJ)$ using the standard trick of constructing a generating function:
\begin{align}
\label{E:Psifull1}
	&\Psi_{n_+, n_-}(\beta_aJ,\beta_mJ) 
	\\
	\nonumber 
	& \quad \simeq \left.\frac{\delta^n}{\delta \xi(\beta_1 J) \cdots \delta \xi(\beta_n J)}\right|_{\xi = 0} \, \exp\left(\sum_{k=1}^\infty \frac{1}{k!} \int_0^\infty \!\! dx_1 \cdots \int_0^\infty \!\! dx_k \, \xi(x_1) \cdots \xi(x_k) \, \Psi_{n_+, n_-}^{\text{conn}}(x_1,...,x_k) \right)\,.
\end{align}

We are now in a position to compute the $Z_{g,n_+,n_-}$'s. In Subsection~\ref{S:partitions} we computed the gravitational path integral on the Hartle-Hawking geometry. This is the analogue of the disc partition function of JT gravity on Euclidean nearly AdS$_2$. Factoring out the Euler term, the gravitational path integral was
\beq
	Z_{0,1,0} = \int\frac{[df]\text{Pf}(\omega)}{PSL(2;\mathbb{R})} e^{iS_{\text{sch}}} \,, \quad S_{\rm sch} = \frac{1}{g}\int_0^{2\pi} d\theta\left( \{f(\theta),\theta\} + \frac{f'(\theta)^2}{2}\right)\,,
\eeq
with $g = 4 G_2 \beta J$, and the result was
\beq
	Z_{0,1,0} = \frac{1}{\sqrt{2\pi}(-2i\beta J)^{3/2}} \, e^{\frac{i\pi}{g}}\,.
\eeq
The path integral for the version with one past asymptotic boundary, $Z_{0,0,1}$, is merely the complex conjugate of $Z_{0,1,0}$.

 To obtain the higher genus contributions we divide up the path integral into the contributions from the Lorentzian dS$_2$ regions, and from the surface $\Sigma$. To get the former we need the path integral on the dS$_2$ version of the future and past ``trumpets'' at fixed $\alpha$. We computed these in Subsection~\ref{S:partitions} with the result
\begin{equation*}
	Z_F(\beta J,\alpha) = Z_P(\beta J,\alpha)^* = Z_T(-i \beta J,2\pi i \alpha) = \frac{1}{\sqrt{2\pi}(-2i \beta J)^{1/2}} \, e^{\frac{i \pi \alpha^2}{4 G_2 \beta J}}\,.
\end{equation*}
Here $Z_T$ is the path integral over the Schwarzian mode on a trumpet for JT gravity on Euclidean AdS$_2$~\cite{Saad:2019lba} (after matching normalizations of the symplectic measure).

In terms of the trumpet $Z$'s, the path integral on global dS$_2$ which is $Z_{0,1,1}$ in the notation here, is given by a stitching of future and past trumpets in Eq.~\eqref{E:Zglobal}. This stitching depends on the symplectic measure for $\alpha$ and the twist $\gamma$. Up to a sign, the measure we get from the symplectic form on flat connections is
\begin{equation*}
	\mp \frac{d\alpha d\gamma \, \alpha}{2G_2}\,.
\end{equation*}
The minus convention is what one finds under the continuation under $b\to 2\pi i \alpha$. Indeed we found that with this convention $Z_{0,1,1}$ may be consistently interpreted as arising from a matrix integral, and further it is the continuation of the ``double trumpet'' path integral in Euclidean AdS$_2$~\cite{Saad:2019lba}. In what follows we continue to use the minus convention.

Before considering more complicated surfaces, consider $Z_{0,2,0}$, i.e. the amplitude for two future universes produced from nothing. There is also $Z_{0,0,2}$, given by its complex conjugate $Z_{0,2,0}^*$. Stitching together two future trumpets we have
\beq
	Z_{0,2,0} = - (2\pi)\int_0^{\infty} \frac{d\alpha\,\alpha}{2G_2}Z_F(\beta_1 J,\alpha) Z_F(\beta_2J, \alpha) = \frac{1}{2\pi}\frac{\sqrt{\beta_1\beta_2}}{\beta_1+\beta_2}\,.
\eeq
This is also the continuation of the double trumpet under $Z_{0,2,0}(\beta_1J,\beta_2J) = Z_{0,2}(-i \beta_1J,-i \beta_2J)$.

At the end of Subsection~\ref{S:partitions} we interpreted the global dS$_2$ partition function as coming from a double scaled matrix model. Assuming such a correspondence we thereby extracted the connected two-point function of resolvents at genus 0, namely $R_{0,2}$. We can also extract $R_{0,2}$ from $Z_{0,2,0}$, or $Z_{0,0,2}$. Interpreting
\beq
	Z_{0,2,0}(\beta_1 J,\beta_2J) = \left\langle \text{tr}\left( e^{i \beta_1H}\right) \text{tr}\left( e^{i \beta_2H}\right)\right\rangle_{\overline{\text{MM}},\,\rm conn,\, 0}\,,
\eeq
using the notation explained in Subsection~\ref{S:partitions}, we would then have
\begin{align}
\begin{split}
	R_{0,2}(\lambda_1,\lambda_2)& = - \int_0^{\infty} d\beta_1 d\beta_2 e^{-i \beta_1 \lambda_1 -i \beta_2\lambda_2} Z_{0,2,0}(\beta_1 J,\beta_2J)
	\\
	& = \frac{1}{4\sqrt{-\lambda_1}\sqrt{-\lambda_2}(\sqrt{-\lambda_1}+\sqrt{-\lambda_2})^2}\,,
\end{split}
\end{align}
for negative real $\lambda_i$. (We will discuss how the minus sign arises in the next Subsection.) This is the same result we found from the path integral on global dS in Eq.~\eqref{E:R02}. This is a nice consistency check on our interpretation: assuming that $Z_{0,1,0}$ and $Z_{0,2,0}$ arise from probing the \emph{same} matrix model with different observables, we then ought to find the same correlation functions of resolvents whether the boundaries are in the future or the past.

In more complicated cases there is now an intermediate surface $\Sigma$ with $n$ holes. The surface gives a topological contribution, the volume of the moduli space of flat connections on $\Sigma$ with prescribed elliptic holonomies around the holes. Equivalently, it is the volume of the moduli space of hyperbolic cones with $n$ cone points of angles $2\pi \alpha_i$. This volume is computed with respect to the Weil-Petersson measure. We denote it as
\beq
	\widetilde{V}_{g,n}(\alpha_1,\hdots,\alpha_n).
\eeq
We then arrive at our expression for the path integral on the total geometry
\begin{align}
\label{E:Zgn1n2}
	Z_{g,n_+,n_-}(\beta_a,\beta_m) =& (-1)^n(2\pi)^n\int_0^{\infty} \frac{d\alpha_1\,\alpha_1}{2G_2}\hdots\frac{d\alpha_n\,\alpha_n}{2G_2} \widetilde{V}_{g,n}(\alpha_1,\hdots,\alpha_n) 
	\\
	\nonumber
	&\times Z_F(\beta_1J;\alpha_1) \hdots Z_F(\beta_{n_+}J;\alpha_{n_+}) Z_P(\beta_{n_++1}J;\alpha_{n_++1}) \hdots Z_P(\beta_nJ;\alpha_n)\,.
\end{align}
The factor of $(-1)^n$ comes from the fact that we take the measure for each $\alpha$ to be $-d\alpha\, \alpha$. It may be interpreted as inserting a factor $(-1)^E$ into the path integral, where $E$ counts the number of independent elliptic holonomies integrated over.

Let us make a brief aside, before returning to the computation of the $Z_{g,n_+,n_-}$'s. The leading dependence of $Z_{g,n_+,n_-}$ on $G_2$ as $G_2\to 0$ is
\beq
	G_2^{-n}\times G_2^{-3g-n} \times G_2^{n} = G_2^{-\frac{3}{2}(2g+n-2)+\frac{n}{2}} = G_2^{-\frac{3\chi-n}{2}}\,.
\eeq
It arises in the following way. The first $G_2^{-n}$ is the manifest $G_2^{-n}$ in the measure over the $n$ $\alpha$'s. It comes from the normalization of the Weil-Petersson symplectic form over the space of flat connections, which goes as $1/G_2$, Eq.~\eqref{E:PSLomega}. The precise normalization chosen there is a matter of convention, but the fact that it goes as $1/G_2$ is physical. The second factor comes from the volume $\widetilde{V}_{g,n}$ which is computed with respect to the same symplectic form and so goes as $O(G_2^{-3g-n+3})$, coming from the integral over $3g+n-3$ internal lengths. The final factor of $G_2^{n}$ comes from the integral, after changing variables $\alpha \to \sqrt{G_2} \alpha$ and using that as $G_2\to 0$ the volume $\widetilde{V}_{g,n}(\alpha_1,\hdots,\alpha_n) \to \widetilde{V}_{g,n}(0)$ goes to a nonzero constant, the moduli space of genus $g$ hyperbolic surfaces with $n$ punctures. We interpret this effect as a renormalization of the Euler term in the action,
\beq
	S_0 \longrightarrow S_0 +  \frac{3}{2} \log(\# G_2)\,.
\eeq
(The precise number depends on the choice of convention for the normalization of the symplectic form.) In Eq.~\eqref{E:Scosmo} we found the cosmological entropy of the ``static patch'' of nearly dS$_2$ to be $2S_0$. It is tempting to guess that the effect above also renormalizes the cosmological entropy, although to see if that is the case one would need to look at genus corrections to the static patch entropy.

To compute the genus expansion coefficients we require the volumes $\widetilde{V}_{g,n}$. These are in general unknown. There is significant evidence that they are the analytic continuation of the Weil-Petersson volumes $V_{g,n}$ of the moduli space of bordered surfaces, i.e. of $R=-2$ surfaces with geodesic boundaries of lengths $b_1,b_2,...,b_n$. The latter play a pivotal role both in JT gravity on Euclidean AdS$_2$ and in the topological recursion of a certain double-scaled matrix model. Following previous work~\cite{2011cones}, we conjecture that the two are related by
\beq
\label{E:theConjecture}
	\widetilde{V}_{g,n}( \alpha_1,\hdots,\alpha_n) = V_{g,n}(2\pi i \alpha_1,\hdots 2\pi i \alpha_n)\,.
\eeq 
One piece of evidence for this statement comes from the $\alpha\to 0$ limit. To see this, let us recall a central observation of Mirzakhani \cite{mirzakhani2007simple, mirzakhani2007weil} that the moduli space of bordered Riemann surfaces ${\cal M}_{g,n}(b_1,\dots,b_n)$ is symplectomorphic to the moduli space of punctured Riemann surfaces ${\cal M}_{g,n}$.  This is proven by considering the symplectic reduction of the natural $\left(\mathbb{S}^1\right)^n$ bundle over ${\cal M}_{g,n}(b_1,\dots,b_n)$ defined by the boundary circles. Roughly speaking, one imagines shrinking the boundary lengths $b_i\to 0$; in this limit, a hyperbolic element $\gamma$ of $PSL(2,\mathbb{R})$ with ${\rm tr}(\gamma)  = 2 \cosh\left(\frac{b_i}{2}\right)$ becomes parabolic, and the resulting surfaces has a cusp at which a point is removed.  
This cusped surface is the same one that arises in the $\alpha \to 0$ limit.\footnote{In fact, it was observed by \cite{DoNorbury} that if we take one of the $b_i\to 2\pi i$ then there is a sense (made precise in \cite{DoNorbury}) in which the marked point is removed entirely!  Our proposal can be regarded as an extension of this to $\alpha \ne 1$.}  

A stronger piece of evidence, which in fact subsumes this observation, comes from Tan, Wong, and Zhang~\cite{2004cones}. They proved that the conjecture Eq.~\eqref{E:theConjecture} holds as long as $\alpha_i \leq 1/2$, i.e. the cone angles are less than or equal to $\pi$. Their methods break down for larger cone angles. See~\cite{DoNorbury,2011cones} for some discussion. With this in mind, our conjecture is really the statement that the volume of genus $g$ surfaces with cone points remains analytic as a function of opening angles past $\pi$.

The volumes $V_{g,n}$ are rather difficult to compute directly. See~\cite{nakanishi2001} for the first computation of $V_{1,1}$ (see also the nice computation of~\cite{Kim:2015qoa} in the language of AdS$_3$ gravity). However, Mirzakhani showed that the volumes obey a recursion relation~\cite{mirzakhani2007simple, mirzakhani2007weil} which allows their efficient computation up to any desired $g$ and $n$. One output of this recursion relation is that the volumes are polynomials in the lengths-squared of the geodesic borders. If our conjecture is right, then the volumes $\widetilde{V}_{g,n}$ satisfy a related recursion relation and may be easily calculated.

Assuming this is indeed the case, then as an example of our formulas we compute the connected part of the Hartle-Hawking nearly dS$_2$ wavefunction up through genus two. We require the volumes $V_{1,1}$ and $V_{2,1}$, which in our normalization of the symplectic form (in the convention of~\cite{Saad:2019lba} we have $\alpha_{\rm them} = \frac{1}{4\pi G_2}$) are\footnote{Each genus one surface with one border is related by a $\mathbb{Z}_2$ symmetry to another such surface. This symmetry is a large gauge transformation from the point of view of the topological gauge theory, and so in the path integral we only account for volume of moduli space modulo it. The resulting volume is denoted as $\overline{V}_{1,1}$.}
\beq
	\overline{V}_{1,1}(2\pi i \alpha) = \frac{\pi(1-\alpha^2)}{48G_2}\,, \qquad V_{2,1}(2\pi i \alpha) = \frac{\pi^4(1-\alpha^2)(3-\alpha^2)(435-96\alpha^2+5\alpha^4)}{2211840G_2^4}\,.
\eeq
A striking feature of these volumes is that they become negative for sufficiently large $\alpha$. Recall that the continuation of the Weil-Petersson measure over hyperbolic holonomies, $bdb$, becomes negative for elliptic holonomies, $-(2\pi)^2 \alpha d\alpha$. Assuming that the conjecture~\eqref{E:theConjecture} is correct, it is tempting to speculate that the origin of these negative volumes is that for sufficiently large $\alpha$ the dominant contribution to the volume comes from regions of moduli space where the holonomies around the internal cycles are elliptic. In any case we then find
\begin{align}
	\Psi_{\text{HH}}^{\rm conn}(\beta J) =&  e^{S_0}\sqrt{\frac{-i\beta J}{16\pi}} \left( \frac{e^{\frac{i \pi}{g}}}{(-i\beta J)^2} -e^{-2S_0}\frac{\pi + i g}{12G_2} \right.
	\\
	\nonumber
	&  -\left. e^{-4S_0} \frac{435 \pi^4+676i\pi^3 g-556\pi^2g^2-232i\pi g^3+40g^4}{184320G_2^4} + O(e^{-6S_0})\right)\,.
\end{align}

More generally, there would be a simple analytic continuation that relates the genus expansion coefficients above to those $Z_{g,n}$ of JT gravity on Euclidean AdS$_2$~\cite{Saad:2019lba}. Since the $V_{g,n}$'s are polynomials, the integrand of the $Z_{g,n_+,n_-}$'s in Eq.~\eqref{E:Zgn1n2} would be analytic away from infinity and we can deform the contour of integration from the real $\alpha$ axis to the negative imaginary $\alpha$ axis, i.e. along the positive real $b=2\pi i \alpha$ axis. Matching the normalization of the symplectic form, one can easily show that our expression~\eqref{E:Zgn1n2} for the higher genus integrals is related to that of~\cite{Saad:2019lba} (see their Eq.~(127)) by
\beq
\label{E:analyticZs}
	Z_{g,n_+,n_-}(\beta_a J,\beta_mJ) = Z_{g,n}(-i\beta_a J,i \beta_mJ)\,,
\eeq
which extends our result above for the disk and for global dS$_2$.
\\ \\
\textbf{Historical note added in [v3]:} The conjectural relation~\eqref{E:theConjecture} was later disproved in~\cite{Turiaci:2020fjj, Eberhardt:2023rzz}. As such, the matrix model interpretation in the following subsection, which assumes the veracity of the conjecture, does not hold.  However, a careful treatment of the gravitational path integral for nearly-dS$_2$ JT gravity, which does not rely on any conjectures, was performed in~\cite{Cotler:2024xzz}. In this treatment, the sum over spacetimes was performed over hyperbolic metrics in $(-,-)$ signature, with the result being dual to a new matrix model for dS JT gravity which is a certain analytic continuation of the Saad-Shenker-Stanford matrix model for AdS JT gravity. The new matrix model bears some resemblance to the discussion below.

%***********************************************
\subsection{Matrix model interpretation}
\label{S:MMdescription}
%***********************************************

In the analysis above we have written a genus expansion for the path integral of JT gravity with positive cosmological constant. The expansion coefficients depend on the Weil-Petersson volumes of hyperbolic cones. At the end we made a conjecture that these volumes are for all cone angles $\alpha_i$ the continuation of the Weil-Petersson volumes of moduli spaces of bordered Riemann surfaces. If that is indeed the case, then the genus expansion coefficients of nearly dS$_2$ gravity are the continuation of those of nearly Euclidean AdS$_2$ gravity. In this Subsection, following~\cite{Saad:2019lba} and assuming that the above conjecture is true, we show that the nearly dS$_2$ expansion comes from a double-scaled matrix integral. Much like the case of AdS$_2$ gravity, this matrix integral gives a non-unique, non-perturbative completion of nearly dS$_2$ gravity.

To begin let us consider a one-matrix model following \cite{Saad:2019lba}. We start with a model of a single $L\times L$ Hermitian matrix $H$ characterized by a real potential $V$. Matrix averages are given by
\beq
	\langle O(H)\rangle_{\rm MM}  = \frac{\int dH \, e^{-L\text{tr}(V(H))}O(H)}{\int dH\, e^{-L\text{tr}(V(H))}}\,,
\eeq
and we take the large $L$ limit.  A fundamental property of a matrix model is its density of eigenvalues $\rho(E)$, computed by
\beq
	\rho(E) = \left\langle \sum_{i=1}^L \delta(E - \lambda_i) \right\rangle_{\text{MM}}\,,
\eeq
where $\lambda_i$ is the $i$th eigenvalue, for $i=1,...,L$.  Since there are $L$ eigenvalues, we must have $\int dE \, \rho(E) = L$.  Suppose we consider a matrix model with a density of eigenvalues given by
\beq
	\rho(E) = \kappa \,\frac{ \gamma e^{S_0'}}{2 \pi^2} \sinh\left(2 \pi \sqrt{2 \gamma}\sqrt{\frac{a^2 - (E-a)^2}{2a}}\right) + \cdots\,,
\eeq
for $E \in [0,2a]$ where $a$ is fixed by the normalization condition, and $\kappa$ is some constant. The ellipses denote terms which go to zero as $L$ goes to infinity, and also terms subleading in powers of $1/e^{S_0'}$ where $e^{S_0'} \gg 1$.  We then take $L \to \infty$, which sends $a \to \infty$. Normally the double scaling limit is the combination of sending $L\to\infty$ along with a redefinition of energy so that the spectrum begins at $E=0$. Since we have shifted the spectrum, the double scaling limit here is merely $L\to\infty$, giving
\beq
\label{E:leadingAdSrho}
	\rho(E) = \kappa\,\frac{\gamma e^{S_0'}}{2 \pi^2} \sinh\left(2 \pi \sqrt{2 \gamma E}\right) \,.
\eeq
This is the form of the density of states one obtains from nearly AdS$_2$ gravity in Eq.~\eqref{E:Zdisc3} with
\beq
\label{E:rhoParamsforAdS}
	\gamma = \frac{1}{8\pi G_2 J'}\,, \qquad \kappa = 8 (\pi G_2)^{3/2}\,.
\eeq
Since we are considering $e^{S_0'} \gg 1$, we have that $e^{S_0'}$ becomes the genus expansion parameter of the double-scaled matrix model.

More generally it is known that a double-scaled matrix model has a genus expansion of the form
\beq
\label{E:MMgenus}
	\left\langle \text{tr}(e^{-\beta_1 H}) \cdots \text{tr}(e^{-\beta_n H}) \right\rangle_{\overline{\text{MM}},\, \text{conn}} \simeq \sum_{g=0}^\infty \frac{Z_{g,n}^{\kappa, \gamma}(\beta_1,\hdots,\beta_n)}{ (e^{S_0'})^{2g + n - 2}}\,,
\eeq
The subscript $\overline{\text{MM}}$ indicates that we are taking expectation values in the double-scaled limit of the matrix model, the subscript ``conn'' that we taking the connected part, and the $\simeq$ indicates that this series is asymptotic. In the above equation, $Z_{g,n}^{\kappa, \gamma}(\beta_i)$ depends on its inputs as well as $\kappa$ and $\gamma$. One of the results of~\cite{Saad:2019lba} is that the genus expansion coefficients $Z_{g,n}(\beta_i)$ of JT gravity on Euclidean AdS$_2$ are simply the $Z_{g,n}^{\kappa,\gamma}$'s evaluated on the parameters in Eq.~\eqref{E:rhoParamsforAdS} that specify the density of states one finds from nearly AdS$_2$ gravity. Soon it will be useful that we can analytically continue to complex $\beta_i$ for $\text{Re}(\beta_i)>0$.

%% HERE HERE

A fact of crucial importance is that the $Z_{g,n}^{\kappa,\gamma}$ are not all independent. They are determined via the Eynard-Orantin topological recursion relations~\cite{eynard2004all, eynard2007invariants, eynard2007weil}, which for the double-scaled density of states above are related to the Mirzakhani recursion relation for the $V_{g,n}$'s~\cite{mirzakhani2007simple, mirzakhani2007weil}. We refer the reader to the nice summary of both in Section 2 of~\cite{Saad:2019lba}. Indeed, the genus expansion of AdS JT gravity obeys these topological recursion relations with the leading density of states~\eqref{E:leadingAdSrho}. It may then be understood as the genus expansion of the double-scaled matrix model above. Further, the coefficients are analytic functions of $\beta_i$ in the domain $\text{Re}(\beta_i)>0$.

Now we turn to the de Sitter expansion. Our approach is to posit that it comes from a double-scaled matrix model, and to then show that this assumption is consistent and leads to the same matrix model one encounters in AdS JT gravity. We begin with the dictionary that maps gravitational observables to matrix integrals. Following our discussion in Subsection~\ref{S:partitions}, we posit that``inserting'' a future boundary of size $\beta$ corresponds to probing the matrix model with $\text{tr}(e^{i \beta H})$, and a past boundary to inserting $\text{tr}(e^{-i \beta H})$. Further, recall that with this interpretation, the ``disc'' partition function of de Sitter JT gravity, the path integral on either the future or past Hartle-Hawking geometry, encodes the same density of states~\eqref{E:leadingAdSrho} as AdS JT gravity under $S_0'\to S_0$ and $J'\to J$.\footnote{In Euclidean AdS JT gravity, the value of the dilaton on the boundary determines $J'$. The analytic continuation from Euclidean AdS JT gravity to de Sitter JT gravity with a future boundary takes $J'\to - i J$, and $\beta' \to \beta$. However, the genus expansion only depends on the combination $\beta J$, not $\beta$ or $J$ separately. We are exploiting this fact here.} 

Given $n_+$ future boundaries of sizes $\beta_a$ and $n_-$ past boundaries of sizes $\beta_m$, with $\text{Im}(\beta_a)>0$ and $\text{Im}(\beta_m)<0$, the genus expansion~\eqref{E:MMgenus} implies that we may rotate the $\beta_i$'s in the complex plane to obtain
\beq
\label{E:MMgenus2}
	\left\langle \text{tr}\left( e^{i \beta_1 H}\right) \hdots \text{tr}\left( e^{i \beta_{n_+}H}\right) \text{tr}\left( e^{-i \beta_{n_++1}H}\right)\hdots \text{tr}\left( e^{-i \beta_n H}\right)\right\rangle_{\rm \overline{MM},conn} \simeq \sum_{g=0}^{\infty} \frac{Z_{g,n}^{\kappa,\gamma}(- i \beta_a, i \beta_m)}{(e^{S_0'})^{2g+n-2}}\,.
\eeq
Note that the arguments $-i\beta_a$ and $i\beta_m$ still have positive real part, so that we remain in the domain where the $Z_{g,n}$'s are analytic functions of their arguments. Using Eqs.~\eqref{E:Psiconn1} and~\eqref{E:analyticZs}, we see that the connected transition amplitudes lead to the series
\beq
	\Psi_{\rm conn}(\beta_aJ,\beta_mJ) \simeq \sum_{g=0}^{\infty} \frac{Z_{g,n_+,n_-}(\beta_aJ,\beta_mJ)}{(e^{S_0})^{2g+n-2}} \simeq \sum_{g=0}^{\infty}\frac{Z_{g,n}(-i\beta_aJ,i\beta_mJ)}{(e^{S_0})^{2g+n-2}}\,.
\eeq
Since the $Z_{g,n}$'s coincide with the matrix model expansion coefficients $Z_{g,n}^{\kappa,\gamma}$, we see that the genus expansion of $\Psi$ is precisely that of the matrix model above with $S_0'\to S_0$ and $J'\to J$, so that to all orders in the genus expansion
\beq
	\Psi_{\rm conn}(\beta_a,\beta_m) \simeq \left\langle \text{tr}\left( e^{i \beta_1 H}\right) \hdots \text{tr}\left( e^{i \beta_{n_+}H}\right) \text{tr}\left( e^{-i \beta_{n_++1}H}\right)\hdots \text{tr}\left( e^{-i \beta_n H}\right)\right\rangle_{\rm \overline{MM},\, conn}\,.
\eeq

That is, the genus expansion of de Sitter JT gravity coincides with the genus expansion of the observables in~\eqref{E:MMgenus2} of the double-scaled matrix model above. Further, these expansions are asymptotic, while the matrix integral is in principle well-defined. From this point of view, the matrix model provides a non-unique, non-perturbative completion of de Sitter JT gravity.

For another perspective more in line with our analysis in Subsection~\ref{S:partitions}, we can also extract the genus expansion for the correlation functions of resolvents. In the matrix model these are given by
\beq
	\left\langle R(\lambda_1) \hdots R(\lambda_n)\right\rangle_{\rm \overline{MM},\,conn} \simeq \sum_{g=0}^{\infty} \frac{R_{g,n}(\lambda_n)}{(e^{S_0})^{2g+n-2}}\,,
\eeq
where the $R_{g,n}$'s obtained from AdS JT gravity are given by the integral transform
\beq
\label{E:Rgns}
	R_{g,n}(\lambda_i) = (-1)^n \int_0^{\infty} d\beta_1\hdots d\beta_n \,e^{\beta_1 \lambda_1 + \hdots + \beta_n\lambda_n} Z_{g,n}(\beta_i J')\,,
\eeq
where  the $\lambda_i$ are real and negative. Interpreting the genus expansion coefficients of de Sitter JT gravity as above,
\beq
	Z_{g,n_+,n_-}(\beta_aJ,\beta_mJ) = \left\langle \text{tr}\left( e^{i \beta_1H}\right) \hdots \text{tr}\left( e^{i\beta_{n_+}H}\right) \text{tr}\left(e^{-i \beta_{n_++1}H}\right)\hdots \text{tr}\left( e^{-i\beta_nH}\right)\right\rangle_{\rm\overline{MM},\,conn,\,g}\,,
\eeq
the $R_{g,n}$'s should also be given by the integral transform
\beq
\label{E:Rgns2}
	R_{g,n}(\lambda_i) = i^{n_+}(-i)^{n_-}\int_0^{\infty} d\beta_1\hdots d\beta_n \,e^{\beta'_1 \lambda_1  + \hdots +  \beta'_n \lambda_n} Z_{g,n_+,n_-}(\beta_aJ,\beta_mJ)\,,
\eeq
where $\beta'_a =- i \beta_a$ and $\beta'_m = i \beta_m$. The factors of $i$ and $-i$ arise in the following way. Given the dictionary above, we express $Z_{g,n_+,n_-}$ as 
\beq
	Z_{g,n_+,n_-} = \int_0^{\infty} dE_1 \hdots dE_n \,e^{i \beta_1 E_1 + \hdots + i \beta_{n_+}E_{n_+} - i \beta_{n_++1}E_{n_++1} - \hdots - i \beta_n E_n} \, \langle \rho(E_1) \hdots \rho(E_n)\rangle_{\rm \overline{MM},\,conn,\,g}\,.
\eeq
A single integral over a future $\beta$ produces
\beq
	i \int_0^{\infty} d\beta \, e^{-i \beta (\lambda-E)} = \frac{1}{\lambda-E}\,,
\eeq
and similarly for integrals over the past $\beta$'s. To see that~\eqref{E:Rgns2} should give $R_{g,n}$ we simply recall the definition of the resolvent,
\begin{equation*}
	R(\lambda) = \int dE \frac{\rho(E)}{\lambda-E}\,.
\end{equation*}
Using~\eqref{E:analyticZs}, we then have
\begin{align}
\begin{split}
	R_{g,n}(\lambda_i) &= i^{n_+}(-i)^{n_-}\int_0^{\infty} d\beta_1\hdots d\beta_n \, e^{\beta'_1 \lambda_1 + \hdots + \beta'_n \lambda_n}Z_{g,n}(-i\beta_aJ,\beta_mJ)
	\\
	& = (-1)^n \int_0^{\infty} d\beta'_1 \hdots d\beta'_n \,e^{\beta'_1 \lambda_1 + \hdots + \beta'_n \lambda_n}Z_{g,n}(\beta'_iJ)\,,
\end{split}
\end{align}
where in the second equality we rotated the contour of integration. This last expression coincides with~\eqref{E:Rgns}, which demonstrates that one finds the same correlation functions of resolvents from Euclidean AdS and dS JT gravity.

In Subsection~\ref{S:partitions} we noted that the symplectic measure over $\alpha$ could be taken to be either minus or plus $\frac{d\alpha d\gamma\,\alpha}{2G_2}$. In our discussion above we consistently chose the minus sign, which is the sign for which the global dS$_2$ path integral is the continuation of the path integral on the hyperbolic cylinder. Assuming a matrix model description, a suitable integral transform of the global dS$_2$ path integral gave $R_{0,2}$. For that choice we found an $R_{0,2}$ which matches the universal form of a double-scaled matrix model with a single cut, while for the other choice one finds minus that universal form. It is not clear if there is a matrix model interpretation for this other sign. However, we note in passing that if we were to take this other choice of sign for the symplectic measure and still conjecture~\eqref{E:theConjecture}, then the genus expansion coefficients of de Sitter JT gravity would be related to those of AdS JT gravity as
\begin{equation*}
	Z_{g,n_+,n_-}(\beta_aJ,\beta_mJ) = (-1)^n Z_{g,n}(-i\beta_aJ,i\beta_mJ)\,,
\end{equation*}
apart from the special cases of genus 0 with either one or two boundaries. Translating these expansion coefficients into correlation functions of resolvents, it is not too hard to verify that these coefficients are consistent with the topological recursion relations of Eynard and Orantin, once we account for the fact that in this setting $R_{0,2}$ is changed by a sign.

%%***********************************************
%\subsection{Non-perturbative effects}
%***********************************************
%
%[[ comments on non-perturbative contributions to no-boundary wave function and propagator ]]

%***********************************************
\section{dS$_3$}
\label{S:dS3}
%***********************************************

In this Section we study various aspects of pure gravity in three-dimensional de Sitter spacetime. While our main focus is to set up a framework to compute scattering and loop corrections in global dS$_3$, along the way we obtain corresponding results for the Hartle-Hawking geometry and an inflating patch. We rely throughout on the Chern-Simons description of three-dimensional gravity \cite{Achucarro:1987vz, Witten:1988hc, Witten:2007kt}, which in many ways resembles the topological gauge theory description of Jackiw-Teitelboim gravity on nearly dS$_2$ spacetime. Our results for dS$_3$ gravity parallel those obtained for Lorentzian AdS$_3$ gravity in~\cite{Cotler:2018zff} (which in turn are related to an effective field theory for large $c$ CFT with sparse spectrum~\cite{Haehl:2018izb}). The dS$_3$ results can also be obtained by an analytic continuation from Euclidean AdS$_3$ gravity, as we discuss along the way.

Pure three-dimensional gravity with positive cosmological constant is described by the Einstein-Hilbert action
\beq
	S= \frac{1}{16\pi G_3}\int d^3x \sqrt{-g} (R-2)\,,
\eeq
up to a boundary term. Its equations of motion possess dS$_3$ solutions with unit radius of curvature.  The action is classically equivalent to a non-compact Chern-Simons theory with algebra $\mathfrak{sl}(2;\mathbb{C})$~\cite{Achucarro:1987vz,Witten:1988hc}. The gauge field of the Chern-Simons description is related to the dreibein $e^A_M$ and spin connection $\omega^A{}_{BM}$ as
\beq
	A^A_M = \omega^A{}_M + i e^A_M\,,
\eeq
where
\beq
	\omega^A{}_M = \frac{1}{2}\epsilon^{ABC}\omega_{BCM}\,.
\eeq
Here $M=0,1,2$ is a spacetime index and $A,B,C=0,1,2$ are flat indices, which are raised and lowered with the Minkowski metric $\eta^{AB}$. We then define
\beq
	A_M = A^A_M J_A\,,
\eeq
where $J_A$ are the generators of $SL(2;\mathbb{C})$ in the fundamental representation, satisfying
\beq
	[J_A,J_B] = \epsilon_{ABC} J^C\,, \qquad \text{tr}(J_AJ_B) = \frac{1}{2}\eta_{AB}\,, \qquad \text{tr}(J_AJ_BJ_C)=\frac{1}{4}\epsilon_{ABC}\,.
\eeq
Explicitly, we take
\beq
\label{E:reps1}
	J_0 =\frac{1}{2} \begin{pmatrix} -i& 0 \\ 0 & i\end{pmatrix}\,, 
	\qquad 
	J_1 =\frac{1}{2} \begin{pmatrix} 0 & - i \\ i& 0 \end{pmatrix}\,,
	\qquad
	J_2 = \frac{1}{2} \begin{pmatrix} 0 & 1 \\ 1 & 0 \end{pmatrix}\,.
\eeq 
Evaluated on a solution of Einstein's equations, infinitesimal diffeomorphisms and local Lorentz transformations act on $A_M$ as infinitesimal $\mathfrak{sl}(2;\mathbb{C})$ gauge transformations. It is in this sense that $A$ is a connection. In terms of $A$, the action is a difference of Chern-Simons terms with imaginary level,
\beq
	S = S_{CS}[A] - S_{CS}[\bar{A}]\,,
\eeq
with
\beq
\label{E:dS3CS1}
	 S_{CS}[A] = \frac{ik}{4\pi}\int \text{tr}\left( A\wedge dA+\frac{2}{3}A\wedge A \wedge A\right)\,, \qquad k = \frac{1}{4G_3}\,,
\eeq
again up to a boundary term. Here $\bar{A}$ may be understood to be either the complex conjugate of $A$, i.e. $\bar{A}_M = (A^A_M)^* (J_A)^*$, or instead $(A^A_M)^* J_A$, since both assignments lead to the same action and equations of motion. In what follows we use $\bar{A}_M = (A^A_M)^* (J_A)^*$.  In terms of the Chern-Simons gauge field, the Einstein's equations and torsion-free constraint are simply the equations of motion that follows from the Chern-Simons action, namely that the field strength vanishes.

There are subtleties that arise in the study of Chern-Simons theory with non-compact gauge group, complex gauge group, and complex level (for instance, see \cite{witten1991quantization, witten2011analytic, gukov2016resurgence, Porrati:2019knx}). Na\"{i}vely all of those subtleties arise here. However, as we will see, the asymptotically dS$_3$ boundary conditions ameliorate the situation, leaving us with a rather benign quantum mechanical model on the boundary of topologically simple spacetimes like inflating or global dS$_3$. This model has ``boundary graviton'' degrees of freedom living on each connected component of the boundary, as well as monodromy degrees of freedom on global dS$_3$.

A particularly useful feature of the Chern-Simons description of gravity is that it allows us to compute the measure of the path integral. See e.g.~\cite{Killingback:1990hi,Verlinde:1989ua,Cotler:2018zff,Maloney:2015ina,Kim:2015qoa,Saad:2019lba} for discussions in the context of AdS$_3$ gravity and Jackiw-Teitelboim gravity. Below we deduce the measure for both the boundary gravitons, as well as for the monodromies, each of which is crucial for the computation of the path integral on dS$_3$ spacetimes.

%***********************************************
\subsection{Preliminaries}
%***********************************************

%***********************************************
\subsubsection{Global dS$_3$}
%***********************************************

\begin{figure}[t]
\begin{center}
\includegraphics[width=2.25in]{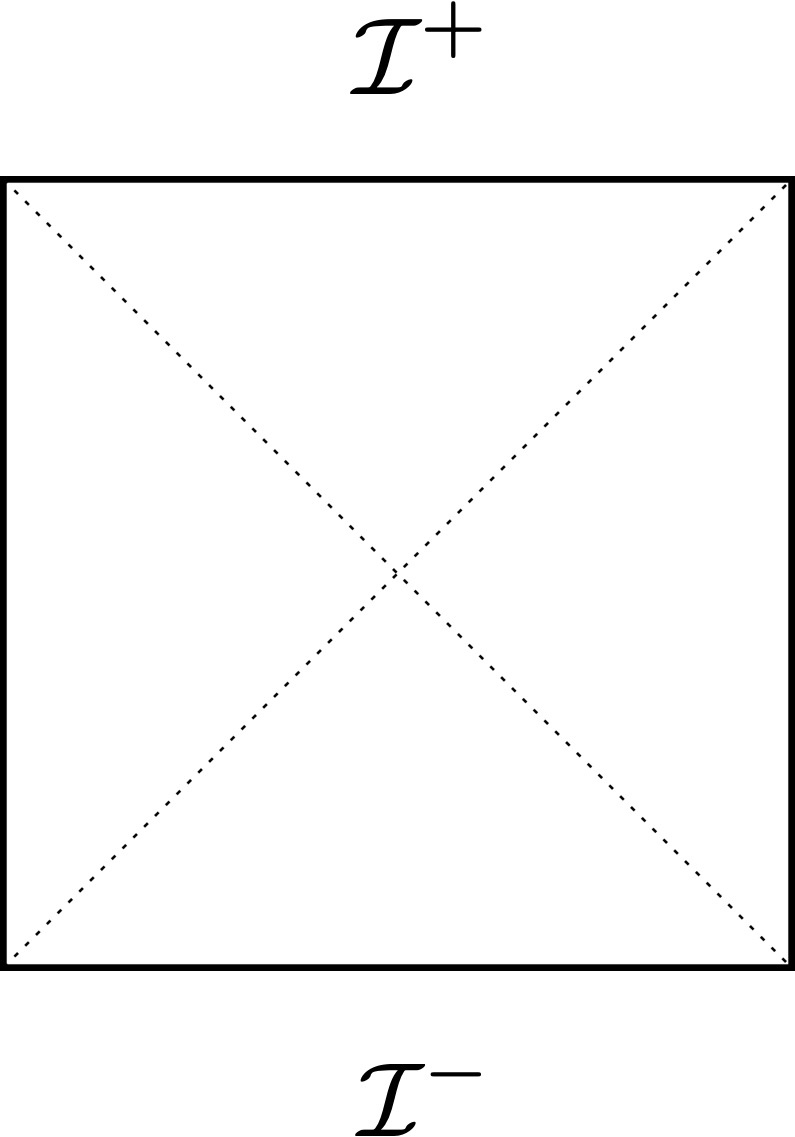}
\end{center}
\caption{
\label{F:dS3Penrose}
The Penrose diagram for dS$_3$.  The fluctuating spatial boundaries at past and future infinity are depicted in blue.}
\end{figure}

Global dS$_3$ space is described by the metric
\beq
\label{E:gdS3}
	ds^2 = -\frac{dt^2}{t^2+1}+(t^2+1)\text{sech}^2(y) (dy^2 + d\theta^2)\,,
\eeq
where $(y,\theta)$ are cylindrical coordinates on a round unit-sphere. Here $t,y\in \mathbb{R}$ and $\theta\sim \theta+2\pi$ is the longitudinal angle on the $\mathbb{S}^2$. This spacetime has conformal boundaries at $t\to \pm \infty$, given by a past sphere and a future sphere. See Fig.~\ref{F:dS3Penrose} for the Penrose diagram.

Using the dreibein
\beq
	e^0 = - \frac{dt}{\sqrt{t^2+1}}\,, \qquad e^1 = \sqrt{t^2+1}\,\text{sech}(y)dy\,, \qquad e^2 = \sqrt{t^2+1} \,\text{sech}(y) d\theta\,,
\eeq
one finds
\begin{align}
\begin{split}
	A &= \left( \frac{i dt}{\sqrt{t^2+1}} + \tanh(y) d\theta\right)J_0 + \text{sech}(y)\left( i\sqrt{t^2+1} \,dy + t d\theta\right) J_1 
	\\
	& \qquad \qquad \qquad \qquad \qquad \qquad  \qquad \qquad \qquad + \text{sech}(y)\left( -t dy+i \sqrt{t^2+1}\,d\theta\right)J_2\,,
\end{split}
\end{align}
or equivalently,
\beq
\label{E:AdS3}
 	A = \begin{pmatrix} \frac{dt}{2\sqrt{t^2+1}}-\frac{i}{2}\tanh(y)d\theta & \frac{i\,\text{sech}(y)}{2}(\sqrt{t^2+1}-t)d\bar{w} \\ \frac{i\,\text{sech}(y)}{2}(\sqrt{t^2+1}+t)dw & - \frac{dt}{2\sqrt{t^2+1}}+\frac{i}{2}\tanh(y)d\theta\end{pmatrix}\,,
\eeq
where $w=\theta+i y$ is a complex coordinate on the sphere at constant time. This connection is flat and therefore may locally be written as
\beq
	A = \tilde{U}^{-1} d\tilde{U}\,,
\eeq
and one representative for $\tilde{U}$ is
\beq
\label{E:uRep}
	\tilde{U} = \sqrt{\text{sech}(y)}\
	\begin{pmatrix} 
	\rho \cos\left( \frac{w}{2}\right) & i\rho^{-1} \sin\left( \frac{\bar{w}}{2}\right)
	\\ 
	i\rho \sin\left( \frac{w}{2}\right) & \rho^{-1}\cos\left( \frac{\bar{w}}{2}\right)
	\end{pmatrix}\,,
\eeq
with
\beq
\label{E:defrho}
	\rho = \sqrt{\sqrt{t^2+1}+t}\,.
\eeq

Observe that this $\tilde{U}$ is double-valued, with $\tilde{U}(\theta+2\pi,y,t) = -\tilde{U}(\theta,y,t)$. Correspondingly the holonomy of $A$ around the contractible $\theta$-circle (the $\theta$-circle shrinks to zero size at the north and south poles of the sphere, i.e. as $y\to\pm \infty$) in the fundamental representation is non-trivial,
\beq
	\mathcal{P}\left(e^{i \int_0^{2\pi}d\theta\,A_{\theta}}\right) = -I\,,
\eeq
and $A$ is singular as an $SL(2;\mathbb{C})$ connection, or for that matter as a connection for any cover of $SL(2;\mathbb{C})$. However, $A$ is non-singular as a $PSL(2;\mathbb{C})=SL(2;\mathbb{C})/\mathbb{Z}_2$ connection, wherein we identify $U\in SL(2;\mathbb{C})$ with $-U$. Thus, in order for global dS$_3$ to be a non-singular configuration, we require the gauge group to be precisely $PSL(2;\mathbb{C})$. This is analogous to pure gravity with negative cosmological constant, which, upon requiring that global AdS$_3$ be a non-singular configuration, can be described in the Chern-Simons formulation as a gauge theory with gauge group either $PSL(2;\mathbb{R})\times PSL(2;\mathbb{R})$~\cite{Castro:2011iw} or $SO(2,2) = \faktor{\left( SL(2;\mathbb{R})\times SL(2;\mathbb{R})\right)}{\mathbb{Z}_2}$.

While $\tilde{U}$ is single-valued around the $\theta$-circle, its logarithm is not. To see this, we write out $\tilde{U}$ in a type of Gauss parameterization,
\beq
	\tilde{U} =\exp\left( \lambda \theta\right) U\,,
\eeq
where
\beq
\label{E:uRep2}
	\lambda = -iJ_1\,, \quad U =  \begin{pmatrix} \cosh\left( \frac{y}{2}\right) & -i\sinh\left( \frac{y}{2}\right) \\ i \sinh\left(\frac{y}{2}\right) & \cosh\left( \frac{y}{2}\right)\end{pmatrix}\begin{pmatrix} i \rho\sqrt{\sech(y)} & 0\\ 0 & (i\rho\sqrt{\sech(y)})^{-1}\end{pmatrix} \begin{pmatrix} 1 & \rho^{-2}\sinh(y) \\ 0 & 1 \end{pmatrix}\,.
\eeq
There is a monodromy characterized by $\lambda$, and both $U$ and its logarithm are single-valued. These values for $\lambda$ and $U$ are to be understood as representatives. By parameterizing $A$ this way, we have introduced a $PSL(2;\mathbb{C})$ redundancy under $\lambda \to h \lambda h^{-1}$, $U\to hU$, for $h\in PSL(2;\mathbb{C})$. The invariant data in the monodromy is given by
\beq
\label{E:lambdaBC}
	\text{tr}(\lambda^2) = - \frac{1}{4}\,.
\eeq

One notable feature of global dS$_3$ is that its on-shell action is divergent. Regulating the divergence as one does in AdS holography, by integrating in $t$ up to cutoff slices at $t = \mp \ln \epsilon$ with $\epsilon \ll 1$, one finds that the on-shell action has a quadratic divergence $\sim \epsilon^{-2}$ and a logarithmic divergence. The logarithmic divergence is physical, and reads
\beq
\label{E:logDiv}
	S_{\rm classical} = -\frac{1}{G_3} \ln \epsilon \,.
\eeq
This divergence encodes the fact that, after suitably canceling the divergence (which can be done at the expense of introducing a scale $\mu$), the classical gravitational approximation to the bulk path integral takes the form
\beq
\label{E:sphereZ}
	e^{iS_{\rm classical}}= (R \mu)^{\frac{2C}{3}}\,, \qquad C = i\frac{3}{2G_3}\,,
\eeq
where $R$ is the radius of the boundary sphere. This is the form of the partition function of a two-dimensional CFT with central charge $C$ on two copies of $\mathbb{S}^2$. Thus we see that the central charge of asymptotically dS$_3$ gravity is pure imaginary, which can also be deduced from a Brown-Henneaux-inspired analysis~\cite{Brown:1986nw, Strominger:2001pn}.

%***********************************************
\subsubsection{Hartle-Hawking}
%***********************************************

There is also a dS$_3$ version of the Hartle-Hawking geometry. It is the union of the future half of global dS$_3$ above, described by the connection~\eqref{E:AdS3}
\begin{equation*}
	A = \begin{pmatrix} \frac{dt}{2\sqrt{t^2+1}}-\frac{i}{2}\tanh(y)d\theta & \frac{i\,\text{sech}(y)}{2}(\sqrt{t^2+1}-t)d\bar{w} \\ \frac{i\,\text{sech}(y)}{2}(\sqrt{t^2+1}+t)dw & - \frac{dt}{2\sqrt{t^2+1}}+\frac{i}{2}\tanh(y)d\theta\end{pmatrix}\,,
\end{equation*}
with $w = \theta+iy$ for $t\geq 0$, glued to a Euclidean hemisphere described by the metric
\beq
	ds^2 = \frac{d\tau^2}{1-\tau^2} + \frac{1-\tau^2}{\cosh^2y}(dy^2 + d\theta^2)\,, \qquad \tau \in [0,1]\,,
\eeq
with $\tau = i t$. The connection $A$ in the Euclidean segment is
\beq
	A = \begin{pmatrix} -\frac{i d\tau}{2\sqrt{1-\tau^2}} -\frac{i}{2}\tanh(y)d\theta & \frac{\sech(y)}{2}(i\sqrt{1-\tau^2}-\tau)d\bar{w} \\ \frac{\sech(y)}{2}(i\sqrt{1-\tau^2}+\tau)dw & \frac{id\tau}{2\sqrt{1-\tau^2}} + \frac{i}{2}\tanh(y)d\theta\end{pmatrix}\,.
\eeq
The on-shell action has a logarithmic divergence that is half of that in global dS$_3$,
\beq
	S_{\rm classical} = -\frac{1}{2G_3}\ln\epsilon\,.
\eeq
Supposing that there is a dual CFT on the future sphere, we infer that its sphere partition function is
\beq
	\mathcal{Z}_{\mathbb{S}^2}= (R\mu)^{\frac{C}{3}} \approx e^{i S_{\rm classical}} \,, \qquad C = i \frac{3}{2G_3}\,,
\eeq
which gives the same central charge we found for global dS$_3$.

%***********************************************
\subsubsection{Inflating patch}
%***********************************************
\begin{figure}[t]
\begin{center}
\includegraphics[width=2.5in]{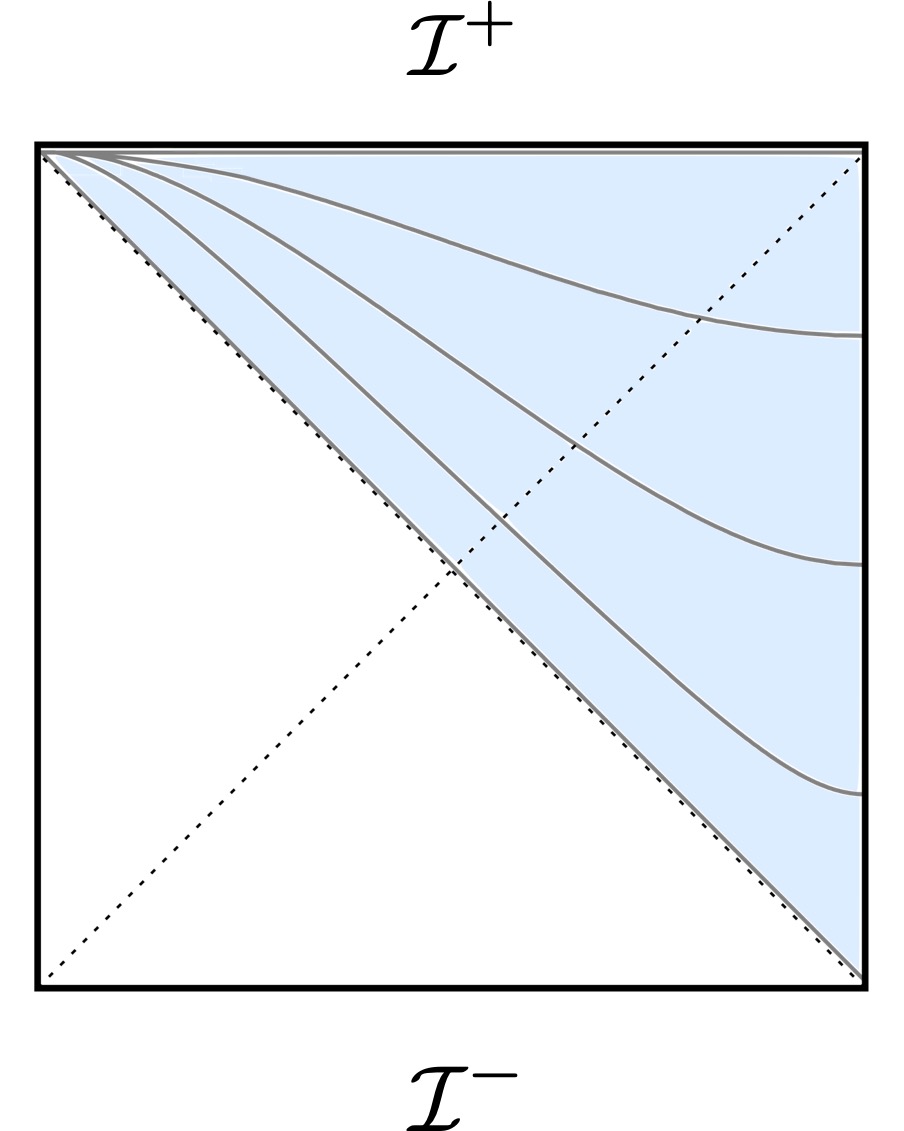}
\end{center}
\caption{
\label{F:InflatingPatch}
The Penrose diagram for the inflating patch of dS$_3$, depicted in light blue.  The constant-time slices (i.e., ``flat slicings'') are shown in gray.}
\end{figure}

One may also consider an inflating patch of global dS$_3$, as pictured in Fig.~\ref{F:InflatingPatch}.  (The inflating patch is also called the flat slicing.) This patch has only one boundary, the one-point uncompactification of the future sphere, which shrinks to zero size in the far past. The metric describing the patch is
\beq
\label{E:inflatingPatch}
	ds^2 = - \frac{dt^2}{t^2} + t^2 dz d\bar{z}\,,
\eeq
where $t>0$, and $z = x + i y$ is a complex coordinate on $\mathbb{R}^2$. Here $x$ is a non-compact variable. The relation between these coordinates and those before can be expressed by way of embedding coordinates. The global dS$_3$ hyperboloid is described as a surface in $\mathbb{R}^{1,3}$, by points $X^M$ satisfying $- (X^0)^2 + (X^1)^2 + (X^2)^2 + (X^3)^2 = 1$. To go from this hyperboloid to the global spacetime~\eqref{E:gdS3} we take 
\beq
\label{E:firstcoords}
	X^0 = t\,, \,\,\,  X^1 = \tanh(y) \sqrt{t^2+1}\,,\,\,\, X^2 = \sech(y) \cos(\theta) \sqrt{t^2+1}\,, \,\,\, X^3 = \sech(y) \sin(\theta)\sqrt{t^2+1}\,.
\eeq
The inflating patch~\eqref{E:inflatingPatch} corresponds instead to
\begin{equation}
\label{E:secondcoords}
X^0 = -\frac{(1+|z|^2)t - t^{-1}}{2}\,, \qquad X^1 =x\, t\,, \qquad X^2 = y\, t\,, \qquad X^3 = \frac{(1-|z|^2)t +t^{-1}}{2} \,,
\end{equation}
where $t > 0$. Note that the coordinates $t$ and $y$ in~\eqref{E:firstcoords} are different from the coordinate $y$ in~\eqref{E:secondcoords}.  The inflating patch does not globally cover de Sitter space, and only covers the upper right half of the Penrose diagram as depicted in blue in Fig.~\ref{F:InflatingPatch}.

Using the dreibein
\beq
	e^0 = \frac{dt}{t}\,, \qquad e^1 = tdy\,, \qquad e^2 = t dx\,,
\eeq
we arrive at a gauge configuration
\beq
	A = \frac{idt}{t}J_0 + t dz\, J_1 + i t dz\, J_2\,,
\eeq
or equivalently
\beq
\label{E:Ainflating}
	A = \begin{pmatrix} \frac{dt}{2t} & 0 \\ i t \,dz & -\frac{dt}{2t}\end{pmatrix}\,.
\eeq
This may be parameterized as
\beq
	A = U^{-1} dU\,,
\eeq
where a representative for $U$ is
\beq
	U=\begin{pmatrix} \sqrt{it} & 0 \\ \sqrt{it} \,z & \frac{1}{\sqrt{it}}\end{pmatrix}\,. 
\eeq
In fact, this $U$ has a winding property which can be made manifest by the decomposition
\beq
	U = \begin{pmatrix} \cos\left(\frac{\phi}{2}\right) & -\sin\left( \frac{\phi}{2}\right) \\ \sin\left(\frac{\phi}{2}\right) & \cos\left(\frac{\phi}{2}\right) \end{pmatrix} \begin{pmatrix} \sqrt{i t} & 0 \\ 0 & \frac{1}{\sqrt{it}}\end{pmatrix}\begin{pmatrix} 1 & (2t)^{-1} \\ 0 & 1 \end{pmatrix}\,, \qquad \phi = 2\arctan(z)\,.
\eeq
In this parameterization, $\phi$ is an angular variable, $\phi \sim \phi + 2\pi$. We see that at fixed $y$, the line $x$ is mapped to the circle $\phi$, with $\lim_{x\to\pm \infty} \phi = \pm \pi$. 

%***********************************************
\subsection{The map to Euclidean AdS$_3$ gravity}
\label{S:AdS3map}
%***********************************************

In Subsection~\ref{S:EAdS2} we explicitly demonstrated a mapping between JT gravity on nearly dS$_2$ spacetime to Euclidean JT gravity on nearly AdS$_2$ spacetime. This mapping holds at arbitrary genus, and also for the non-perturbative completions of the dS$_2$ and AdS$_2$ theories via a double-scaled matrix model. There is a similar such mapping between pure dS$_3$ gravity and pure Euclidean AdS$_3$ gravity, which we elucidate in this Subsection.  This map was discussed previously by \cite{Castro:2012gc}, where it was used to formulate a sum over geometries to compute the wave function of the universe for geometries which asymptote to a torus at future infinity.

As is well-known, Euclidean AdS$_3$ gravity may be recast as a Chern-Simons theory \cite{Achucarro:1987vz, Witten:1988hc}. Our starting point is the Einstein-Hilbert action
\beq
	S_E = -\frac{1}{16 \pi G_3} \int d^3 x \sqrt{g'} (R' + 2)\,,
\eeq
augmented with a suitable boundary term. Here we denote the metric, curvature, and so on with primes, to distinguish them from the de Sitter quantities discussed in the last Subsection. The Einstein-Hilbert action may be rewritten in terms of linear combinations of the first-order variables $(e'^A_M, \omega'^A{}_{BM})$,
\beq
\label{E:EAdS3As}
	(A_M')^A = i(\omega'^A{}_M + i \, e'^A_M)\,, \qquad (\bar{A}')_M^A = -i (\omega'^A{}_M - i \, e'^A_M)\,,
\eeq
with
\beq
	\omega'^A{}_M = \frac{1}{2} \epsilon^{ABC} \omega'_{BCM}
\eeq
where $M=1,2,3$ is a spacetime index and $A,B,C = 1,2,3$ are flat indices that are raised and lowered by $\delta^{AB}$. Our convention for the epsilon tensor with flat indices is $\epsilon_{123} = 1$. We decompose the gauge fields as
\beq
	A'_M = A_M'^A J_A'
\eeq
where the $J_A'$ are the generators of $\mathfrak{su}(2)$ in the fundamental representation. Because the $A'^A_M$ are complex, $A'_M$ is an adjoint vector of the complexification of $\mathfrak{su}(2)$, $\mathfrak{sl}(2;\mathbb{C})$. Below we will use the explicit basis
\beq
	J_1' =\frac{1}{2} \begin{pmatrix} 1 & 0 \\ 0 & -1 \end{pmatrix}\,, 
	\qquad 
	J_2' = \frac{1}{2} \begin{pmatrix} 0 & 1 \\ 1 & 0 \end{pmatrix}\,,
	\qquad
	J_3' = \frac{1}{2} \begin{pmatrix} 0 &  i \\ - i& 0 \end{pmatrix}\,.
\eeq
These are related to the $\mathfrak{sl}(2;\mathbb{C})$ generators in the Lorentzian dS$_3$ analysis in~\eqref{E:reps1} by
\beq
	J_1' = - i J_0\,, \qquad J_2' = J_2\,, \qquad J_3' = -J_1\,.
\eeq
In any case the Einstein-Hilbert action may be rewritten in terms of $A', \bar{A}'$ in~\eqref{E:EAdS3As} as a difference of Chern-Simons actions
\beq
	S_E = S_{CS}[A'] - S_{CS}[\bar{A}']\,,
\eeq
where
\beq
\label{E:EAdSCS1}
	 S_{CS}[A'] =- \frac{ik}{4\pi}\int \text{tr}\left( A'\wedge dA'+\frac{2}{3}A'\wedge A' \wedge A'\right)\,, \qquad k = \frac{1}{4G_3}\,,
\eeq
up to a boundary term.

Let us consider the particular case of the Euclidean AdS$_3$ geometry with $\mathbb{S}^2$ boundary, which we will map to the future half of global dS$_3$.  The corresponding metric is
\beq
	ds'^2 = \left(r^2 \, d\Omega_2^2 + \frac{dr^2}{r^2+ 1} \right)\,.
\eeq
As in our dS$_3$ analysis we write the sphere metric in cylindrical coordinates,
\beq
	d\Omega_2^2 = \sech^2(y) ( dy^2 + d\theta^2)\,.
\eeq
To compute $A$ and $\bar{A}$ for this metric, we use the dreibein
\beq
	e'^1 = \frac{dr}{\sqrt{r^2 + 1}}\,, \qquad e'^2 = r \, \sech(y) dy\,, \qquad e'^3 = r \, \sech(y) d\theta\,,
\eeq
which gives us
\begin{align}
	A' &= \left(\frac{i \, dr}{\sqrt{r^2 + 1}} + \tanh(y) d\theta \right) i J_0' + \sech(y)\left(\sqrt{r^2 + 1} \,d\theta + i r \, dy \right) i J_1' \nonumber \\
& \qquad \qquad \qquad \qquad \qquad \qquad \qquad \qquad + \sech(y) \left(-\sqrt{r^2 + 1} \,dy + i r \, d\theta\right) i J_2'\,.
\end{align}
In matrix form, $A'$ reads
\begin{equation}
	A' = \begin{pmatrix} \frac{dr}{2\sqrt{r^2+1}}-\frac{i}{2}\tanh(y)d\theta & \frac{i\,\text{sech}(y)}{2}(\sqrt{r^2+1}-r)d\bar{w} \\ \frac{i\,\text{sech}(y)}{2}(\sqrt{r^2+1}+r)dw & - \frac{dr}{2\sqrt{r^2+1}}+\frac{i}{2}\tanh(y)d\theta\end{pmatrix}\,,
\end{equation}
where $w = \theta + i y$.  This precisely agrees with the connection in~\eqref{E:AdS3} describing the future half of global dS$_3$ upon replacing $r$ with $t$.  Since the connections in the two analyses are identical, they evidently satisfy the same boundary conditions as $r \to \infty$ and $t \to \infty$ upon interchanging $r \leftrightarrow t$.

Furthermore, we have that in the fundamental representation,
\beq
	\mathcal{P}\left(e^{i \int_0^{2\pi} d\theta \, A'_\theta} \right) = - I\,.
\eeq
The $\theta$ circle is contractible in the total geometry, shrinking to zero size at the poles of the sphere at constant time. The flatness condition implies that this Wilson loop ought to be trivial in all representations. Thus $A'$ is singular as a $SL(2;\mathbb{C})$ connection or as a cover thereof, but non-singular as a $PSL(2;\mathbb{C})=SL(2;\mathbb{C})/\mathbb{Z}_2$ connection. Thus we see that both for Euclidean AdS$_3$ gravity and Lorentzian dS$_3$ gravity the gauge group of the Chern-Simons formulation is $PSL(2;\mathbb{C})$ (see~\cite{Banados:1998ta,Castro:2011iw}). 

To match the Euclidean AdS$_3$ action in~\eqref{E:EAdSCS1} with the dS$_3$ action in~\eqref{E:dS3CS1}, we send $L_{\rm AdS} \to i L_{\rm dS}$.  This takes $- S_{\text{AdS}_3} \to i \, S_{\text{dS}_3}$, as needed.  So indeed, analogous to the dS$_2$ analysis above, dS$_3$ gravity is precisely an analytic continuation of Euclidean AdS$_3$ gravity, at least for global dS$_3$ and Hartle-Hawking, and their various slicings.  We anticipate that such a correspondence via analytic continuation will hold on spacetimes of non-trivial topology (as in e.g. \cite{Castro:2012gc}).

We have phrased the analytic continuation in terms of the first order formalism, but can likewise make contact with the second order formalism.  Since we have defined the gauge field in terms of first-order data in two different ways depending on whether we are in Lorentzian dS$_3$ or Euclidean AdS$_3$, the two cases have different protocols to translate back to the metric formalism.  In particular, for dS$_3$ we obtain the dreibein via
\beq
	(e^0, e^1, e^2) =  ( 2i \, \text{tr}(A \, J_0), \,  - 2i \,\text{tr}(A \, J_1),\,  - 2i \,\text{tr}(A \, J_2))  \,.
\eeq
Likewise, for Euclidean AdS$_3$,
\beq
	(e'^1, e'^2, e'^3) = ( - 2 \,\text{tr}(A' \, J_0'), \,  - 2 \,\text{tr}(A' \, J_1'),\,  - 2 \,\text{tr}(A' \, J_2')) \,.
\eeq

%***********************************************
\subsection{Boundary actions}
%***********************************************

We proceed with a path integral quantization of the $PSL(2;\mathbb{C})$ Chern-Simons theory. We take the ``constrain first'' approach~\cite{Elitzur:1989nr}, separating the direction $y$ from the others $x^i$ as
\beq
	A = A_y dy + \tilde{A}_i dx^i\,.
\eeq
Now including boundary terms, the bulk action on a spacetime $\mathcal{M}$ may then be written as
\beq
\label{E:3dS}
	S_{\rm grav} = S[A] - S[\bar{A}] + S_{\rm bdy}\,,
\eeq
with\footnote{We choose an orientation so that $\epsilon^{ty\theta} = \frac{1}{\sqrt{-g}}$ in global dS$_3$ and $\epsilon^{tyx} = \frac{1}{\sqrt{-g}}$ in an inflating patch.} 
\begin{align}
\begin{split}
	S[A] &=  \frac{ik}{2\pi}\int_{\mathcal{M}} dy \wedge \text{tr}\left( - \frac{1}{2}\tilde{A}\wedge \partial_y \tilde{A} + A_y \tilde{F}\right)\,,
	\\ \\
	S_{\rm bdy} &=- \frac{ k}{4\pi}\int_{\partial\mathcal{M}}d^2x \,n\,\text{tr}(\tilde{A}^2)+(\text{c.c.})\,,
\end{split}
\end{align}
where $\tilde{F} = \tilde{d}\tilde{A} + \tilde{A} \wedge \tilde{A}$ and $\tilde{d}=dx^i\frac{\partial}{\partial x^i}$ is the exterior derivative in the $x$-directions. The number $n = \pm 1$ appearing in the boundary term is $+1$ when $\partial_t$ is outward-pointing on a component of the boundary, and $-1$ when $\partial_t$ is inward-pointing. The term $S[A]$ is the ordinary Chern-Simons functional after addition of a suitable boundary term. The boundary term is required to enforce a variational principle consistent with the asymptotically dS$_3$ boundary conditions. These boundary conditions depend on whether we have global dS$_3$ or an inflating patch, and we will describe them shortly.

In what follows we integrate out $A_y$, which enforces the constraint $\tilde{F} = 0$, leaving us with an integral over the moduli space of flat connections on constant-$y$ slices. The space of flat connections depends on the topology of the constant-$y$ slice. For an inflating patch, those slices are topological discs, and quantization leads to a boundary chiral WZW action. For global dS$_3$ the slice is a topological  annulus, and quantization leads to a sector of a WZW model. In each case, we then impose asymptotically dS$_3$ boundary conditions, which introduces current constraints on the boundary WZW model. The resulting boundary path integral is closely related to the quantization of the first exceptional coadjoint orbit of the Virasoro group. It is an analogue of the Schwarzian action, in one dimension higher.

%***********************************************
\subsubsection{Boundary action for inflating patch}
%***********************************************

We begin with deriving a boundary action for the inflating patch, which is conceptually simpler and technically easier than the global case.  The derivation amounts to integrating over fluctuations around an inflating patch of dS$_3$, meaning the spacetime described by the metric~\eqref{E:inflatingPatch}. This is closely related to the quantization of AdS$_3$ gravity around a massless BTZ geometry, and we refer the reader to~\cite{Cotler:2018zff} for more details.

In the background of the inflating patch, the asymptotically dS$_3$ boundary conditions are that $A$ asymptotes to
\beq
\label{E:inflatingBC}
	A = \begin{pmatrix} \frac{dt}{2t} + O(t^{-2}) & O(t^{-1})\\ i t dw + O(t^{-1}) & -\frac{dt}{2t} + O(t^{-2}) \end{pmatrix}\,,
\eeq
as $t\to\infty$, and $A$ is allowed to fluctuate at the indicated powers in $1/t$. The connection~\eqref{E:Ainflating} clearly respects these boundary conditions. The on-shell variation of the total action, including the boundary term, is
\beq
	\delta S_{\rm grav} = - \frac{k}{\pi} \int dx\,dy\,\text{tr}(A_{\bar{z}}\delta A_{x}) + (\text{c.c.})\,.
\eeq
We gauge-fix $A_y$ to its value for the inflating patch,
\beq
	A_y = \begin{pmatrix} 0 &0 \\ -t & 0 \end{pmatrix} \,,
\eeq
and the residual integral is taken over the moduli space of flat connections $\tilde{A}$ on constant-$y$ slices. These are topological disks parameterized by $t\geq 0$ and the non-compact variable $x$. The flat connections may be decomposed as
\beq
\label{E:inflatingFlatA}
	\tilde{A} = U^{-1} \tilde{d}U\,.
\eeq
The map $U$ has a winding property as we discussed above. To take care of the winding property and to impose boundary conditions, it is convenient to pass over to an explicit parameterization of $PSL(2;\mathbb{C})$ elements. We decompose $U$ as
\beq
\label{E:Uinflating}
	U = \begin{pmatrix} \cos\left( \frac{\phi}{2}\right) & -\sin\left(\frac{\phi}{2}\right) \\ \sin\left(\frac{\phi}{2}\right) & \cos\left(\frac{\phi}{2}\right) \end{pmatrix} \begin{pmatrix} \Lambda & 0 \\ 0 & \Lambda^{-1} \end{pmatrix} \begin{pmatrix} 1 & \Psi \\ 0 & 1 \end{pmatrix}\,,
\eeq
where $\text{Re}(\Lambda) >0$, and components $\phi, \Psi \in \mathbb{C}$ and $\phi \sim \phi + 2\pi$. We account for the winding property by demanding that, at fixed $t,y$, the map $\phi$ takes the line $x$ to the circle, with 
\beq
	\lim_{x\to\pm\infty} \phi(x) = \pm \pi\,.
\eeq
We accomplish this by parameterizing $\phi = 2\arctan(F)$, with $F$ a diffeomorphism of the line at fixed $y$.

Evaluating the bulk action~\eqref{E:3dS} on a flat connection~\eqref{E:inflatingFlatA} gives a chiral WZW model on the boundary,
\beq
	S = S_+ + S_{\rm WZ}\,,
\eeq
where
\begin{align}
\begin{split}
\label{E:WZWsplit}
	S_+ &= \frac{k}{2\pi}\int_{\partial\mathcal{M}} dx dy\,\text{tr}\left( (U^{-1})' \bar{\partial} U\right) +(\text{c.c.})
	\\ \\
	S_{\rm WZ} & = -\frac{ik}{12\pi}\int_{\mathcal{M}} \text{tr}\left( U^{-1}dU\wedge U^{-1}dU\wedge U^{-1}dU\right) + (\text{c.c.})
\end{split}
\end{align}
are the kinetic and Wess-Zumino terms respectively. Here $\bar{\partial} = \frac{1}{2}(\partial_{x} + i \partial_y)$ and $'=\partial_{x}$. The boundary term leads to the $\text{tr}((U^{-1})'U')$ contribution to the kinetic term. Plugging in the parameterization~\eqref{E:Uinflating} we find
\begin{align}
\begin{split}
	S_+ & = -\frac{k}{2\pi} \int dx dy \left( \frac{2\Lambda'\bar{\partial}\Lambda}{\Lambda^2} - \frac{1}{2}\phi'\bar{\partial}\phi + \frac{\Lambda^2}{2}\Psi' \phi' + \frac{i\Lambda^2}{4}(\partial_y \Psi \phi' + \Psi' \partial_y \phi)\right) + (\text{c.c.})
	\\ \\
	S_{\rm WZ} & = -\frac{ik}{8\pi} \int_{\mathcal{M}}d\Lambda_+^2 \wedge d\Psi\wedge d\phi + (\text{c.c.})
\end{split}
\end{align}
so that the total action reads
\beq
\label{E:inflatingPreS}
	S = -\frac{k}{2\pi}\int dx dy\left( \frac{2\Lambda'\bar{\partial}\Lambda}{\Lambda^2} - \frac{1}{2}\phi'\bar{\partial}\phi + \Lambda^2\phi' \bar{\partial}\Psi\right) + (\text{c.c.})\,.
\eeq
The next stop is to solve the asymptotically dS$_3$ boundary conditions~\eqref{E:inflatingBC}. Solving the bottom left component of $A$ for $\Lambda$, and solving the diagonal components for $\Psi$, we find that as $t\to\infty$
\beq
\label{E:inflatingConstraints}
	\Lambda = \sqrt{\frac{2it}{\phi'}}\,, \qquad \Psi = \frac{i}{2t}\frac{\phi''}{\phi'}\,,
\eeq
with $\phi$ finite. Plugging these constrained values into the action~\eqref{E:inflatingPreS} and integrating by parts, we arrive at the boundary effective action
\beq
\label{E:inflatingS}
	S = \frac{iC}{24\pi}\int dx dy\left( \frac{\phi''\bar{\partial}\phi'}{\phi'^2}-\phi'\bar{\partial}\phi\right) + (\text{c.c.}) = \frac{iC}{24\pi}\int dx dy\, \frac{F''\bar{\partial}F'}{F'^2} + (\text{c.c.})\,,
\eeq
where $C = i\frac{3L_{\rm dS}}{2G_3}$ is the (imaginary) classical central charge of dS$_3$ gravity, where we have restored the de Sitter radius.

Eq.~\eqref{E:inflatingS} gives us the action for the path integral over the boundary gravitons of the inflating patch, represented by the $\phi$ field. The path integral has weight $\exp( i S)$, i.e.
\beq
\label{E:dS3pathint1}
	\mathcal{Z} = \int [d\phi] \exp\left( i S\right) = \int [d\phi] \exp\left( - \frac{C}{24\pi} \int dx dy\left( \frac{\phi''\bar{\partial}\phi'}{\phi'^2}-\phi'\bar{\partial}\phi\right) - (\text{c.c.})\right)\,,
\eeq
with a suitable measure for $\phi$ which we will derive shortly. This model is the analytic continuation of Euclidean AdS$_3$ gravity in the Poincar\'e patch, per our discussion in Subsection~\ref{S:AdS3map}. 

Recently, two of us~\cite{Cotler:2018zff} obtained the boundary path integral for Lorentzian AdS$_3$. That theory has the same form as~\eqref{E:dS3pathint1} with two differences.  For the AdS$_3$ theory (i) $C$ is real, given by $C = \frac{3L_{\rm AdS}}{2G_3}$, and (ii) $\phi$ is real. The AdS$_3$ theory is in fact the path integral quantization of the first exceptional coadjoint orbit of the Virasoro group, Diff$(\mathbb{S}^1)/PSL(2;\mathbb{R})$~\cite{Alekseev:1988ce}. As recently emphasized in~\cite{Porrati:2019knx}, the Chern-Simons formulation of Euclidean AdS$_3$ gravity differs from that of Lorentzian AdS$_3$. The Euclidean version is a constrained $PSL(2;\mathbb{C})$ Chern-Simons theory, and the Lorentzian theory is a constrained $PSL(2;\mathbb{R})\times PSL(2;\mathbb{R})$ Chern-Simons theory. The Euclidean analysis differs slightly from the Lorentzian version. (In~\cite{Cotler:2018zff} various Euclidean computations were done by first taking the Lorentzian boundary path integral, Wick-rotating, and then computing, rather than starting from the gravitational theory in Euclidean signature.)

Repeating the analysis of~\cite{Cotler:2018zff} for the $PSL(2;\mathbb{C})$ Chern-Simons description of Euclidean AdS$_3$ gravity, we find exactly the same path integral above except with $C= \frac{3L_{\rm AdS}}{2G_3}$. We then see that the Euclidean AdS$_3$ and Lorentzian dS$_3$ results are related by the simple analytic continuation $L_{\rm AdS} \to i L_{\rm dS}$. 

Moreover, this dS$_3$ model is a path integral quantization of a coadjoint orbit of the complexified Virasoro group $\widehat{\text{Diff}}_{\mathbb{C}}(\mathbb{S}^1)$, i.e. the centrally extended group of complex diffeomorphisms of the circle satisfying $\phi(\theta+2\pi) = \phi(\theta)+2\pi$ with $\phi \in \mathbb{C}$. The relevant orbit is $\text{Diff}_{\mathbb{C}}(\mathbb{S}^1)/PSL(2;\mathbb{C})$.

%***********************************************
\subsubsection{Boundary action for global dS$_3$}
%***********************************************

We now consider the quantization of fluctuations around global dS$_3$. Here there are two boundaries and constant-$y$ slices are topological annuli, which degenerate at the poles of the sphere. As a result we find boundary graviton degrees of freedom on each boundary, along with a quantum mechanical monodromy which is pinned down by boundary conditions at the poles. 

It seems reasonable that this quantization may be performed by gluing together past and future patches, along the lines of the multi-boundary analysis for Jackiw-Teitelboim gravity in~\cite{Saad:2019lba}. Here we elect to perform a direct quantization in order to treat subtleties that arise in the integral over the monodromy, similar to our treatment of nearly dS$_2$ gravity at integer values of $\alpha$.

In global dS$_3$ there are two boundaries, and thus two boundary terms and two boundary conditions. The boundary conditions are that near the $t\to\infty$ future conformal boundary, $A$ asymptotes to
\beq
\label{E:bc1}
	A = \begin{pmatrix} -\frac{i}{2}\tanh(y)d\theta + \frac{dt}{2t} + O(t^{-2}) & O(t^{-1}) \\ i t\,\text{sech}(y)dw +O(t^{-1})& \frac{i}{2}\tanh(y)d\theta - \frac{dt}{2t}+O(t^{-2}) \end{pmatrix}\,,
\eeq
and near the $t\to-\infty$ past conformal boundary, $A$ asymptotes to
\beq
\label{E:bc2}
	A = \begin{pmatrix} -\frac{i}{2}\tanh(y)d\theta - \frac{dt}{2t} + O(t^{-2}) & -i t\,\text{sech}(y) d\bar{w} + O(t^{-1}) \\ O(t^{-1}) & \frac{i}{2}\tanh(y)d\theta + \frac{dt}{2t}+O(t^{-2}) \end{pmatrix}\,,
\eeq
where $A$ is allowed to fluctuate at the indicated powers in $1/t$. The connection~\eqref{E:AdS3} describing global dS$_3$ clearly respects these boundary conditions. The on-shell variation of the total action, including the boundary term, is then given by
\beq
	\delta S_{\rm grav} = -\frac{ k}{\pi}\left( \int_{\partial\mathcal{M}_+} dyd\theta\, \text{tr}(A_{\bar{w}} \delta A_{\theta}) -\int_{\partial\mathcal{M}_-} dyd\theta \,\text{tr}(A_{w}\delta A_{\theta})\right) + (\text{c.c})\,.
\eeq
We gauge-fix $A_y$ to its value for global dS$_3$,
\beq
	A_y = -\frac{\text{sech}(y)}{2}\begin{pmatrix} 0 & \sqrt{t^2+1}-t \\ \sqrt{t^2+1} + t & 0 \end{pmatrix}\,,
\eeq
and then the residual integral is taken over the moduli space of flat connections $\tilde{A}$ on slices of constant $y$. Away from the poles of the sphere $y\to \pm \infty$, constant-$y$ slices are topologically global dS$_2$, i.e. annuli. The $\theta$-circle is contractible only at the poles. Consequently the most general flat connection has monodromy parameterized by
\beq
	\tilde{A} = \tilde{U}^{-1} \tilde{d} \tilde{U}\,, \qquad \tilde{U} = \exp\left( \lambda(y) \theta\right) U\,,
\eeq
where $\lambda(y) \in \mathfrak{sl}(2;\mathbb{C})$ and $U\in PSL(2;\mathbb{C})$ is single-valued. Further, because the $\theta$-circle is contractible at the poles, we enforce the boundary condition that the monodromy goes to the value 
\beq
	\lim_{y\to\pm \infty} \text{tr}(\lambda(y)^2) = -\frac{1}{4}
\eeq
that we found above for global dS$_3$.

The integral over flat connections then becomes a functional integral over $U$ and $\lambda$. In terms of these fields the total action becomes
\beq
\label{E:totalS}
	S = S_+[U] - S_-[U] +S_{\rm WZ}[U] + S_{\lambda}[U;\lambda]\,,
\eeq
where $S_+$ and $S_{\rm WZ}$ were given in~\eqref{E:WZWsplit} (upon the replacement $x\to \theta$), $S_-$ takes the same form as $S_+$ except on the past boundary, and the coupling $S_{\lambda}$ is given by
\beq
\label{E:Slambda}
	S_{\lambda} = -\frac{k}{\pi}\int d\theta dy \,\text{tr}\Big( \lambda(y)  \left( (\bar{\partial}U_+)U_+^{-1} - (\partial U_-)U_-^{-1}\right) \Big)+(\text{c.c.})\,,
\eeq
where $U_{\pm}$ are the asymptotic values of $U$ as $t\to \pm \infty$. In order to write out the action in terms of component fields and to impose the asymptotically dS$_3$ boundary conditions, we pass to an explicit parameterization of $PSL(2;\mathbb{C})$ elements. A general element $\tilde{U}_+ \in PSL(2;\mathbb{C})$ (we use a $+$ subscript here since this parameterization is the natural one near future infinity) with monodromy may be parameterized as
\beq
\label{E:gauss1}
	 U_+ = \begin{pmatrix} \cos\left(\frac{\phi_+}{2}\right) & -\sin\left( \frac{\phi_+}{2}\right) \\ \sin\left(\frac{\phi_+}{2}\right) & \cos\left(\frac{\phi_+}{2}\right) \end{pmatrix} \begin{pmatrix} \Lambda_+ & 0 \\ 0 & \Lambda_+^{-1} \end{pmatrix} \begin{pmatrix} 1 & \Psi_+ \\ 0 & 1 \end{pmatrix}\,,
\eeq
with $\phi_+,\Psi_+\in \mathbb{C}$ and Re$(\Lambda_+)\geq 0$. For example, the representative in Eqs.~\eqref{E:uRep},~\eqref{E:uRep2} describing global dS$_3$ is parameterized by
\begin{align}
\begin{split}
	\Lambda_+ & =\rho\sqrt{i\,\sech(y)}\,,
	\\
	\Psi_+ & = \rho^{-2}\sinh(y)\,,
	\\
	\phi_+ & = iy\,,
\end{split}
\end{align}
and $\lambda = - i J_1$. Near past infinity it is more convenient to employ an alternate parameterization
\beq
\label{E:gauss2}
	 U_- =  \begin{pmatrix} \cos\left(\frac{\phi_-}{2}\right) & -\sin\left(\frac{\phi_-}{2}\right) \\ \sin\left(\frac{\phi_-}{2}\right) & \cos\left(\frac{\phi_-}{2}\right) \end{pmatrix} \begin{pmatrix} \Lambda_-^{-1} & 0 \\ 0 & \Lambda_-\end{pmatrix} \begin{pmatrix} 1 & 0 \\ - \Psi_- & 1\end{pmatrix}\,.
\eeq
As an example, the representative $U$ in~\eqref{E:uRep2} is parameterized by
\begin{align}
\begin{split}
	\Lambda_- & = \rho^{-1}\sqrt{-i \,\sech(y)}\,,
	\\
	\Psi_- & = \rho^{2}\sinh(y)\,,
	\\
	\phi_- & = -i y\,.
\end{split}
\end{align}
The $-$ fields are related to the $+$ fields by
\beq
	\Lambda_- = \sqrt{\frac{1}{\Lambda_+^2}+\Lambda_+^2\Psi_+^2}\,, \quad \Psi_- = -\frac{\Lambda_+^4\Psi_+}{1+ \Lambda_+^4 \Psi_+^2}\,, \quad \tan\left(\frac{\phi_-}{2}\right) = \frac{\tan\left( \frac{\phi_+}{2}\right) - \Lambda_+^2\Psi_+}{\tan\left(\frac{\phi_+}{2}\right)\Lambda_+^2\Psi_++1}\,.
\eeq

In terms of the component fields we find that the total action~\eqref{E:totalS} reads 
\begin{align}
\begin{split}
\label{E:preGlobalS}
	S = -\frac{k}{2\pi}\int d\theta dy &\left( \frac{2\Lambda_+'\bar{\partial}\Lambda_+}{\Lambda_+^2}-\frac{1}{2}\phi_+'\bar{\partial}\phi_+ + \Lambda_+^2(\gamma_++\phi_+') \bar{\partial}\Psi_+ - \frac{2\gamma_+'}{\phi_+'}\frac{\bar{\partial}\Lambda_+}{\Lambda_+}-i\lambda_1 \bar{\partial}\phi_+ \right.
	\\
	& \qquad \left. \phantom{\frac{1}{2}}- (+\to -)\right) + (\text{c.c.})\,,
\end{split}
\end{align}
where $\lambda=\lambda^A J_A$ and
\beq
	\gamma_{\pm} = i\lambda_1(y) \mp i \lambda_0(y) \sin(\phi_{\pm}) \pm \lambda_2(y) \cos(\phi_{\pm})\,.
\eeq
After imposing the boundary conditions we find that as $t\to \infty$ the components $\Lambda_+$ and $\Psi_+$ are constrained in terms of $\phi_+$ as
\begin{align}
\begin{split}
	\Lambda_+&= \sqrt{\frac{2i t\,\sech(y)}{\gamma_+ + \phi_+'}}\,, 
	\\
	\Psi_+ &=\frac{1}{2t}\left( \sinh(y) +i\cosh(y)\left( \frac{\gamma_+'}{\phi_+'}+ \frac{(\gamma_++\phi_+')'}{\gamma_++\phi_+'}\right)\right)\,,
\end{split}
\end{align}
with $\phi_+$ finite. Similarly, as $t\to-\infty$ the components $\Lambda_-$ and $\Psi_-$ are constrained in terms of $\phi_-$ as
\begin{align}
\begin{split}
	\Lambda_- & = \sqrt{\frac{2it\,\sech(y)}{\gamma_-+\phi_-'}}\,,
	\\
	\Psi_- & = \frac{1}{2t}\left( -\sinh(y) +i\cosh(y)\left( \frac{\gamma_-'}{\phi_-'}+ \frac{(\gamma_-+\phi_-')'}{\gamma_-+\phi_-'}\right)\right)\,,
\end{split}
\end{align}
with $\phi_-$ finite. Substituting these expressions into~\eqref{E:preGlobalS} and after suitably integrating by parts, we arrive at the relatively simple expression for the boundary effective action
\begin{align}
\begin{split}
\label{E:globaldS3S}
	S = \frac{iC}{24\pi}\int d\theta dy \left( \frac{\Phi_+'\bar{\partial}\Phi_+}{\Phi_+^2} - (\phi_+'+2i\lambda_1)\bar{\partial}\phi_+ + \frac{2\gamma_+'}{\phi_+'}\frac{\bar{\partial}\Phi_+}{\Phi_+}  - (+\to -) - 2\right) + (\text{c.c.})\,,
\end{split}
\end{align}
where 
\beq
	\Phi_{\pm} = \phi_{\pm}' + \gamma_{\pm}\,.
\eeq

In this path integral quantization the $\phi_{\pm}$ are periodic in $\theta$, and the monodromy fluctuates around its value for global dS$_3$, which satisfies $\text{tr}(\lambda^2) = -1/4$. It is convenient to repackage this monodromy into a winding boundary condition for the $\phi_{\pm}$,
\beq
	\phi_{\pm}(\theta+2\pi,y) = \phi_{\pm}(\theta,y) + 2\pi\,,
\eeq
and for $\lambda$ to fluctuate around zero. With this choice, one parameterization for global dS$_3$ is
\beq
	\lambda = 0\,, \qquad \phi_{\pm} = \theta\,,
\eeq
which is indeed a critical point of the action. The on-shell action has a logarithmic divergence, coming from the $-2$ in the Lagrangian, corresponding to the classical central charge of dS$_3$ gravity. In this form we regulate it by integrating within a distance $\epsilon$ of the poles of the sphere, $y\in [\ln \epsilon,-\ln\epsilon]$, so that the on-shell action is, after using $k=\frac{1}{4G_3}$,
\begin{equation*}
	S = -\frac{1}{G_3}\ln\epsilon\,.
\end{equation*}
This precisely matches the logarithmic divergence we noted in Eq.~\eqref{E:logDiv}.

%***********************************************
\subsubsection{Hartle-Hawking}
%***********************************************

The quantization on the Hartle-Hawking background is essentially the same as for global dS$_3$, or at least its future half, upon taking the monodromy to be trivial. The end result is that the effective action is
\beq
	S = \frac{iC}{24\pi} \int d\theta dy\left( \frac{\phi''\bar{\partial}\phi'}{\phi'^2} - \phi'\bar{\partial}\phi \right) + (\text{c.c.})\,,
\eeq
where at fixed $y$ the $\phi$ field is an element of Diff$_\mathbb{C}(\mathbb{S}^1)$,
\beq
	\phi(\theta+2\pi,y) = \phi(\theta,y) + 2\pi\,,
\eeq
and moreover we have chosen a convention so that $\phi$ is finite at the poles of the sphere $y\to \pm \infty$. The critical point of the model corresponding to the classical background is simply
\beq
	\phi = \theta\,,
\eeq
and the on-shell action evaluated on this configuration has a logarithmic divergence
\begin{equation*}
	S = -\frac{1}{2G_3}\ln\epsilon\,,
\end{equation*}
as it ought to.

%***********************************************
\subsubsection{$PSL(2;\mathbb{C})$ quotient}
%***********************************************

In the path integral quantizations presented above we have introduced a redundancy. In the inflating patch, by writing $\tilde{A} = U^{-1} \tilde{d} U$, the configurations $U(x^i,y)$ and $h(y)U(x^i,y)$ give the same $\tilde{A}$ for any $h(y)\in PSL(2;\mathbb{C})$. Consequently we identify those configurations in the path integral over $U$. With $h=\begin{pmatrix} d & c \\ b & a \end{pmatrix}$, we identify $\phi$ as
\beq
\label{E:inflatingGauge}
	\tan\left( \frac{\phi}{2}\right) \sim \frac{a \tan\left(\frac{\phi}{2}\right)+b}{c\tan\left( \frac{\phi}{2}\right) + d}\,, \qquad ad-bc=1\,,
\eeq
or equivalently
\beq
	F \sim \frac{aF+b}{cF+d}\,,
\eeq
where $a,b,c,d$ are functions of $y$. There is also a complicated identification on the components $\Lambda$ and $\Psi$, which is consistent with the identification on $\phi$ and the asymptotic constraints~\eqref{E:inflatingConstraints}.

There is a similar story for global dS$_3$. There we decomposed flat connections as $\tilde{A} = \tilde{U}^{-1} \tilde{d}\tilde{U}$ with $\tilde{U} = \exp\left( \lambda(y)\theta\right) U$. In this case we have introduced a redundancy under
\beq
	\lambda (y) \to h(y) \lambda(y) h^{-1}(y) \,, \qquad U \to h(y) U\,.
\eeq
This leads to the identification
\beq
	\lambda \sim h\lambda h^{-1}\,, \quad \tan\left(\frac{\phi_+}{2}\right) \sim \frac{a\tan\left(\frac{\phi_+}{2}\right)+b}{c\tan\left(\frac{\phi_+}{2}\right)+d}\,, \quad -\cot\left(\frac{\phi_-}{2}\right) \sim \frac{a \left(-\cot\left(\frac{\phi_-}{2}\right)\right)+b}{c\left(-\cot\left(\frac{\phi_-}{2}\right)\right)+d}\,.
\eeq
Observe that the transformation law of $\phi_-$ is that its $S$-transform, $\tan\left(\frac{\phi_-}{2}\right) \to -\frac{1}{\tan\left(\frac{\phi_-}{2}\right)}=-\cot\left(\frac{\phi_-}{2}\right)$, is identified with its image under fractional linear transformations.

In the Hartle-Hawking geometry $\phi$ is subject to the same $PSL(2;\mathbb{R})$ quotient as $\phi_+$.

%***********************************************
\subsubsection{The boundary graviton measure}
%***********************************************

We would like to know the correct measure for the boundary graviton degrees of freedom, the reparameterization field $\phi$ appearing for the inflating patch of dS$_3$ and the future/past reparameterizations $\phi_{\pm}$ living on the boundaries of global dS$_3$.

The derivation of the measure proceeds in almost the same way as in our analysis of nearly dS$_2$ gravity. There we considered $BF$ theory, which after integrating out $B$ reduces to an integral over flat connections $A$. The flat measure on the space of flat connections reduces to the boundary measure we studied for nearly dS$_2$.

Consider Chern-Simons theory with compact, connected gauge group $G$ and Chern-Simons level $k$ on the Lorentzian cylinder. This model reduces to a boundary $G$ chiral WZW model at level $k$, which can be thought of as the path integral quantization of a coadjoint orbit of the corresponding Kac-Moody group~\cite{Alekseev:1988ce}. Separating the time component $A_t$ from the others $\tilde{A}=A_i dx^i$, and after integrating out $A_t$, one is left with a residual integral over the moduli space of flat spatial connections $\tilde{A} = U^{-1} \tilde{d}U$. At fixed time, this space of flat connections is symplectic, with a symplectic form identical to what one finds in $BF$ theory (with $A\to \tilde{A}$)
\beq
\label{E:CSmeasure}
	\omega= \frac{k}{4\pi} \int d^2x \, \epsilon^{ij} \text{tr}\left( d\tilde{A}_i\wedge d\tilde{A}_j\right)\,.
\eeq
The total integration space of the Chern-Simons theory amounts to promoting the flat connections under consideration to functions of time. As in our nearly dS$_2$ analysis, we parameterize a variation of $\tilde{A}$ through a variation of $U$, namely $dU = (dX) U$ with $X\in\mathfrak{g}$ a vector. We then have
\beq
\label{E:KMmeasure}
	\omega = \frac{k}{4\pi} \int d^2x \,\epsilon^{ij} \text{tr}\left( d\partial_i X \wedge d \partial_j X\right) = \frac{k}{4\pi} \int_0^{2\pi} d\theta \,\text{tr}(dX \wedge dX')\,,
\eeq
as before. Here $'$ indicates the angular derivative on the cylinder.

As in our discussion of $BF$ theory we have to account for a gauge redundancy. Doing so in a local way by taking $U(0) =1$, or equivalently $X(0) = 0$, the gauge-fixed measure at fixed time is identical to that in the $BF$ case, namely $\left( \prod_{\theta} dX(\theta)\right) \, \delta(X(0)) \text{Pf}(\omega)$. This measure is equivalent to $\prod_{\theta>0} d\mu(U(\theta))$ with $d\mu$ the Haar measure on $G$. Because $G$ is compact, this measure is positive-definite.

Now let us restore time. We now have a gauge redundancy under $U(\theta,t) \sim h(t) U(\theta,t)$, which can be still be fixed in a local way, e.g. $U(0,t) = 1$. The gauge-fixed measure may be denoted as $[dX(t)]\text{Pf}(\omega)$, or more precisely $\prod_t \left[ (\prod_{\theta>0} dX(\theta,t) )\text{Pf}(\omega)\right] = \prod_{\theta>0,t} d\mu(U(\theta,t))$.

Next we review the case of pure AdS$_3$ gravity on the Lorentzian cylinder, which is a little bit more involved than the WZW example above.  This AdS$_3$ setting was discussed by two of us in~\cite{Cotler:2018zff}. Using the $PSL(2;\mathbb{R})\times PSL(2;\mathbb{R})$ Chern-Simons description of three-dimensional gravity, one arrives at a $PSL(2;\mathbb{R})\times PSL(2;\mathbb{R})$ chiral WZW model on the boundary, supplemented with constraints that encode  the asymptotically AdS$_3$ boundary conditions. The boundary degrees of freedom are reparameterization fields $\phi(\theta,t)$ and $\bar{\phi}(\theta,t)$, where at fixed time both $\phi$ and $\bar{\phi}$ are elements of the quotient space Diff($\mathbb{S}^1)/PSL(2;\mathbb{R})$. This space is the first exceptional coadjoint orbit of the Virasoro group, with Kirillov-Kostant symplectic form~\cite{Bowick:1987pw,Witten:1987ty,Alekseev:1988ce}
\beq
\label{E:AdS3KK}
	\omega = \frac{C_{\rm AdS}}{48\pi} \int_0^{2\pi} d\theta\left( \frac{d\phi' \wedge d\phi''}{\phi'^2} - d\phi \wedge d\phi'\right)\,,
\eeq
where $C_{\rm AdS} = \frac{3}{2G_3}$ is the classical central charge of AdS$_3$ gravity. (There is a corresponding term for $\bar{\phi}$.) In this setting, there is a $PSL(2;\mathbb{R})$ gauge symmetry under
\beq
	\tan\left(\frac{\phi}{2}\right) \sim \frac{a\tan\left(\frac{\phi}{2}\right)+b}{c\tan\left(\frac{\phi}{2}\right)+d}\,.
\eeq
It may be fixed locally, say by taking $\phi(0)=0$, $\phi'(0)=1$, and $\phi''(0) = 0$. The measure at fixed time is $\left(\prod_{\theta}d\phi(\theta)\right)\delta(\phi(0))\delta(\phi'(0)-1)\delta(\phi''(0))  \text{Pf}(\omega)$. Stanford and Witten~\cite{Stanford:2017thb} showed that the gauge-fixed measure is local away from $\theta=0$, given by $\prod_{\theta>0} \frac{d\phi(\theta)}{\phi'(\theta)}$ (this measure was used in previous work \cite{bagrets2016sachdev} on the Schwarzian path integral) by explicit computation. In the path integral quantization of the coadjoint orbit, the field $\phi(\theta)$ is promoted to a function of time, and after fixing the $PSL(2;\mathbb{R})$ gauge symmetry locally (say via $\phi(0,t) = 0, \phi'(0,t) = 1, \phi''(0,t) = 0$), the measure becomes $\prod_{\theta>0,t} \frac{d\phi}{\phi'}$.

In~\cite{Cotler:2018zff} it was shown (inspired by~\cite{Alekseev:1988ce}) that the Haar measure on $PSL(2;\mathbb{R})$, after imposing the AdS$_3$ boundary conditions, reduces to precisely $\prod_{\theta>0,t} \frac{d\phi}{\phi'}$. However, using the methods below, one can get the Kirillov-Kostant symplectic form~\eqref{E:AdS3KK} directly from the flat Chern-Simons measure~\eqref{E:CSmeasure}.

Now let us attack the case of interest, dS$_3$ gravity in the $PSL(2;\mathbb{C})$ Chern-Simons formulation. Let us consider an inflating patch for simplicity. After integrating out $A_y$, the space of flat connections $\tilde{A}$ at constant $y$ is symplectic with
\beq
	\omega = \frac{ik}{4\pi}\int dt dx \,\epsilon^{ij} \text{tr}\left( d\tilde{A}_i \wedge d\tilde{A}_j\right) +(\text{c.c.})\,, \qquad k = \frac{1}{4G_3}\,.
\eeq
Parameterizing $\tilde{A} = U^{-1} \tilde{d}U$ and writing out a variation of $U$ as above, $U^{-1}dU = dX$, we find
\beq
\label{E:dS3omega}
	\omega = \frac{ik}{4\pi} \int dx\, \text{tr}\left( dX \wedge dX'\right) + (\text{c.c.})\,.
\eeq
Using the constraints~\eqref{E:inflatingConstraints},  we find that asymptotically a $d\phi$ induces a variation $dX=(dU)U^{-1}$ so that after some integration by parts Eq.~\eqref{E:dS3omega} becomes
\beq
\label{E:inflatingomega}
	\omega = \frac{C}{48\pi}\int dx \left( \frac{d\phi'\wedge d\phi''}{\phi'^2} - d\phi\wedge d\phi'\right) + (\text{c.c.})\,, \qquad C = i\frac{3}{2G_3}\,.
\eeq
In this form we recognize $\omega$ as the Kirillov-Kostant symplectic form. Because of the boundary condition on $\phi$, the field $\phi$ is at constant time an element of $\text{Diff}_{\mathbb{C}}(\mathbb{R})/PSL(2;\mathbb{C})$, with imaginary central charge. This space may be viewed as a coadjoint orbit of $\widehat{\text{Diff}}_{\mathbb{C}}(\mathbb{R})$. 

%***********************************************
\subsubsection{Monodromies}
%***********************************************

As we have seen, global dS$_3$ is more difficult to analyze than the inflating patch. In the global case we decompose $\tilde{A} = \tilde{U}^{-1} \tilde{d} \tilde{U}$ with $\tilde{U} = \exp(\lambda(y)\theta) U$. The symplectic form on the space of flat connections at fixed $y$ is now
\begin{align}
\begin{split}
\label{E:globalMeasure}
	\omega &= \frac{ik}{4\pi}\int d^2x \,\epsilon^{ij}\text{tr}\left(d \tilde{A}_i \wedge d\tilde{A}_j\right) +(\text{c.c.})
	\\
	& = \frac{ik}{4\pi}\int_0^{2\pi} d\theta\, \text{tr}\left( d\tilde{X}_+ \wedge d\tilde{X}_+' - d\tilde{X}_- \wedge d\tilde{X}_-'\right)+(\text{c.c.})\,,
\end{split}
\end{align}
where $d\tilde{X} = (d\tilde{U})\tilde{U}^{-1} $. 

We have not found an enlightening presentation for this measure in terms of the boundary reparameterization modes $\phi_{\pm}$ and the monodromy $\lambda$.  Let us focus on a few simple cases. The simplest is to set the monodomy $\lambda$ and its variations to vanish. Using the same methods as above for the inflating patch, we find after some integration by parts that
\beq
	\omega = \frac{C_{\rm dS}}{48\pi}\int_0^{2\pi} d\theta\left( \frac{d\phi_+'\wedge d\phi_+''}{\phi_+'^2} - d\phi_+\wedge d\phi_+'- \left(\frac{d\phi_-'\wedge d\phi_-''}{\phi_-'^2} - d\phi_-\wedge d\phi_-' \right) \right) + (\text{c.c.})\,.
\eeq
The second simplest case is to expand perturbatively around the dS$_3$ critical point.  Specifically, we expand around the dS$_3$ solution
\beq
	\phi_{\pm} = \theta + \sum_n \epsilon^{\pm}_{n} e^{in\theta}\,,
\eeq
and take $\epsilon_{\pm n}$ and $\lambda$ to be of the same infinitesimal order. We readily find the symplectic form to quadratic order in fluctuations,
\begin{align}
\label{E:globalomegaQuad}
	\omega = \frac{C_{\rm dS}}{24} \Big[& i \sum_{n\neq -1,0,+1}  n(n^2-1)\left( d\epsilon^+_{-n} \wedge d\epsilon^+_{n} - d\epsilon^-_{-n}\wedge d\epsilon^-_{n} \right) 
	\\
	\nonumber
	&\qquad +4i  d\lambda_1\wedge d\epsilon^{(0)} -4 d\lambda^{(+)} \wedge d\epsilon^{(-)} + 4d\lambda^{(-)} \wedge d\epsilon^{(+)}-\frac{3i}{2}d\lambda^{(-)}\wedge d\lambda^{(+)}\Big] + (\text{c.c.})\,.
\end{align}
where we have defined
\beq
	\epsilon^{(0)} \equiv \epsilon_0^+ - \epsilon_0^- \,, \qquad \epsilon^{(\pm)} \equiv\frac{ \epsilon^+_{\pm 1} + \epsilon^-_{\pm 1}}{2}\,, \qquad \lambda^{(\pm)} = \lambda_2 \pm  \lambda_0\,.
\eeq
There is a lot of information here to unpack. In this perturbative setting, infinitesimal $PSL(2;\mathbb{C})$ transformations act as
\beq
	\delta\epsilon_0^+ =\delta\epsilon_0^- \,, \qquad \delta \epsilon_{\pm 1}^+ = - \delta\epsilon_{\pm 1}^- \,,
\eeq
while the $\lambda$'s and the combinations $\epsilon_{-1,0,1}$ defined above are gauge-invariant. We then see that the perturbative symplectic form~\eqref{E:globalomegaQuad} is gauge-invariant, as the first line receives no contribution from the $n= -1,0,1$ modes, and the second line entails gauge-invariant quantities. Further, the gauge-invariant combinations of $n=-1,0, 1$ modes, $\epsilon^{(0),(+),(-)}$, are conjugate to the monodromy.

The gauge-invariant degrees of freedom are then the $n\neq -1,0,1$ fluctuations of the reparameterization fields $\epsilon^{\pm}_n$, and the conjugate $(\epsilon^{(0)},\lambda^{(1)})$, $(\epsilon^{(+)},\lambda^{(-)})$ and $(\epsilon^{(-)},\lambda^{(+)})$. 

The final simple case we can analyze is the nonlinear measure on the ``twist'' $\epsilon^{(0)}$ and $\lambda_1$. Let us choose a convention whereby $\phi_{\pm}$ are initially periodic around the circle with $\lambda = - i \alpha J_1$. The monodromy and $\phi_{\pm}$ appear through the combination $\alpha \theta + \phi_{\pm}$, and so we perform the field redefinition $\alpha \theta + \phi_{\pm} \to \alpha \phi_{\pm}$. The new fields $\phi_{\pm}$ obey $\phi_{\pm}(\theta+2\pi,y) = \phi_{\pm}(\theta,y) + 2\pi $. Consider the particular fluctuation $d\alpha$, $d\phi_+ = d\gamma$, $d\phi_- = 0$. We readily compute the contribution to the measure to be
\beq
\label{E:omegaWP}
	\omega_{\rm WP} = \frac{C_{\rm dS}}{24}\alpha d\alpha \wedge d\gamma + (\text{c.c.})\,.
\eeq
Up to a constant, this is the Weil-Petersson measure, as in our nearly dS$_2$ analysis. In this parameterization one has 
\beq
	\tilde{U}_+ = \begin{pmatrix} \cos\left( \frac{\alpha \phi_+}{2}\right) & - \sin\left( \frac{\alpha  \phi_+}{2}\right) \\ \sin\left(\frac{\alpha\phi_+}{2}\right) & \cos\left(\frac{\alpha \phi_+}{2}\right) \end{pmatrix} \begin{pmatrix} \Lambda_+ & 0 \\ 0 & \Lambda_+^{-1}\end{pmatrix} \begin{pmatrix} 1 & \Psi_+ \\ 0 & 1 \end{pmatrix}\,,
\eeq
so that $\gamma$ lives on the complex cylinder with $\gamma \sim \gamma + 2\pi$.

The gravitational interpretation of this field configuration is the following. It is topologically an annulus, obtained by gluing a future dS$_3$ patch to a past dS$_3$ patch. When $\gamma$ and $\alpha$ are real, the gluing is performed across a geodesic of length $\sim \alpha$, and where the angle on the past circle is related by a shift $\gamma$ to the angle on the future circle.

%***********************************************
\subsection{Gravitational path integrals}
%***********************************************

In this Subsection we put the pieces together from the last Subsection and compute loop corrections to the gravitational path integral on an inflating patch and for global dS$_3$.  Similar computations, albeit in a slightly different context, were performed in \cite{Castro:2011xb,Castro:2011ke,Castro:2012gc}.

%***********************************************
\subsubsection{Inflating patch}
%***********************************************

We would like to compute loop corrections to the central charge of gravity in an inflating patch of de Sitter, as well as the spectrum of local operators on the boundary. To get both at once, let us use the trick of putting the theory on the boundary of the inflating patch on a torus of complex structure $\tau$ by identifying $z\sim z + 2\pi$ and $z\sim z+2\pi \tau$. On the infinite plane, the field $\phi$ maps the constant-$y$ surface, i.e.\! the line $x$, to the circle. We keep this property intact, so that the boundary conditions on $\phi$ are
\begin{align}
\begin{split}
	\phi(x+2\pi,y) & = \phi(x,y) + 2\pi\,,
	\\
	\phi(x+2\pi \text{Re}(\tau),y+2\pi\text{Im}(\tau)) & = \phi(x,y)\,.
\end{split}
\end{align}
The field $\phi$ is subject to the ``gauge symmetry''~\eqref{E:inflatingGauge}, and the symplectic form on the integration space at fixed $y$ is given by~\eqref{E:inflatingomega}.

The computation of the torus partition function of this model to one-loop order closely imitates the analysis in~\cite{Cotler:2018zff} for pure AdS$_3$ gravity (which reproduces previous results~\cite{Maloney:2007ud,Giombi:2008vd}). The unique critical point of the model (modulo the $PSL(2;\mathbb{C})$ quotient) is
\beq
	\phi_0 = x - \frac{\text{Re}(\tau)}{\text{Im}(\tau)}y\,.
\eeq
Its action is
\beq
	S_0 = -\frac{\pi  C}{12}\left( \tau - \bar{\tau}\right) \,,
\eeq
so that the classical approximation to the partition function is
\beq
	Z_{\rm classical} = e^{iS_0} = |q|^{-\frac{C}{12}}\,, \qquad q=e^{2\pi i \tau}\,.
\eeq
We expand in fluctuations around the critical point $\phi = \phi_0 + \epsilon$, and we then decompose $\epsilon$ into Fourier modes on the torus,
\beq
	\phi = \phi_0 + \sum_{n,m} \epsilon_{m,n} e^{i n \left( x-\frac{\text{Re}(\tau)}{\text{Im}(\tau)}y\right) + \frac{imy}{\text{Im}(\tau)}}\,.
\eeq
Here $\epsilon$ is complex and so we further decompose it into real and imaginary parts, $R\epsilon$ and $I\epsilon$. Letting $R\epsilon_{m,n}$ and $I\epsilon_{m,n}$ denote the Fourier modes of $R\epsilon$ and $I\epsilon$, we have $\epsilon_{m,n} = R\epsilon_{m,n} + i I\epsilon_{m,n}$ and $R\epsilon_{m,n}^* = R\epsilon_{-m,-n}$, $I\epsilon_{m,n}^* = I\epsilon_{-m,-n}$. The quadratic approximation to the action~\eqref{E:inflatingS} is then
\beq
	S = S_0 + \frac{\pi i C}{3}\sum_{n>1,m} n(n^2-1)E_{m,n}^{\dagger} D_{m,n} E_{m,n} + O(\epsilon^3)\,.
\eeq
where
\beq
\label{E:DandE}
	D_{m,n} = \begin{pmatrix} n\text{Im}(\tau) & n\text{Re}(\tau) - m \\ n\text{Re}(\tau) - m & -n \text{Im}(\tau)\end{pmatrix} \,, \qquad E_{m,n} =\begin{pmatrix} R\epsilon_{m,n} \\ I\epsilon_{m,n}\end{pmatrix}\,.
\eeq
The matrix $D_{m,n}$ has the nice feature that
\beq
	\text{det}(D_{m,n}) = -|n \tau-m|^2\,,
\eeq
the eigenvalues of the scalar Laplacian on the torus. We may use the $PSL(2;\mathbb{C})$ freedom to set
\beq
	\epsilon_{m,n=-1,0,1} = 0\,.
\eeq
The quadratic action is completely real and so to path integrate $e^{i S}$, we rotate the contour of field integration. Equivalently we define the Gaussian integral by analytic continuation.

Now we turn to the measure. Evaluating the symplectic form~\eqref{E:inflatingomega} on the critical point, we find
\beq
	\omega = \frac{C}{6} \sum_{n}  n(n^2-1) dR\epsilon_n \wedge dI\epsilon_{-n} \,, \qquad \text{Pf}(\omega) = \prod_{n>1}\left|  \frac{3}{6}n(n^2-1)\right|^2
\eeq
and so the relevant one-loop measure is
\beq
	[d\phi] = \prod_{n>1,m}d^2R\epsilon_{m,n} d^2I\epsilon_{m,n}\left|\frac{C}{3}n(n^2-1)\right|^2\,,
\eeq
and the one-loop partition function reads
\begin{align}
\nn
	Z_{\rm HH} =\,& |q|^{-\frac{C}{12}}\prod_{n>1,m}  \int d^2R\epsilon_{m,n}d^2I\epsilon_{m,n} \, \left|\frac{C}{3}n(n^2-1)\right|^2  \exp\left( - \frac{\pi  C}{3}n(n^2-1) E_{m,n}^{\dagger}D_{m,n}E_{m,n}\right)
	\\
	=\,& |q|^{-\frac{C}{12}} \prod_{n>1,m}\frac{1}{|n\tau-m|^2}= |q|^{-\frac{C+13}{12}} \prod_{n=2}^{\infty}\frac{1}{|1-q^n|^2}\,.
\end{align}
The final result is the character of the vacuum representation of the Virasoro algebra with central charge
\beq
	c_{\text{1-loop}} = C + 13 = i\frac{3}{2G_3}+13\,.
\eeq

We see that the model on the future patch has a spectrum of operators given by the identity and its Virasoro descendants, and that the one-loop renormalization of the central charge is finite and real. In fact, the one-loop renormalization of the central charge by 13 is the same for pure AdS$_3$ gravity~\cite{Cotler:2018zff,Giombi:2008vd}. Restoring the radius, the result there is
\beq
	c = \frac{3L_{\rm AdS}}{2G_3} + 13\,,
\eeq
which matches the one we find above under the simple analytic continuation $L_{\rm AdS} \to i L_{\rm dS}$. One also has the same spectrum of local operators, the identity and its Virasoro descendants.

The torus partition function of the model obtained from Lorentzian AdS$_3$ gravity is one-loop exact~\cite{Cotler:2018zff}, by a similar localization argument used by Stanford and Witten~\cite{Stanford:2017thb} for the Schwarzian path integral. The crucial feature is that one has a phase space path integral, where the phase space is K\"ahler with a metric invariant under the Hamiltonian flow. Those same features persist in this model, since we are dealing with the coadjoint orbit Diff$_{\mathbb{C}}(\mathbb{S}^1)/PSL(2;\mathbb{C})$.

Indeed, a closely related result has appeared in the literature previously \cite{Castro:2012gc}, where the authors studied the wavefunction for universes which asymptote to a torus at future infinity.  The wave function was studied as a function of $\tau$, and both perturbative and non-perturbative contributions were evaluated using the continuation to Euclidean AdS.  The result was a wavefunction that diverged as $\tau\to i \infty$, which in that case arose from the sum over Euclidean AdS saddle points with torus boundary.

%***********************************************
\subsubsection{Global dS$_3$}
%***********************************************

Now we treat global dS$_3$. We take the same approach as in the last Subsection, putting the theory on a torus and dropping the constant term in the action~\eqref{E:globaldS3S}, which encodes the contribution of the classical central charge $C$ to the partition function on $\mathbb{S}^2\cup\mathbb{S}^2$. This procedure is a bit artificial, and so at the end we take a degeneration limit so as to find ourselves back on the sphere.  We note that, as we now have a pair of fields $\phi_\pm$, the result of the computation will be a bit different from that on the inflating patch discussed above.

We start from a convention where $\phi_{\pm}$ are periodic around $\theta$ and $\lambda = - i J_1$. The boundary condition on $\lambda$ implies that $\lambda^1$ is fixed at the poles of the sphere. We then move to a convention where $\phi_{\pm}$ obey the boundary condition
\begin{align}
\begin{split}
	\phi_{\pm}(\theta+2\pi, y)& = \phi_{\pm}(\theta,y) + 2\pi\,, 
	\\
	 \phi_{\pm}(\theta+2\pi \text{Re}(\tau),y+2\pi\text{Im}(\tau)) &= \phi_{\pm}(\theta,y)\,,
\end{split}
\end{align}
and $\lambda$ fluctuates around zero. We take the monodromy to be periodic in $y$, but do not allow $\lambda^1$ to have a constant mode.

There is a manifold of critical points of the model modulo the $PSL(2;\mathbb{R})$ quotient,
\beq
	\phi_{+,-0} = \phi_{-,0} + \gamma\\,, \qquad \phi_{-,0} = \theta-\frac{\text{Re}(\tau)}{\text{Im}(\tau)}y\,, \qquad \lambda = 0\,,
\eeq
parameterized by a twist $\gamma$. We then perturb around this solution with $\phi_+ = \phi_{+,0} + \epsilon^+$ and $\phi_-=\phi_{-,0} +\epsilon^-$, taking $\epsilon^{\pm}$ and $\lambda$ to be of the same order. Decomposing these into Fourier modes as before,
\begin{align}
\begin{split}
	\phi_+ &= \phi_{+,0} + \sum_{n,m} \epsilon^+_{m,n} e^{\frac{i m y}{\text{Im}(\tau)} + i n \left(\theta-\frac{\text{Re}(\tau)}{\text{Im}(\tau)}y\right)}\,,
	\\
	\phi_- & = \phi_{-,0} + \sum_{n,m} \epsilon^-_{m,n}e^{\frac{imy}{\text{Im}(\tau)}+in\left(\theta-\frac{\text{Re}(\tau)}{\text{Im}(\tau)}y\right)}\,,
\end{split}
\end{align}
we further decompose
\beq
	\epsilon^{\pm}_{m,n} = R\epsilon^{\pm}_{m,n} + i I\epsilon^{\pm}_{m,n}\,,  \qquad (R\epsilon^{\pm}_{m,n})^* = R\epsilon^{\pm}_{-m,-n} \,,\qquad (I\epsilon^{\pm}_{m,n})^* = I\epsilon^{\pm}_{-n,-m}\,.
\eeq
We also decompose the fluctuations of the monodromy into Fourier modes
\beq
	\lambda^A = \sum_m \lambda^A_m e^{\frac{i m y}{\text{Im}(\tau)}}\,, \qquad \lambda^A_m = R\lambda^A_m +i I\lambda^A_m \,,
\eeq
with
\beq
	(R\lambda^A_m)^* = R\lambda^A_{-m}\,, \qquad (I\lambda^A_m)^* = I\lambda^A_{-m}\,.
\eeq
The boundary condition on the monodromy amounts to the statement that there is no $m=0$ mode for $\lambda^1$. Infinitesimal $PSL(2;\mathbb{R})$ transformations act as
\beq
	\delta \epsilon^+_{m,0} = \delta \epsilon^-_{m,0}\,, \qquad \delta \epsilon^+_{m,\pm 1} = -\delta \epsilon^-_{m,\pm 1}\,,
\eeq
and the monodromy is invariant. The linearized gauge-invariant combinations are 
\beq
	\epsilon^{(0)}_{m} = \epsilon^+_{m,0} - \epsilon^-_{m,0}\,, \qquad \epsilon^{(\pm)}_{m} = \frac{\epsilon^+_{m,\pm 1} + \epsilon^-_{m,\pm 1}}{2}\,.
\eeq
The $m=0$ modes of $\epsilon_{m}^{(0)}$ are already accounted for with the integral over the manifold of critical points.

The quadratic approximations to the action and symplectic form read
\begin{align}
\nonumber
	S& = \frac{\pi i C}{3}\left[  \sum_{n>1,m} n(n^2-1) \left( E^{+\dagger}_{m,n}D_{m,n} E^+_{m,n} - E^{-\dagger}_{m,n} D_{m,n}E^-_{m,n}\right)\right.
	\\
	&\qquad  \qquad \qquad \left. - \sum_{m}\left[ (\tau-m) \left( \lambda^{(+)}_m \epsilon^{(-)}_{-m}  + \lambda^{(-)}_m \epsilon^{(+)}_{-m}\right) +\frac{i}{2}m \lambda^1_m \epsilon^{(0)}_{-m}+ (\text{c.c.})\right]\right]
	\\
	\nonumber
	\omega & = \frac{C}{6}\left[  \sum_{n} n(n^2-1) \left( dR\epsilon^+_n\wedge dI\epsilon^+_{-n} - dR\epsilon^-_n\wedge dI\epsilon^-_{-n}\right) \right.
	\\
	\nonumber
	& \qquad\qquad  \left. +\left( i  d\lambda^1\wedge d\epsilon^{(0)} -d\lambda^{(+)} \wedge d\epsilon^{(-)} + d\lambda^{(-)} \wedge d\epsilon^{(+)}-\frac{3i}{8}d\lambda^{(-)}\wedge d\lambda^{(+)}- (\text{c.c.})\right)\right]\,,
\end{align}
where
\beq
	\lambda^{(\pm)}_m =\lambda^0_m\pm \lambda^2_m\,,
\eeq
and in the symplectic form we are only considering contributions from a constant-$y$ slice. The matrix $D_{m,n}$ is given in~\eqref{E:DandE}, and
\beq
	E^{\pm}_{m,n}=\begin{pmatrix} R\epsilon_{m,n}^{\pm} \\ I\epsilon_{m,n}^{\pm}\end{pmatrix}\,.
\eeq
For the $|n|>1$ modes, this is just a doubled version of the path integral at future infinity. The remaining reparameterization modes, except for the zero modes, are lifted by the coupling to the monodromy. The remaining zero modes are already parameterized through $\gamma$. Computing the functional determinant we find that the one-loop approximation to the path integral is
\beq
\label{E:Zresultglobal1}
	Z =  \left( |q|^{-\frac{C+1}{12}} \prod_{n=1}^{\infty}\frac{1}{|1-q^n|^2}\right)^2
\eeq
where we have included by hand the classical central charge.  This partition function is the square of a character of an ordinary Verma module.

Let us briefly discuss how the different parts of this result arise. There is an implicit infinite volume prefactor coming from the manifold of critical points labeled by $\gamma$, which we regularize. The determinant that arises from integrating out the $|n|>1$ modes of the reparameterization fields is
\beq
	\prod_{n>1,m} \frac{1}{|n\tau-m|^4}\,,
\eeq
which leads to the part of the infinite product in~\eqref{E:Zresultglobal1} from $n=2$ to infinity. Integrating out the modes of the monodromy leads to an infinite product of delta functions for the gauge-invariant modes $\epsilon^{(0),(+),(-)}_m$, and the integral over those modes picks up a factor
\beq
	\prod_{m} \frac{1}{|\tau-m|^4} \,,
\eeq
which leads to a multiplicative factor of $\frac{1}{|1-q|^4}$.  This accounts for the $n=1$ contribution to the product in~\eqref{E:Zresultglobal1}. 

For computational ease we have put our dS$_3$ model ``by hand'' on boundary tori.  However, the true global dS$_3$ model lives on boundary spheres.  We recover the corresponding sphere partition function from a degeneration of the tori (see \cite{Cotler:2018zff} for details for a similar analysis in AdS$_3$), giving us the global dS$_3$ partition function
\beq
	Z = (R\mu)^{2c/3}\,,\qquad  c = C +1\,,
\eeq
where $R$ is the radius of each of the boundary spheres.

%***********************************************
\section{Concluding remarks}
\label{S:discuss}
%***********************************************

%***********************************************
\subsection{Summary}
%***********************************************

\begin{figure}[t]
\centering
\includegraphics[width=2in]{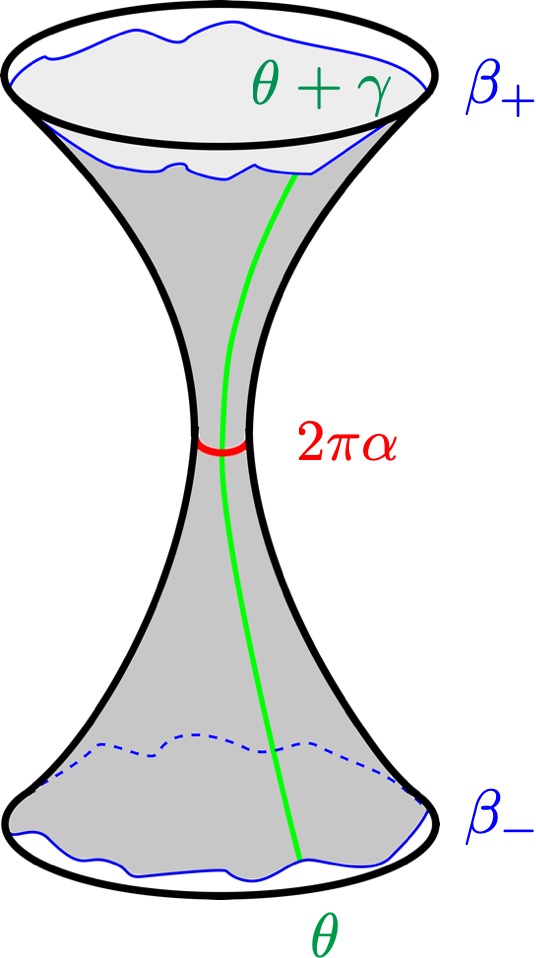}
\vskip.5cm
\captionof{figure}{\label{F:sumglobal} Schematic diagram of global dS$_2$. The asymptotic boundaries have lengths $\beta_+$ and $\beta_-$, there is a minimal length geodesic of length $2\pi \alpha$, and there is a twist $\gamma$.}
\end{figure}

We dedicated most of our attention in this manuscript to the path integral of Jackiw-Teitelboim gravity in two-dimensional nearly de Sitter spacetime. This theory of gravity has a small phase space of connected, purely Lorentzian solutions.  The only saddle points of the model are global nearly dS$_2$ spacetimes which have circle boundaries at future and past infinity.  All have the shape of a Lorentzian hyperboloid. These geometries are characterized by two canonically conjugate quantities: the length $2\pi \alpha$ of the minimal length geodesic around the ``bottleneck,'' and the twist $\gamma$ of the angle on the future circle relative to the angle on $\mathcal{I}^-$. (This twist is the de Sitter analogue of the ``time shift'' in~\cite{Harlow:2018tqv}.) See Fig.~\ref{F:sumglobal}.

As is familiar from worldsheet string theory, to define a genus expansion we must continue to Euclidean signature. One might think of defining a continuation modeled on the Hartle-Hawking construction, where instead of gluing a Euclidean hemisphere to the geodesic $t=0$ slice of global dS$_2$, we glue in some other constant positive curvature smooth Euclidean geometry of higher genus. However, such geometries do not exist. In this work we have exploited another analytic continuation, originally due to Maldacena~\cite{Maldacena:2010un}, of Lorentzian dS to Euclidean AdS. In this scheme, rather than gluing in a Euclidean hemisphere to the $t=0$ slice, one attaches Euclidean AdS directly to $\mathcal{I}^+$. Inspired by this, we showed that Jackiw-Teitelboim gravity in dS$_2$ may be analytically continued to Jackiw-Teitelboim gravity in Euclidean AdS$_2$, provided that one works in the first-order formalism. 

The first-order presentation of the Lorentzian nearly dS$_2$ theory and the Euclidean nearly AdS$_2$ theory may each be written as a $PSL(2;\mathbb{R})$ $BF$ theory.  To two theories are mapped to one other under a permutation of the first-order variables and a suitable analytic continuation of the dilaton boundary condition.\footnote{\label{FN:dilatonBC} 
%In nearly AdS$_2$ spacetime, the dilaton goes to a larger constant value on the boundary cutoff slice than in the nearly dS$_2$ case.  Namely, in 
In nearly AdS$_2$ the dilaton goes to $1/(J'\epsilon)$ with $\epsilon\to 0$ as the boundary is sent to conformal infinity. In nearly dS$_2$ spacetime, the dilaton goes from a large negative value near past infinity, $-1/(J\epsilon)$, to a large positive value near future infinity, $1/(J\epsilon)$. When mapping a Euclidean AdS boundary to a component of $\mathcal{I}^-$ one takes $J' \to i J$, and $J'\to -i J$ for a component of $\mathcal{I}^+$.}  In Euclidean AdS, the path integral in the first order formulation is not quite the same as the corresponding path integral in the metric formulation. The topological gauge theory integrates over metrics continuously connected to a single representative, and so has no sum over topologies, whereas the metric formulation does contain such a sum. This being said, the two formulations, expanded around a given geometry, are expected to agree to all orders in perturbation theory. Thus to match the first order formalism to the metric formulation, one includes an additional sum over topologies in the $BF$ theory ``by hand.''

To define a genus expansion for de Sitter JT gravity, we first carefully studied the Hartle-Hawking geometry and global dS$_2$ from a variety of perspectives. In the metric formulation it is easy to find a Schwarzian-like boundary action for fluctuations around both spacetimes, and we verified that one finds the same action in the first-order formulation. One advantage of the first order formalism is that we can use it to compute the nonlinear measure for the path integral, as is well-known in the context of $BF$ theory~\cite{Witten:1991we}, and which was analyzed recently for AdS$_3$~\cite{Cotler:2018zff} and nearly AdS$_2$ gravity~\cite{Saad:2019lba}. We then relate the Hartle-Hawking and global dS$_2$ backgrounds, and the Schwarzian-like models on their boundaries, to corresponding backgrounds and models of Euclidean nearly AdS$_2$ gravity.  JT gravity on the Hartle-Hawking geometry is mapped to Euclidean AdS$_2$ JT gravity on the Poincar\'e disk, via the 2d version of the Maldacena contour. In the $BF$ formulation this mapping between the two models is consistent, as both the Hartle-Hawking geometry and the Poincar\'e disk are characterized by a trivial holonomy at infinity. JT gravity on global dS$_2$ is mapped to JT gravity on the ``double trumpet'' geometry of~\cite{Saad:2019lba}, which is the Euclidean hyperbolic cylinder. However, this map is not immediate, and involves another ingredient in the continuation from Euclidean AdS to dS. 

Let us recall one of the key points of~\cite{Saad:2019lba}. Smooth Euclidean $R=-2$ geometries with multiple asymptotic regions are characterized by minimal geodesics of lengths $b_i$, along which the asymptotic regions are glued to an intermediate surface. In the $BF$ formulation, there is a hyperbolic $PSL(2;\mathbb{R})$ holonomy around each asymptotic circle, and the path integral includes a suitable integral over these $b$'s. In the present paper, we found that the situation is reversed in de Sitter. The asymptotic de Sitter regions are characterized by \emph{elliptic} holonomies around the asymptotic circles. In mapping the double trumpet to global dS, we must not only analytically continue the dilaton boundary condition, but we must also analytically continue the length of the Euclidean ``bottleneck'' as $b\to 2\pi i \alpha$, where $2\pi \alpha$ is the length of the Lorentzian ``bottleneck.''  We were then confronted by the question of what integration measure to use over $\alpha$. We found that there were two consistent definitions of the path integral, depending on whether we took the measure over $\alpha$ to be either minus or plus $(2\pi)^2 \alpha d\alpha$. The choice with a minus sign amounts to putting a factor of $(-1)^E$ into the path integral, where $E$ counts the number of independent elliptic holonomies. Crucially, for that choice, the integration measure over $b$ (namely $bdb$) continues directly to the measure over $\alpha$, and in fact the path integral for the double trumpet is then mapped exactly to the path integral of global dS$_2$. In what follows we elect to choose the $\alpha$ measure to be $-(2\pi)^2 \alpha d\alpha$.  

After the continuation, the double trumpet becomes a singular $R=-2$ space. It is composed of two hyperbolic cones with cone angles $2\pi \alpha$, glued to each other at the tips. However, we stress that while the space is singular a Riemannian manifold, it is a completely smooth gauge configuration. From the gauge theory point of view it is simply an annulus endowed with a smooth, flat $PSL(2;\mathbb{R})$ bundle.

Indeed, one way to describe the map from Euclidean AdS$_2$ to dS$_2$ is that, by going to the topological gauge theory description, the same $PSL(2;\mathbb{R})$ gauge field corresponds to two different metrics. One is Lorentzian with curvature $R=2$, and the other Euclidean with $R=-2$. When one metric has a cone point, the other is smooth. In the global dS$_2$ case, the $PSL(2;\mathbb{R})$ connection describing the double hyperbolic cone admits a smooth metric everywhere.

With all of these results in hand we can define the de Sitter genus expansion. Take $\mathcal{I}^+$ to have $n_+$ circles of sizes $\beta^+_{a}$, and $\mathcal{I}^-$ to have $n_-$ circles of sizes $\beta^-_{m}$. The gravitational path integral with these boundary conditions computes a transition amplitude, which depends on the number of boundary circles and their sizes. We sum over bulk surfaces that fill in the boundaries. In the topological gauge theory description these surfaces are simply characterized by smooth, flat $PSL(2;\mathbb{R})$ bundles, with elliptic holonomies around the asymptotic circles. Locally, there are two smooth metric descriptions. Inevitably, once the bulk surface has more than two boundaries or has genus $g>1$, neither of these metric descriptions is smooth everywhere. Elliptic holonomies guarantee cone points in the Euclidean $R=-2$ description, while hyperbolic holonomies guarantee the analogue of cone points in the Lorentzian $R=2$ description. For practical purposes we simply define the path integral  at fixed genus using the topological gauge theory, but if we wanted to think in terms of spacetime geometry, then the natural procedure is to use different metric descriptions in different regions as need be.

One way to regard these geometries recalls the Hartle-Hawking construction. Take $n_+$ future de Sitter trumpets and $n_-$ past trumpets, each of which is characterized by a minimal length geodesics with lengths $2\pi \alpha_a$ and $2\pi \alpha_m$, or equivalently by elliptic holonomies. We then glue in an intermediate surface $\Sigma$. From the point of view of the Euclidean continuation, the intermediate surface has $R=-2$ surface with cone points, characterized by the same elliptic holonomies. We then glue $\Sigma$ to the dS trumpets. See Fig.~\ref{F:singularGluing}. While this procedure might appear singular, it is not: the gauge connection is everywhere smooth. In the metric description, we could simply use the Lorentzian $R=2$ description slightly past the gluing, in terms of which the cone point is merely a bottleneck. 

Another way to think about the bulk geometry is analogous to the Maldacena contour, in which one uses the Euclidean $R=-2$ description throughout. We have the same intermediate surface $\Sigma$, but we replace the asymptotic dS$_2$ regions with $n$ hyperbolic disks $\mathcal{D}_i$ attached to $\mathcal{I}^+$ and $\mathcal{I}^-$. The elliptic holonomies guarantee that these disks are in fact hyperbolic cones of angles $2\pi \alpha_i$. The tips of these cones are then glued to the tips of $\Sigma$. Near the tips the Euclidean $R=-2$ description breaks down, but the Lorentzian $R=2$ description is perfectly smooth. 

One might wonder if there is a smooth Lorentzian $R=2$ description throughout. However, there is not. At generic points in the moduli space of $\Sigma$ there are hyperbolic holonomies around the internal cycles. In the Euclidean description these cycles remain finite size, but in the Lorentzian description they shrink to zero size at points, and the Lorentzian metric there has the analogue of conical singularities, locally of the form
\beq
	ds^2 = -dt^2 + b^2 t^2 du^2\,, \qquad u \sim u + 1\,.
\eeq

The transition amplitude then has a genus expansion. The term at genus $g$ with $n_+$ future and $n_-$ past boundaries with $n=n_++n_-$ is, for some convention for volumes (here we have in mind the more general topologies, not the Hartle-Hawking geometries or global dS$_2$, whose path integrals are given by slightly different expressions),
\begin{align}
	Z_{g,n_+,n_-}(\beta_{a}J,\beta_{m}J) &= (-1)^n(2\pi)^{2n}\int_0^{\infty}\alpha_1d\alpha_1\hdots \alpha_n d\alpha_n \widetilde{V}_{g,n}(\alpha_1,\hdots,\alpha_n) 
	\\
	\nonumber
	&\times Z_F(\beta_{1}J,\alpha_1)\hdots Z_F(\beta_{n_+}J,\alpha_{n_+}) Z_P(\beta_{n_++1}J,\alpha_{n_++1}) \hdots Z_P(\beta_{n}J,\alpha_n)\,,
\end{align}
where $Z_F$ is the integral over the Schwarzian mode on one of the circles at $\mathcal{I}^+$, and $Z_P$ is the integral over the Schwarzian mode on one of the circles at $\mathcal{I}^-$.  $Z_P$ is in fact the complex conjugate of $Z_F$. The factor of $\widetilde{V}_{g,n}$ refers to the volume of the moduli space of flat connections on the intermediate surface $\Sigma$. These $\Sigma$ may be described metrically as genus $g$ hyperbolic cones with $n$ cone points of angles $2\pi \alpha_i$. $\widetilde{V}_{g,n}$ is the Weil-Petersson volume of the moduli space of such surfaces. It has been proven~\cite{2004cones} that in the region $\alpha_i \leq 1/2$, the $\widetilde{V}_{g,n}$'s are in fact the continuation of the Weil-Petersson volumes $V_{g,n}$ of moduli spaces of genus $g$ hyperbolic surfaces with geodesic borders of lengths $b_i$, with
\beq
	\widetilde{V}_{g,n}(\alpha_1,\hdots,\alpha_n) = V_{g,n}(b_1 \to 2\pi i \alpha_1, \hdots b_n \to 2\pi i \alpha_n)\,, \qquad \alpha_i \leq \frac{1}{2}\,.
\eeq
However once one of the cone angles opens up past $\pi$, the proof of~\cite{2004cones} breaks down. See also~\cite{DoNorbury,2011cones} for a discussion.

Now as we have seen the natural description of $\Sigma$ is not in terms of a metric, but in terms of a flat $PSL(2;\mathbb{R})$ bundle. From that point of view $\Sigma$ is a genus $g$ surface with $n$ holes, and elliptic holonomies around the holes. What do the volumes $\widetilde{V}_{g,n}$ correspond to in the gauge theory description? To answer this it is worthwhile to back up a step, and revisit the gauge theory description of the more tame $V_{g,n}$'s. We refer the reader to an elegant computation of $V_{1,1}$ by Kim and Porrati~\cite{Kim:2015qoa}. Rather than considering a hyperbolic metric on a one-holed torus, they compute the volume of the moduli space of flat $PSL(2;\mathbb{R})$ connections on a one-holed torus with the condition that the holonomy $H$ around the hole is hyperbolic and fixed via $\text{tr}(H) = 2\cosh(b/2)$. In the metric interpretation the hole is homotopic to a geodesic boundary of length $b$. As in our computations, they integrate using the symplectic measure on flat connections, which is equivalent to the Weil-Petersson measure. The resulting volume precisely reproduces $V_{1,1}$, and indeed it had to be the case.

For our $\widetilde{V}_{g,n}$'s we have in mind the analogous computation where the holonomies around the holes are elliptic rather than hyperbolic, with $\text{tr}(H) = 2\cos(\pi \alpha)$. From this point of view $\alpha = 1/2$ is clearly special, but there does not seem to be an obstruction to continuing past it. There is one subtle point here: from the Lorentzian $R=2$ metric, it is clear that we integrate over all positive $\alpha$. (In global dS$_2$, $\alpha$ is the length of the ``bottleneck'' and can be arbitrarily large.) Thus these elliptic holonomies are valued in the universal cover of $PSL(2;\mathbb{R})$, rather than $PSL(2;\mathbb{R})$. In any case, there seems to be no problem in defining $\widetilde{V}_{g,n}$ beyond for all $\alpha$, and indeed this definition is what the JT path integral hands us. 

In this work, we follow a conjecture that the $\widetilde{V}_{g,n}$'s are analytic functions of their arguments past the $\alpha_i = 1/2$ boundary. If this is indeed the case, then there is a direct analytic continuation between the genus expansion of Lorentzian de Sitter JT gravity and that of~\cite{Saad:2019lba} for Euclidean AdS$_2$. 

In the rest of our discussion let us assume that this continuation holds. Then the Lorentzian dS$_2$ and Euclidean AdS$_2$ genus expansions are related up to a continuation of the dilaton boundary condition $J$, described in Footnote~\ref{FN:dilatonBC}. We then find
\beq
\label{E:summaryZcontinuation}
	Z_{g,n_+,n_-}(\beta_aJ,\beta_mJ) = Z_{g,n}(-i \beta_aJ,i \beta_mJ)\,,
\eeq
where the $Z_{g,n}$'s are the expansion coefficients for Euclidean AdS$_2$ at genus $g$ with $n$ boundaries. Since the $Z_{g,n}$'s may be generated from a matrix integral~\cite{Saad:2019lba}, it follows that the genus expansion de Sitter JT gravity is likewise captured by a matrix integral. Specifically, the matrix integral computes the de Sitter transition amplitudes. The dictionary would be
\begin{equation*}
	Z_{g,n_+,n_-}(\beta_a J,\beta_m J) = \left\langle \text{tr}\left( e^{i \beta_1 H}\right)\hdots \text{tr}\left( e^{i \beta_{n_+}H}\right)\text{tr}\left( e^{-i\beta_{n_++1}H}\right)\hdots \text{tr}\left( e^{-i \beta_nH}\right)\right\rangle_{\rm \overline{MM}, conn, g}\,.
\end{equation*}
The right-hand-side is an average within the putative double-scaled matrix model, ``conn'' denotes the connected part, and we take the genus $g$ contribution in the usual sense of the genus expansion in matrix integrals. See Subsections~\ref{S:partitions} and~\ref{S:HigherGenusContinuation} for further discussion.

Now let us discuss the genus expansion parameter in de Sitter JT gravity. There is a topological term
\beq
	\frac{S_0}{4 \pi }\int d^2x\sqrt{-g} \,R\,, \qquad S_0 = \frac{\varphi_0}{4G_2}\,,
\eeq
in the Lorentzian action. Here $\varphi_0$ is the value of the dilaton on the horizon of the ``static patch.'' It is related to the classical cosmological entropy of the ``static patch'' of nearly dS$_2$ (see Subsection~\ref{S:basicdS2}) as $S_{\rm cosmo} = 2S_0$. Naively this term evaluates to the Euler characteristic of the spacetime. However, as we saw for the Hartle-Hawking geometry, it in fact evaluates to $i \chi_T$ with $\chi_T = 2g+n-2$ the topological characteristic. As a result genus $g$ surfaces with $n$ boundaries contribute to the path integral as $(e^{-S_0})^{2g+n-2}$, with $e^{-S_0}$ the small genus expansion parameter. The connected part of the amplitude between $n_-$ past and $n_+$ future boundaries is then
\beq
	\Psi_{n_+,n_-,\rm conn}(\beta_a J,\beta_m J) \simeq \sum_{g=0}^{\infty} \frac{Z_{g,n_+,n_-}(\beta_aJ,\beta_mJ)}{(e^{S_0})^{2g+n-2}}\,.
\eeq
This series expansion is asymptotic, and so de Sitter JT gravity begs for a non-perturbative completion. Given that the coefficients of its genus expansion are analytic continuations of the coefficients for the genus expansion of Euclidean AdS$_2$, one attractive, (non-unique) non-perturbative completion is the matrix integral of~\cite{Saad:2019lba} which generates the latter coefficients. One would simply probe the double scaled matrix model with insertions of $\text{tr}\left( e^{i \beta H}\right)$ and $\text{tr}\left( e^{-i \beta H}\right)$.

Another way to realize an average over Hamiltonians is with annealed or quenched disorder as in the SYK model \cite{kitaev, Maldacena:2016hyu, Kitaev:2017awl}. It is already known that the SYK model exhibits gravitational physics, in that it has a Schwarzian limit at low temperature and large $N$, as one finds in Euclidean AdS$_2$ \cite{Jensen:2016pah, Maldacena:2016upp, Engelsoy:2016xyb}. So one might wonder whether there is an SYK-inspired model which imitates gravitational physics in nearly dS$_2$. At the level of the matrix model, the continuation suggested by gravity is that a future boundary corresponds to $\text{tr}(e^{- \beta H})$ with the coupling $J$ of SYK analytically continued as $J\to -i J$. This is accomplished by considering
\beq
	\int dJ_{i_1...i_q} \exp\left( -\sum_{i_1< \cdots <i_q}\frac{J_{i_1...i_q}^2}{2\mathcal{J}^2}\right)\left(\int [d\chi] e^{i S_E[\chi;J_{i_1...i_q}]} \right)\,,
\eeq
where the $\chi^i$ are $N$ real quantum mechanical fermions, $S_E[\chi;J]$ is the Euclidean action of SYK on a thermal circle of size $\beta$, and $\mathcal{J}^2 = \frac{(q-1)! J^2}{N^{q-1}}$. This model has a dynamical mean field description with large $N$ equations which can be mapped to those of Euclidean SYK. However, the ensuing model does not seem to bear a resemblance to gravitational physics in nearly dS$_2$. The path integral $Z = e^{-S}$ is purely real, whereas we saw that with a single future boundary the JT path integral is complex, with $Z_{\rm HH} \sim e^{S_0} \int [Df] \exp(i S_{\rm sch}[f])$ and $S_{\rm sch}[f]$ the Schwarzian action. However, it is possible that there is some modification that connects to nearly-dS$_2$ physics.

This all being said, there is significant evidence that the right microscopic completion is in terms of a disordered system like random matrix theory. Consider the transition amplitude with fixed boundary conditions near $\mathcal{I}^{\pm}$, where one might imagine that there is a dual QFT living on $\mathcal{I}^+\cup \mathcal{I}^-$.  Crucially, each connected component of the boundary is endowed with an independently conserved stress tensor. See e.g.~\cite{Balasubramanian:1999re} for a discussion in AdS holography. This fact implies that there is an independent rotation symmetry associated with each component of the boundary, $U(1)^n$ in two dimensions with $n$ boundaries, which is broken down to at most a single $U(1)$ symmetry by the spacetime which connects the $n$ components of the boundary. This in turn implies that, if there is a dual QFT description, there are no local interactions between degrees of freedom on one boundary component and degrees of freedom on another.

However, from the bulk spacetime we infer the existence of nonzero correlations between different components of the boundary. As we just saw, these correlations cannot be mediated by local interactions. For a theory in Euclidean signature, the only options we know concretely are to (i) sacrifice locality (or perhaps for it to be broken at a high energy scale compared to the dS radius), or (ii) consider an ensemble of theories, or perhaps some combination of (i) and (ii). To see that an ensemble may generate such correlations, consider a QFT with two sets of degrees of freedom, $\chi_1$ and $\chi_2$, with a simultaneous coupling $\lambda$ for an operator $\mathcal{O}$,
\beq
	S \supset \lambda \int d\theta_1\, \mathcal{O}_1(\theta_1) + \lambda \int d\theta_2 \, \mathcal{O}_2(\theta_2)\,.
\eeq
Averaging over $\lambda$ with annealed Gaussian disorder\footnote{One can also implement quenched disorder, but the corresponding calculation is more subtle.} and integrating it out, the combined effective action now has non-local interactions
\beq
	S \supset \left( \int d\theta_1\, \mathcal{O}_1(\theta_1)\right)\left( \int d\theta_2\, \mathcal{O}_2(\theta_2)\right)\,.
\eeq
Said another way, certain nonlocal interactions may be understood as coming from joint disorder averages over ensembles of local theories. Furthermore, an appealing feature of an ensemble is that it seems broad enough to encapsulate not just a boundary theory on say a single circle, or two circles, but instead to describe an arbitrary number of components for $\mathcal{I}^{\pm}$ all at once, for instance through averages of the form $\langle \text{tr}(e^{i\beta_1 H}) \cdots \text{tr}(e^{-i\beta_nH})\rangle_{\rm MM}$.

There have been previous hints (see e.g.~\cite{anninos2016cosmic}) in the literature that the holographic dual to de Sitter gravity is a disordered system. Certainly, when one computes cosmological correlators one averages over boundary conditions on $\mathcal{I}^+$, which can be thought of as integrating over sources for a dual CFT. The indications for disorder here are of a related, but somewhat different vein, since they are visible even before performing an average over boundary conditions. 

In the last part of the present paper we studied pure gravity in three-dimensional de Sitter spacetime. Three-dimensional gravity is also topological, with the Chern-Simons formulation playing the same role as the $BF$ theory in two dimensions. In Lorentzian AdS$_3$ it was recently appreciated in~\cite{Cotler:2018zff} (see also~\cite{barnich2017geometric}), correcting previous work~\cite{Coussaert:1995zp}, that one can obtain the path integral for the boundary gravitons of AdS$_3$. The resulting theory is, in a sense, the quantization of the Schwarzian theory that arises in nearly Euclidean AdS$_2$. The integration space of the latter is a coadjoint orbit of the Virasoro group, Diff$(\mathbb{S}^1)/PSL(2;\mathbb{R})$, which is symplectic, and the Schwarzian path integral is morally of the form $\int [dq\, dp] e^{-H}$. The integration space on the boundary of global AdS$_3$ is, at constant time, (two copies of) the space Diff($\mathbb{S}^1)/PSL(2;\mathbb{R})$, and its path integral is morally of the form $\int [dq(t)dp(t)] e^{i \int dt (p\dot{q} - H)}$~\cite{Alekseev:1988ce}. Theories of this kind, built from a phase space $\mathcal{M}$, are sometimes called the path integral quantization of $\mathcal{M}$. (Indeed, for $\mathcal{M}$ that admit quantization, this is the path integral version of geometric quantization (see e.g.~\cite{woodhouse1997geometric,Verlinde:1989ua,Verlinde:1989hv}).)

The story for dS$_3$ parallels that of Lorentzian AdS$_3$. The model on the boundary of dS$_3$ is a path integral quantization of a complexified version of the Schwarzian model on the boundary of nearly dS$_2$, where ``time'' is one of the Euclidean directions at future infinity. Borrowing heavily from the techniques two of us used to study AdS$_3$~\cite{Cotler:2018zff}, above we obtained boundary path integrals for an inflating patch of dS$_3$ as well as global dS$_3$, and computed the one-loop approximation to the path integral for each. 

A nice technical result is the path integral for global dS$_3$, Eq.~\eqref{E:Zresultglobal1}. This result appears to factorize into contributions coming from each boundary, which may be somewhat surprising. The theory on the boundary of global dS$_3$ has a reparameterization mode on each boundary, corresponding to the boundary gravitons, along with a monodromy degree of freedom. (The path integral quantization resembles that of Chern-Simons theory on the annulus~\cite{Elitzur:1989nr}, since the space at fixed latitude on the constant-time slice sphere is a topological annulus.) The reparameterization modes are not coupled by local interactions, but they are coupled to each other through the monodromy. Further, all fields are subject to a redundancy that simultaneously involves the degrees of freedom living on $\mathcal{I}^{\pm}$. Accordingly, this model fails to ``factorize'' between the two boundaries, in a related way as nearly AdS$_2$ gravity~\cite{Harlow:2018tqv}.

However, this non-factorization was previously known.  It is a general feature in Chern-Simons theory with compact gauge group $G$ and level $k$ on the annulus, which is equivalent to a $G$ WZW model at level $k$~\cite{Elitzur:1989nr}. The Hilbert space of the model is $\bigoplus_{\lambda} \mathcal{H}_{\lambda}\otimes \mathcal{H}_{\bar{\lambda}}$, where $\lambda$ labels the unitary highest-weight Kac-Moody representations with group $G$ and level $k$, and $\bar{\lambda}$ is the conjugate representation. This Hilbert space clearly does not factorize: left-movers in the representation $\lambda$, which live on one boundary, are tied to right-movers in the conjugate representation, living on the other boundary.

This non-factorization is important to keep in mind when interpreting the path integral for global dS$_3$ Eq.~\eqref{E:Zresultglobal1}. There, we effectively restricted ourselves to a single representation through our boundary condition on the monodromy, which in the language of our dS$_2$ analysis would correspond to fluctuating around $\alpha = 1$. If we had restricted to a more general $\alpha$, we would have instead found
\beq
	Z = \left( |q|^{-\frac{C\alpha^2 + 1}{12}} \prod_{n=1}^{\infty}\frac{1}{|1-q^n|^2}\right)^2\,,
\eeq
which is the character of a highest-weight representation of two copies of Virasoro with central charge $c = C+1$ and dimension $h =\bar{h}= \frac{C(1-\alpha^2)}{24}$. This is of course consistent with the result for compact $G$ mentioned above.

We have emphasized that nearly dS$_2$ and dS$_3$ gravity can be regarded as suitable analytic continuations of Euclidean nearly-AdS$_2$ and Euclidean AdS$_3$ gravity. We extensively used this continuation to interpret the backgrounds we summed over in the nearly dS$_2$ genus expansion. However, we stress that the de Sitter transition amplitudes may \emph{not} be automatically obtained by continuation of the Euclidean path integral. Indeed, while the ``disk'' amplitude of dS$_2$ JT gravity follows from continuation, we saw that subtleties arise beyond the disk. Already for global dS$_2$, the path integral depends on a choice of integration measure for the variable $\alpha$, and depending on that choice the de Sitter transition amplitude matches the continuation from Euclidean signature up to a sign. The situation is even more intricate at higher genus or with more boundaries. Our implicit point of view throughout is that to compute dS observables we must still perform the honest dS computation. This is in accord with the analysis in \cite{witten2011analytic, harlow2011analytic}.

%***********************************************
\subsection{Discussion}
%***********************************************

Let us conclude with an outlook on some directions for future study.

One striking feature of our boundary path integrals is that the bulk Lorentzian time is emergent. While the boundary theories are Euclidean, they describe real-time gravitational contributions to scattering. This is particularly emphatic in global de Sitter, in which we can compute scattering from past to future infinity.  Of course, emergent time is a prominent feature of various proposals for the dS/CFT correspondence \cite{Strominger:2001pn, Witten:2001kn, Alishahiha:2004md,susskind2009census, sekino2009census}.  In light of recent developments in AdS/CFT which explore the emergence of space from entanglement in the dual CFT (see e.g. \cite{van2010building, ryu2006holographic, czech2012gravity, maldacena2013cool, faulkner2013quantum, lashkari2014gravitational, faulkner2014gravitation, almheiri2015bulk, jafferis2016relative, dong2016reconstruction, gao2017traversable}), it should be fruitful to import these ideas to study the emergence of time in de Sitter. There have been suggestions over the years that time in global de Sitter arises from some appropriate notion of entanglement between the past and future boundary theories (see, for instance, \cite{balasubramanian2003exploring, de2016holographic, cotler2018superdensity, cotler2018quantum}).  The basic intuition is that the global de Sitter Penrose diagram resembles the AdS BTZ black hole Penrose diagram rotated by 90 degrees, and that one may study the emergence of space in the latter through AdS/CFT \cite{van2010building}. In the absence of an example of a dS/CFT duality where the gravitational theory lives on global dS, sharper analyses may be premature. Nevertheless, the boundary path integrals in nearly dS$_2$ and dS$_3$ give a direct boundary description of the gravitational degrees of freedom and so may provide a fertile testing ground in the interim. In any case, it appears there are immediate opportunities to import recent work at the intersection of quantum information and quantum gravity into de Sitter. See~\cite{dong2018sitter} for some recent steps in this direction.

Our nearly-dS$_2$ and dS$_3$ boundary graviton theories possess a diagrammatic expansion which computes the gravitational corrections to scattering. We discussed the dictionary for nearly dS$_2$ gravity in Subsection~\ref{S:scattering}. There is a corresponding story for dS$_3$, although we did not present it in the main text. Recall that the model on the boundary of an inflating patch of dS$_3$ is just the analytic continuation of a model on the boundary of the Poincar\'e patch of Euclidean AdS$_3$ under $L_{\rm AdS} \to i L_{\rm dS}$. The model arising from Euclidean AdS$_3$ computes Virasoro identity blocks \cite{fitzpatrick2014universality, fitzpatrick2015virasoro, perlmutter2015virasoro, fitzpatrick2016conformal, beccaria2016virasoro, chen2017degenerate, Cotler:2018zff,collier2018quantum}, and provides another method to calculate Virasoro Wilson lines~\cite{Bhatta:2016hpz,Besken:2016ooo,Fitzpatrick:2016mtp,Besken:2017fsj,Hikida:2018dxe,Besken:2018zro}. Under analytic continuation, these blocks encode gravitational corrections to correlation functions on $\mathcal{I}^+$. One could also couple our theories to matter fields, as in \cite{blommaert2019clocks}.  It would also be interesting to study quantum supergravity in nearly-dS$_2$ and dS$_3$ with a similar approach as we used above.  We expect SUSY to be broken, as in \cite{bergshoeff2015pure}, although perhaps not so badly. The natural guess is that for nearly dS$_2$ supergravity one finds the continuation of the super-Schwarzian theory~\cite{Fu:2016vas} under $C\to i C$ with $C$ the coupling in front of the Schwarzian action. In dS$_3$ one would expect to find a continuation of a path integral quantization of the super-Virasoro group~\cite{Bakas:1988mq,Aratyn:1989qq,Delius:1990pt,Cotler:2018zff}. If that is the case, then these boundary models would secretly be supersymmetric, but under new non-Hermitian supercharges on account of the continuation. 

It appears promising that our higher genus analysis of dS$_2$ gravity may shed light on an analogous genus expansion of dS$_3$.  Of course, the latter is markedly more complicated -- for instance, the asymptotic spacelike surfaces can be higher-genus Riemann surfaces, and the intermediate interpolating regions are characterized by an appropriate moduli space of 3-manifolds.  Nonetheless, several lessons from the dS$_2$ case may carry over.  For example, a suitable analytic continuation of Euclidean AdS$_3$ higher genus partition functions, along the lines explained in Section~\ref{S:dS3} above, should yield corresponding higher genus dS$_3$ wave functions of the universe.  However, almost all of the work on Euclidean AdS$_3$ higher genus partition functions (see e.g.~\cite{krasnov2000holography, gaiotto2007genus, yin2007partition, Giombi:2008vd, skenderis2011holography, Maloney:2015ina}) examines higher genus asymptotic 2-surfaces, but not higher genus interpolating 3-manifolds.  It is as of yet unclear how to incorporate such 3-manifolds systematically.  One may envision such a genus expansion as coming from the genus expansion of some (possibly scaled) matrix quantum mechanics, which describes a non-unique, non-perturbative completion of both pure gravity in AdS$_3$ and dS$_3$, akin to the 2d case explained in this paper.

Coadjoint orbits of infinite dimensional groups and their quantization provide a unified description of the boundary models for nearly dS$_2$ and dS$_3$ gravity obtained in this work, as well as for nearly AdS$_2$ and pure AdS$_3$ gravity \cite{Stanford:2017thb, barnich2017geometric, Mertens:2018fds, Cotler:2018zff}. In these settings the gravitational dynamics has no bulk modes, only having boundary and topological degrees of freedom. The boundary modes have an integration space which, in two dimensions, is precisely a coadjoint orbit of an infinite dimensional group. It is well-known that Hamiltonian flow on coadjoint orbits can be interpreted as fluid equations \cite{khesin2008geometry, arnold1999topological}. The connection to fluid dynamics is no accident.  Fluid dynamics captures conservation laws, and these coadjoint orbit path integrals are ultimately theories of the boundary stress tensor. 

It is sometimes said that classical gravity can be thought of as the hydrodynamics of some microscopic degrees of freedom comprising spacetime. In this manuscript we have arrived at path integrals for the ``hydrodynamics'' of low-dimensional de Sitter spacetimes. These theories may perhaps be viewed as lying between the macroscopic and microscopic, in a realm we can refer to as ``mesoscopic,'' to borrow terminology from condensed matter physics. While the Schwarzian path integral and the coadjoint orbit quantizations are UV-finite, and so do not require UV-completion in the usual sense of effective field theory, they clearly do not give a complete accounting of the physics one expects to find in a consistent theory of quantum gravity. For example, in JT gravity one finds a different Schwarzian model on the boundary of spacetimes of different topology, and the genus expansion sums over these different theories. Further, the spectrum of states one finds from these theories is continuous, and so does not resolve the black hole spectrum into discrete microstates. Rather, these boundary path integrals compute loop corrections around a fixed topology, giving a course-grained description at energies well below the Planck scale. 

Beyond perturbation theory we have a proposed matrix model, which would give a non-perturbative completion of nearly dS$_2$ gravity. It encapsulates topology change among other features. Crucially, the matrix model we arrived at is the same as the one relevant for nearly AdS$_2$ gravity in~\cite{Saad:2019lba}. In the AdS$_2$ setting one ``inserts'' a boundary of size $\beta$ by probing the matrix model with $\text{tr}(e^{-\beta H})$. To make contact with de Sitter, inserting a future boundary corresponds to probing the matrix model with $\text{tr}(e^{i\beta H})$ and inserting a past boundary corresponds to probing with $\text{tr}(e^{-i \beta H})$. 

There is some evidence~\cite{Saad:2019lba} that this matrix model is dual to a minimal string, given by a non-unitary $(2,p)$ minimal model coupled to Liouville theory and worldsheet gravity in the $p\to \infty$ limit. The claim of~\cite{Saad:2019lba} is that insertions of $\text{tr}(e^{-\beta H})$ in the matrix model, i.e.~insertions of asymptotically AdS$_2$ boundaries, correspond to studying the minimal string on surfaces with boundary components of fixed lengths $\beta_i$. With this in mind one might wonder whether the putative matrix model description of nearly dS$_2$ gravity has a minimal string dual. We intend to return to this question in future work.

%***********************************************

\subsection*{Acknowledgements}

We would like to thank A. Altland, A.~Belin, P. Saad, S. Shenker, G. Turiaci, E.~Witten, and especially D. Stanford for useful and enlightening discussions. K.~J. would like to thank the organizers of the program ``Chaos and Order'' at the Kavli Institute for Theoretical Physics for their hospitality during which significant work on this project was done. K.~J. would like to thank the organizers of the workshop ``Advances in QFT'' at CERN for the same. J. C. is supported by the Fannie and John Hertz Foundation and the Stanford Graduate Fellowship program.  K.~J. is supported in part by the Department of Energy under grant number DE-SC0013682.  A.M. acknowledges the support of the Natural Sciences and Engineering Research Council of Canada (NSERC), funding reference number SAPIN/00032-2015. This work was supported in part by DOE award DE-SC0019380 (J.C.), and also supported in part by a grant from the Simons Foundation (385602, A.M.).

%***********************************************

\bibliographystyle{JHEP}
\bibliography{refs}

\end{document}